\documentclass{aa}
\newcommand\arcdeg{\mbox{$^\circ$}\xspace} 
\usepackage{latexsym}		
\usepackage{graphicx}		
\usepackage{rotating}		
\usepackage{natbib}  
\usepackage{savesym}
\usepackage{pdflscape}
\usepackage{amssymb}
\usepackage{morefloats}
\savesymbol{doublespace}
\usepackage{xspace}
\usepackage{color}
\usepackage{multicol}
\usepackage{mdframed}
\usepackage{url}
\usepackage{subfigure}
\usepackage{lscape}
\usepackage{grffile}
\usepackage{standalone}
\standalonetrue
\usepackage{import}
\usepackage[utf8]{inputenc}
\usepackage{longtable}
\usepackage{supertabular}
\usepackage{booktabs}
\usepackage[yyyymmdd,hhmmss]{datetime}
\usepackage{fancyhdr}

\newcommand{\msun}{\ensuremath{M_{\odot}}\xspace}			

\newcommand{\lsun}{\ensuremath{L_{\odot}}\xspace}			
\newcommand{\hh}{\ensuremath{\textrm{H}_{2}}\xspace}			

\newcommand{\formaldehyde}{\ensuremath{\textrm{H}_2\textrm{CO}}\xspace}

\newcommand{\methanol}{\ensuremath{\textrm{CH}_3\textrm{OH}}\xspace}
\newcommand{\ortho}{\ensuremath{\textrm{o-H}_2\textrm{CO}}\xspace}
\newcommand{\para}{\ensuremath{\textrm{p-H}_2\textrm{CO}}\xspace}
\newcommand{\oneone}{\ensuremath{1_{1,0}-1_{1,1}}\xspace}
\newcommand{\twotwo}{\ensuremath{2_{1,1}-2_{1,2}}\xspace}

\newcommand{\water}{H$_{2}$O\xspace}		
\newcommand{\uchii}{\ion{UCH}{ii}\xspace}

\newcommand{\hchii}{\ion{HCH}{ii}\xspace}

\newcommand{\hii}{\ion{H}{ii}\xspace}

\newcommand{\kms}{\textrm{km~s}\ensuremath{^{-1}}\xspace}	

\newcommand{\pers}{\ensuremath{\mathrm{s}^{-1}}\xspace}

\newcommand{\percc}{\ensuremath{\textrm{cm}^{-3}}\xspace}

\newcommand{\persc}{\ensuremath{\textrm{cm}^{-2}}\xspace}

\newcommand{\um}{\ensuremath{\mu \textrm{m}}\xspace}    
\newcommand{\microjy}{\ensuremath{\mu\textrm{Jy}}\xspace}    

\newcommand{\ammonia}{NH\ensuremath{_3}\xspace}

\newcommand{\ceighteeno}{\ensuremath{\textrm{C}^{18}\textrm{O}}\xspace}
\def\ee#1{\ensuremath{\times10^{#1}}}

\def\eqref#1{Equation \ref{#1}}

\renewcommand\arcdeg{\mbox{$^\circ$}\xspace} 
 
\renewcommand\arcsec{\mbox{$^{\prime\prime}$}\xspace} 

\newcommand{\todo}[1]{\textcolor{red}{#1}}

\def\Figure#1#2#3#4#5{
\begin{figure*}[!htp]
\includegraphics[scale=#4,width=#5]{#1}
\caption{#2}
\label{#3}
\end{figure*}
}

\def\FigureTwo#1#2#3#4#5#6{
\begin{figure*}[!htp]
\subfigure[]{ \includegraphics[scale=#5,width=#6]{#1} }
\subfigure[]{ \includegraphics[scale=#5,width=#6]{#2} }
\caption{#3}
\label{#4}
\end{figure*}
}

\def\FigureTwoAA#1#2#3#4#5#6{
\begin{figure*}[!htp]
\subfigure[]{ \includegraphics[scale=#5,width=#6]{#1} }
\subfigure[]{ \includegraphics[scale=#5,width=#6]{#2} }
\caption{#3}
\label{#4}
\end{figure*}
}

\newenvironment{rotatepage}%
{}{}

\def\FigureThreeAA#1#2#3#4#5#6#7{
\begin{figure*}[!htp]
\subfigure[]{ \includegraphics[scale=#6,width=#7]{#1} }
\subfigure[]{ \includegraphics[scale=#6,width=#7]{#2} }
\subfigure[]{ \includegraphics[scale=#6,width=#7]{#3} }
\caption{#4}
\label{#5}
\end{figure*}
}

\begin{document}

\title{Toward gas exhaustion in the W51 high-mass protoclusters}
\titlerunning{JVLA observations of W51}
\authorrunning{Ginsburg et al}
\newcommand{\eso}{$^{1}$}
\newcommand{\nrao}{$^{2}$}
\newcommand{\radboud}{$^{3}$}
\newcommand{\allegro}{$^{4}$}
\newcommand{\morelia}{$^{5}$}
\newcommand{\excellence}{$^{6}$}
\newcommand{\casa}{$^{7}$}
\newcommand{\cfa}{$^{8}$}
\newcommand{\lasp}{$^{9}$}
\newcommand{\jodrell}{$^{11}$}
\newcommand{\sofia}{$^{12}$}
\newcommand{\mpia}{$^{13}$}
\newcommand{\iah}{$^{14}$}

\author{
Adam Ginsburg{\eso},
W.~M. Goss{\nrao}, Ciriaco Goddi{\radboud$^{,}$\allegro}, Roberto Galv{\'a}n-Madrid{\morelia}, James E. Dale\excellence, John Bally{\casa}, Cara D.
Battersby{\cfa}, Allison Youngblood{\lasp}, Ravi Sankrit{\sofia}, Rowan Smith{\jodrell}, Jeremy Darling\casa, J.~M.~Diederik Kruijssen{\mpia$^{,}$\iah},
Hauyu Baobab Liu{\eso}
        }

\institute{
    {\eso}{
           \it{
               European Southern Observatory, Karl-Schwarzschild-Stra{\ss}e 2, D-85748 Garching bei München, Germany\\
                      \email{Adam.Ginsburg@eso.org}
               }
           } \\ 
    {\nrao}{\it{National Radio Astronomy Observatory, Socorro, NM 87801 USA}}\\
    {\radboud}{\it{Department of Astrophysics/IMAPP, Radboud University Nijmegen, PO Box 9010, 6500 GL Nijmegen, the Netherlands}} \\
    {\allegro}{\it{ALLEGRO/Leiden Observatory, Leiden University, PO Box 9513, NL-2300 RA Leiden, the Netherlands}} \\
    {\morelia}{\it{Instituto de Radioastronom{\'i}a y Astrof{\'i}sica, UNAM, A.P. 3-72, Xangari, Morelia, 58089, Mexico}} \\
    {\excellence}{\it{University Observatory/Excellence Cluster `Universe' Scheinerstra{\ss}e 1, 81679 M{\"u}nchen, Germany}} \\
    {\casa}{\it{CASA, University of Colorado, 389-UCB, Boulder, CO 80309}} \\ 
    {\cfa}{\it{Harvard-Smithsonian Center for Astrophysics, 60 Garden
               Street, Cambridge, MA 02138, USA}} \\ 
    {\lasp}{\it{LASP, University of Colorado, 600 UCB, Boulder, CO 80309}}\\
    {\sofia}{\it{SOFIA Science Center, NASA Ames Research Center, M/S 232-12, Moffett Field, CA 94035, USA}}\\
    {\jodrell}{\it{Jodrell Bank Centre for Astrophysics, School of Physics and Astronomy, University of Manchester, Oxford Road, Manchester M13 9PL, UK}} \\
    {\mpia}{\it{Max-Planck Institut f{\"u}r Astrophysik, Karl-Schwarzschild-Stra{\ss}e 1, 85748 Garching, Germany}} \\
    {\iah}{\it{Astronomisches Rechen-Institut, Zentrum f{\"u}r Astronomie der Universit{\"a}t Heidelberg, M{\"o}nchhofstra{\ss}e 12-14, 69120 Heidelberg, Germany}}
    }

\date{Date: \today ~~ Time: \currenttime}

\abstract
{
We present new JVLA observations of the high-mass cluster-forming region W51A
from 2 to 16 GHz with resolution
$\theta_{fwhm}\approx0.3-0.5\arcsec$.  The data reveal a wealth of
observational results:
(1) Currently-forming, very massive (proto-O) stars are traced by \ortho
\twotwo emission, suggesting that this line can be used efficiently as a
massive protostar tracer.
(2) There is a spatially distributed population of $\lesssim$mJy continuum sources,
including hypercompact \hii regions and candidate colliding wind binaries,
in and around the W51 proto-clusters.  
(3) There are two clearly detected protoclusters, W51e and W51 IRS2, that are
gas-rich but may have most of their mass in stars within their inner $\lesssim0.05$
pc.  The
majority of the bolometric luminosity in W51 most likely comes from a third
population of OB stars between these clusters.
The presence of a substantial population of exposed O-stars coincident with
a population of still-forming massive stars, along with a direct measurement
of the low mass loss rate via ionized gas outflow from W51 IRS2, together imply
that feedback is ineffective at halting star formation in massive
protoclusters.  Instead, feedback may shut off the large-scale accretion of
diffuse gas onto the W51 protoclusters, implying that they are evolving towards
a state of gas exhaustion rather than gas expulsion. Recent theoretical models
predict gas exhaustion to be a necessary step in the formation of
gravitationally bound stellar clusters, and our results provide an
observational validation of this process.
}

\maketitle

\section{Introduction}
\footnote{
This paper and all related analysis code are available on the web at
https://github.com/adamginsburg/paper\_w51\_evla.
}
The highest-mass clusters forming in our Galaxy have masses
$M\gtrsim\textrm{few}\ee{4}$ \msun \citep{Portegies-Zwart2010a}.
These rare regions mark the high end of the protocluster mass function.
Recent surveys have provided a complete census of all protoclusters in this
high mass range \citep{Ginsburg2012a,Urquhart2013b,Urquhart2014b}.  As the
most massive and dense star-forming regions in the Galaxy, these locations
provide an important local analogue to the early universe, where stars more
commonly formed in clusters and generally formed in denser gas
\citep{Kruijssen2012a,Madau2014a}.

In such environments, feedback from massive stars is an important factor.  Most star
formation within high-mass clusters is likely to occur after at least one
high-mass star has ignited.  Many open questions remain about the effects of
this feedback, as the physical complexity of such regions renders them
difficult to simulate.  A crucial question is whether the clusters complete the
star formation process by consuming most of their gas (exhaustion) or
by removing it via feedback 
\citep[expulsion;][]{Kruijssen2012b,Longmore2014a,Matzner2015a}.
The balance between the two processes is expected to depend on the gas volume
density because higher-density gas clumps are capable of evolving through a
larger number of free-fall times before the gas supply is shut off by feedback,
and thus higher-density clumps attain larger star formation efficiencies
\citep{Kruijssen2012a}.
This distinction therefore governs whether the forming stars will be observable as a
bound cluster or an expanding OB association.
When integrated over the density probability distribution function of the
interstellar medium in a galaxy, this argument predicts that the fraction of
star formation occurring in bound clusters increases with the gas pressure and
surface density, as is indeed observed \citep{Goddard2010a,Kruijssen2012a,Adamo2015a}.

W51 contains two of the best candidates for protoclusters that will experience
gas exhaustion.  This region is already extremely luminous both bolometrically
and in the radio continuum.  Of all the star forming regions included in the
Red MSX Survey, which is complete to sources with infrared luminosity $L>10^4$
\lsun, W51 is the most luminous \citep{Urquhart2014a} and among the closest at
a parallax-measured distance of 5.41 kpc \citep{Sato2010a,Xu2009a}.  The total
luminosity of the W51 protocluster complex has been estimated using IRAS and
KAO, $L_{bol}\sim9.3\ee{6} (D/5.4\mathrm{kpc})^2$ \lsun
\citep{Harvey1986a,Sievers1991a}.  The pair of clusters in W51 providing this
luminosity, W51 Main and W51 IRS2, each contain $\sim5\ee{4}$ \msun of
molecular gas \citep{Ginsburg2012a}.

As one of the most luminous star-forming regions in our Galaxy, W51 has been
well-studied in the radio and millimeter.  
Previous investigations have shown that the protoclusters within W51 contain many
forming massive stars. Five high-mass protostars have been identified:
W51 North, W51e8, W51e2w, W51e2e, W51e2nw, and W51d2
\citep{Zhang1997a,Keto2008b,Zapata2008a,Zapata2009a,Zapata2010a,Shi2010b,Shi2010a,Surcis2012a,Goddi2015a,Goddi2016a}.
Another handful have already reached the main sequence and are visible in the
near infrared
\citep{Goldader1994a,Okumura2000a,Kumar2004a,Barbosa2008a,Figueredo2008a}.

W51 contains many distinct \hii regions.  These are classified as extended
\hii regions ($\sim10$ pc), ultracompact \hii regions
(\uchii's; $r\sim0.1$ pc), or hypercompact \hii regions \citep[\hchii's;
$r<0.05$ pc][]{Kurtz2002a}.  About ten \hchii regions, which are the
youngest \hii region class and likely contain accreting proto-OB stars
\citep{Peters2010c}, have been found in W51 \citep{Mehringer1994a}.  However,
the census of both massive stars and protostars in W51 is incomplete.

We present observations of the inner few parsecs of the W51 protocluster
forming region.  We refer to the whole region, including W51 Main (which
contains the W51e sources) and W51 IRS2 as `W51A'.  There is a shell
centered on W51 Main that has a particularly bright northwest edge that is 
known as W51 IRS1.  Other names are explained
when they are used.  An overview of the nomenclature used, including labels, is
in Figures \ref{fig:coverview_pointsrcs} and \ref{fig:coverview_diffuse}.

In Section \ref{sec:observations}, we describe the observations and data
reduction.  Section \ref{sec:results} presents the observational results,
summarizing key features identified in the images and describing the
point-source photometry (\S \ref{sec:pointsources}).  Section
\ref{sec:analysis} provides analysis and interpretation of the observational
results.  Section \ref{sec:discussion} discusses the implications of the
observations.  Section \ref{sec:conclusion} states the conclusions.

Finally, we include the following appendices: Appendix
\ref{sec:appendix_labels} provides figures labeling named objects,
\ref{sec:SEDs} provides cutout images and SED plots for all point sources,
\ref{sec:obsmeta} includes tables of observational metadata, and
\ref{sec:vfield} shows the velocity field of the molecular gas.

\section{Observations}
\label{sec:observations}
We used the Karl G. Jansky Very Large Array (JVLA) in multiple bands and
configurations. 
The observations and resulting images are summarized in Table
\ref{tab:observations}.  The complete observational metadata, including times
of observation and program IDs, are in Table \ref{tab:obs_meta}.  In project
12B-365, we observed in A-array in S and C bands (3 and 5 GHz, approximately)
with 2 GHz total bandwidth.
In project 13A-064, we observed in C-band in C (1h) and A (5h) arrays and in
Ku-band in D (1h) and B (5h) arrays.
We also include continuum data from project 12A-274
\citep{Goddi2015a,Goddi2016a} covering the range 25-36 GHz using the JVLA B
array configuration.  

Our spectral coverage included \ortho
\oneone at 4.82966 GHz and \twotwo at 14.488 GHz with 0.3 \kms resolution and the
radio recombination lines (RRLs) H77$\alpha$ (14.1286 GHz) and H110$\alpha$
(4.8741 GHz) at 1 \kms resolution.  The H110$\alpha$ line had lower S/N than
the H77$\alpha$ line but was otherwise similar; it is not discussed further
in this paper.

Data reduction was performed using CASA\footnote{\url{http://casa.nrao.edu}}
\citep{McMullin2007a}.  The pipeline-calibrated products were
used, then imaging was performed using CLEAN.  For most images discussed here, we
used uniform weighting.  The reduction scripts are included in a repository
\url{https://github.com/adamginsburg/w51evlareductionscripts}.

\begin{table*}[htp]
\caption{Observations}
\begin{tabular}{cccccccc}
\label{tab:observations}
Epoch & Frequency & BMAJ & BMIN & BPA & Noise Estimate & Dynamic Range & Jy-Kelvin \\
 & $\mathrm{GHz}$ & $\mathrm{{}^{\prime\prime}}$ & $\mathrm{{}^{\prime\prime}}$ & $\mathrm{{}^{\circ}}$ & $\mathrm{mJy\,beam^{-1}}$ &  &  \\
\hline
2 & 2.5 & 0.54 & 0.53 & -175 & 0.20 & 54 & 6.8\ee{5} \\
2 & 3.5 & 0.41 & 0.39 & 0 & 0.06 & 190 & 6.3\ee{5} \\
1 & 4.9 & 0.45 & 0.38 & 88 & 0.18 & 150 & 3\ee{5} \\
2 & 4.9 & 0.32 & 0.29 & -84 & 0.04 & 320 & 5.6\ee{5} \\
3 & 4.9 & 0.33 & 0.26 & 68 & 0.06 & 200 & 5.9\ee{5} \\
3 & 5.5 & 0.29 & 0.22 & 80 & 0.03 & 420 & 6.4\ee{5} \\
2 & 5.9 & 0.27 & 0.23 & -81 & 0.03 & 460 & 5.7\ee{5} \\
3 & 5.9 & 0.31 & 0.19 & 74 & 0.03 & 420 & 6\ee{5} \\
1 & 8.4 & 0.47 & 0.39 & 82 & 0.08 & 860 & 9.5\ee{4} \\
2 & 12.6 & 0.38 & 0.35 & 83 & 0.08 & 1200 & 5.7\ee{4} \\
2 & 13.4 & 0.34 & 0.33 & 14 & 0.05 & 1800 & 6\ee{4} \\
2 & 14.1 & 0.34 & 0.33 & 68 & 0.09 & 1200 & 5.6\ee{4} \\
1 & 22.5 & 0.32 & 0.29 & -84 & 0.57 & 280 & 2.6\ee{4} \\
2 & 25.0 & 0.28 & 0.24 & -4 & 0.64 & 300 & 2.8\ee{4} \\
2 & 27.0 & 0.25 & 0.22 & 56 & 1.13 & 140 & 3.1\ee{4} \\
2 & 29.0 & 0.23 & 0.21 & 63 & 1.32 & 86 & 3\ee{4} \\
2 & 33.0 & 0.21 & 0.18 & 54 & 1.11 & 130 & 2.9\ee{4} \\
2 & 36.0 & 0.18 & 0.17 & 74 & 1.23 & 72 & 3.1\ee{4} \\
\hline
\end{tabular}
\par
Jy-Kelvin gives the conversion factor from Jy to Kelvin given the synthesized beam size and observation frequency.
\end{table*}

We did not achieve a thermal-noise-limited image in any of the 13A-064 data.
For example, in the C-band A-array data, the thermal noise is $\sim5$ \microjy,
while the achieved RMS is $\sim30-60$ \microjy depending on location in the
image.  There is a large amount of resolved-out structure in the images that
hinders cleaning.  The images are generally artifact-dominated at the
low-signal end, where inadequately cleaned sidelobes of bright features provide
significant non-gaussian noise.  Some of these artifacts can be seen in the C-band
image, Figure \ref{fig:coverview_pointsrcs} (or Figure \ref{fig:coverview}
without annotations).

Despite the deficiency in the image quality, the new observations are
$\sim5\times$ more sensitive than any previous observations.  The new
observations also cover a much broader range of frequencies than those
performed previously with the VLA.  Table \ref{tab:observations} gives the
estimated noise in each image and an estimate of the dynamic range, which is
the peak intensity in the image divided by the noise measured in a signal-free
region of the image.  

Figures \ref{fig:kuoverview} and \ref{fig:coverview_pointsrcs} show the
best-quality
images in the Ku-band and C-band combining the short and long array
configurations.  These images were made from a combination of the full
bandwidth in their respective bands. They were used in the figures in the main
text of the paper and for source identification but not for photometry.
Instead, for the 13A-064 data, individual 1 GHz bands including only the
longer-baseline configuration were used to measure the intensity of the
identified sources.

For the spectral line analysis, we used the cubes listed in Table
\ref{tab:cubes}.  For the Ku-band naturally-weighted cube, the noise is higher
than in the robust-weighted cube because large angular scale artifacts are
prevalent, especially at the image edges.

In some figures, we show unsharp-masked versions of the data.  These are
equivalent to images made excluding short baselines, but we instead performed
unsharp masking (subtracting a gaussian-smoothed version of the image from
itself) in the image domain for computational convenience.

\begin{table*}[htp]
\caption{Spectral Cubes}
\begin{tabular}{lllllllll}
\label{tab:cubes}
Cube ID & Frequency & Channel Width & RMS & RMS$_K$ & BMAJ & BMIN & BPA & Jy-Kelvin \\
 & $\mathrm{GHz}$ & $\mathrm{km\,s^{-1}}$ & $\mathrm{mJy\,beam^{-1}}$ & $\mathrm{K}$ & $\mathrm{{}^{\prime\prime}}$ & $\mathrm{{}^{\prime\prime}}$ & $\mathrm{{}^{\circ}}$ &  \\
\hline
1-1 Uniform & 4.829660 & 0.50 & 1.03 & 63 & 0.51 & 0.35 & 84 & 6.1\ee{4} \\
1-1 Natural & 4.829660 & 0.50 & 0.56 & 1.6\ee{2} & 1.06 & 0.81 & 52 & 2.9\ee{5} \\
H$77\alpha$ Briggs 0 & 14.128610 & 1.33 & 0.33 & 12 & 0.41 & 0.42 & 13 & 3.6\ee{4} \\
2-2 Natural & 14.488479 & 0.50 & 1.28 & 7.6 & 1.13 & 0.87 & 57 & 5.9\ee{3} \\
2-2 Briggs 0 & 14.488479 & 0.50 & 0.50 & 17 & 0.43 & 0.42 & 20 & 3.3\ee{4} \\
\hline
\end{tabular}
\par
Jy-Kelvin gives the conversion factor from Jy to Kelvin given the synthesized beam size and observation frequency.
\end{table*}

\Figure{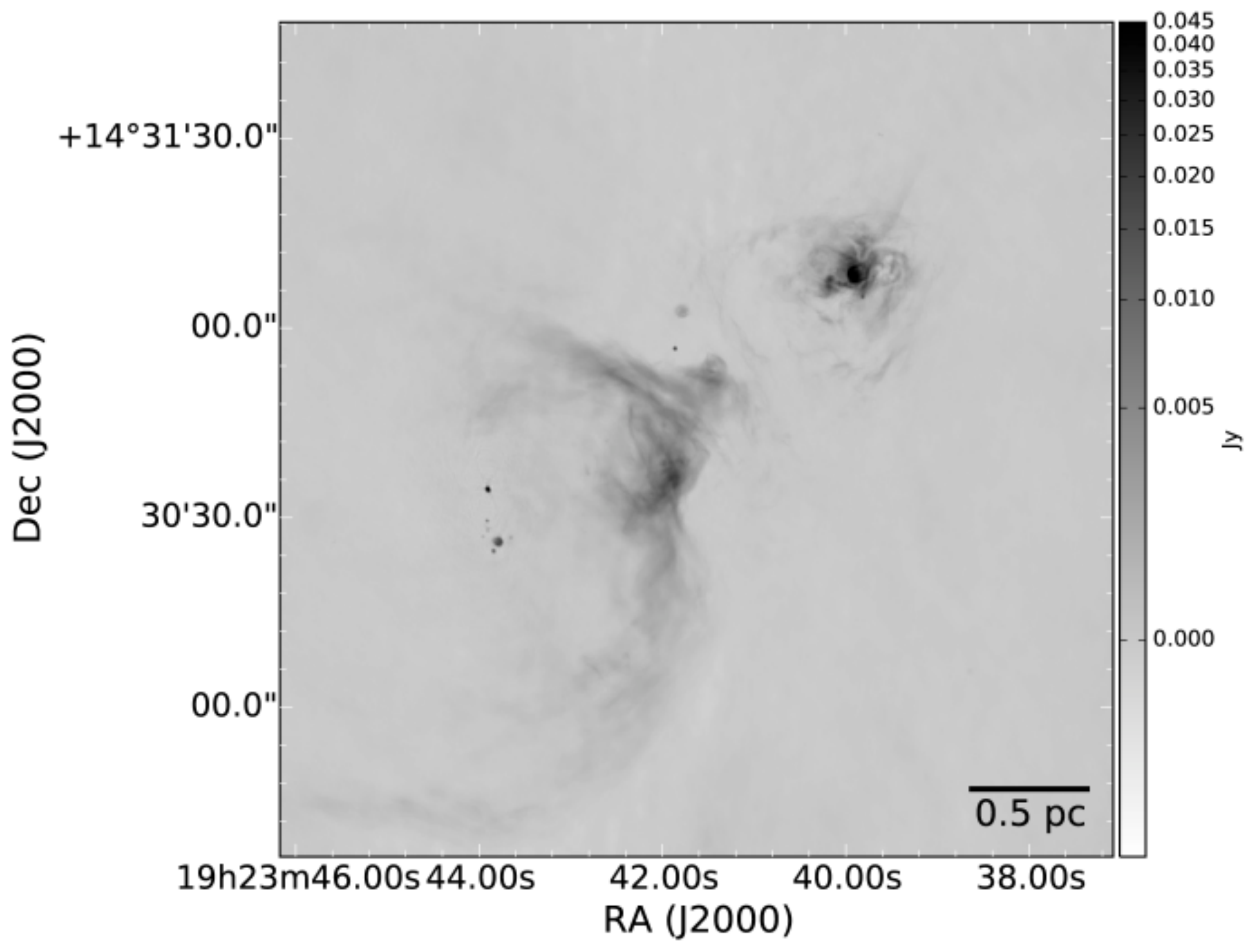}
{The Ku-band (14.5 GHz, tracing ionized gas) image of the W51 region produced
with a combination of JVLA B and D array data using uniform weighting.  The
synthesized beam size is
$\theta_{FWHM}\approx0.33\arcsec$.
The compact, circular region in the upper-right corner is IRS2, while the more
diffuse bright region in the center is IRS1, marking the edge of the W51 Main
shell.  Compact sources are identified
in Figure \ref{fig:coverview_pointsrcs}.
}
{fig:kuoverview}{0.9}{6.5in}

\Figure{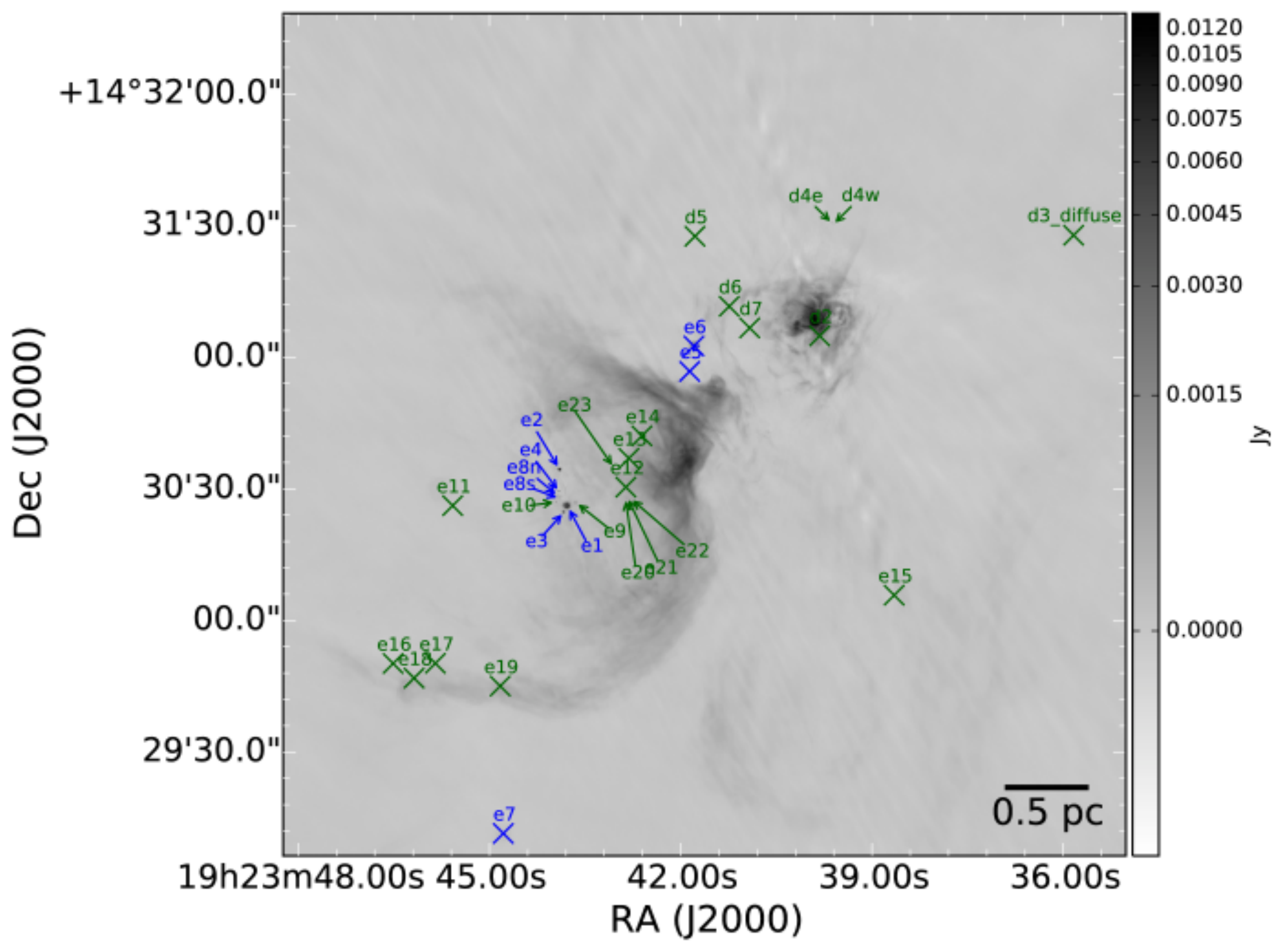}
{The C-band (4.9 GHz, tracing ionized gas) image of the W51A region with point sources labeled.  The
blue sources are previously known; the green are new additions in this paper.
The synthesized beam size is $\theta_{FWHM}\approx0.3\arcsec$.
}
{fig:coverview_pointsrcs}{0.9}{6.5in}

\section{Observational Results}
\label{sec:results}
We report seven distinct observational results: 
\begin{enumerate}
    \item Section \ref{sec:pointsources}: The detection of new continuum
        sources, including hypercompact \hii regions and probable
        colliding wind binaries. 
    \item Section \ref{sec:variability}: The detection of variability in
        some of the faint point sources over a 20 year timescale.
    \item Section \ref{sec:associations}: The detection of radio continuum
        sources associated with near-infrared sources, most likely un-embedded
        O-stars.
    \item Section \ref{sec:twotwoemission}: The detection of \formaldehyde
        \twotwo emission around sources e2, e8, and W51 North.
    \item Section \ref{sec:LOSvelo}: Measurements of line-of-sight velocities
        toward many ultracompact \hii regions using H77$\alpha$ and/or
        \formaldehyde.
    \item Section \ref{sec:diffuseemission}: Detection of Orion-bar-like sharp
        edges to the W51 Main / W51 IRS1 \hii region, most likely tracing 
        photon-dominated regions (PDRs).
    \item Section \ref{sec:lacyjet}: Detection of a $\sim0.1$ pc ionized
        outflow in H77$\alpha$ that had previously been identified in
        mid-infrared ionized line emission.
\end{enumerate}

\subsection{Continuum sources and photometry}
\label{sec:pointsources}
We report new detections of several sources and concrete
identifications of others that were detected in previous data sets but not
reported.

We follow the naming scheme introduced by \citet{Mehringer1994a}.  For the
compact ($r<1\arcsec$) sources within 1 arcminute of W51e2, we use the name
W51e followed by a number.  We identify two new sources, e9  and e10, which
were previously detected but never named.  We split source e8 into a north and
south component, plus a more extended molecular component e8mol.  We also
identify a molecular component between e1, e8, and e10: e10mol.  The source
positions and approximate radii for resolved continuum sources are shown in Figure
\ref{fig:coverview_pointsrcs} and listed in Table \ref{tab:positions}.
The positions and radii are estimated by eye; they are not fits to the data.

\begin{table*}[htp]
\caption{Source Positions}
\begin{tabular}{lllrrll}
\label{tab:positions}
Source Name & RA & Dec & Radius & Phys. Radius & SED Class & Classification \\
 &  &  & $\mathrm{{}^{\prime\prime}}$ & $\mathrm{pc}$ &  &  \\
\hline
Between & 19:23:41.47 & 14:30:51.0 & 4.5 & 0.11 & $E$ & - \\
G49.46-0.36 & 19:23:35.12 & 14:29:55.8 & 34.4 & 0.85 & $E$ & - \\
G49.46-0.38 & 19:23:40.18 & 14:29:39.1 & 19.1 & 0.47 & $E$ & - \\
W51 IRS 2 & 19:23:39.89 & 14:31:08.3 & 1.3 & 0.03 & $E$ & - \\
W51 IRS 2 bubble & 19:23:39.90 & 14:31:08.5 & 21.4 & 0.53 & $E$ & - \\
W51 Main 40 kms RRL Shell & 19:23:42.23 & 14:30:40.2 & 7.8 & 0.19 & $E$ & - \\
W51 Main 65 kms RRL shell & 19:23:42.55 & 14:30:43.1 & 10.8 & 0.27 & $E$ & - \\
W51 Main Shell & 19:23:44.68 & 14:30:23.5 & 41.5 & 1.03 & $E$ & - \\
W51G 8 & 19:23:50.69 & 14:32:50.1 & 7.1 & 0.18 & $E$ & - \\
arc & 19:23:41.00 & 14:30:37.3 & 6.0 & 0.15 & $E$ & - \\
d2 & 19:23:39.82 & 14:31:04.9 & - & - & $b,d,n$ & HCHII \\
d3 & 19:23:35.87 & 14:31:27.9 & 1.9 & 0.05 & $E$ & UCHII \\
d4e & 19:23:39.64 & 14:31:30.6 & - & - & $f,v$ & cCWB \\
d4w & 19:23:39.60 & 14:31:30.4 & - & - & $f$ & cCWB \\
d5 & 19:23:41.78 & 14:31:27.6 & - & - & $w$ & cCWB \\
d6 & 19:23:41.24 & 14:31:11.6 & - & - & $f,n$ & cCWB \\
d7 & 19:23:40.92 & 14:31:06.7 & - & - & $f$ & cCWB \\
e1 & 19:23:43.78 & 14:30:26.1 & 0.7 & 0.02 & $E$ & UCHII \\
e2 & 19:23:43.91 & 14:30:34.6 & - & - & $a,d$ & HCHII \\
e3 & 19:23:43.84 & 14:30:24.7 & - & - & $b,d$ & HCHII \\
e4 & 19:23:43.91 & 14:30:29.5 & - & - & $b,d$ & HCHII \\
e5 & 19:23:41.86 & 14:30:56.8 & - & - & $b,d$ & HCHII \\
e6 & 19:23:41.79 & 14:31:02.7 & 1.0 & 0.03 & $E$ & UCHII \\
e7 & 19:23:44.79 & 14:29:11.3 & 1.8 & 0.04 & $E$ & UCHII \\
e8n & 19:23:43.91 & 14:30:28.2 & - & - & $b/w$ & HCHII \\
e8s & 19:23:43.91 & 14:30:27.9 & - & - & $b$ & HCHII \\
e9 & 19:23:43.65 & 14:30:26.8 & - & - & $c,f$ & HCHII \\
e10 & 19:23:43.96 & 14:30:27.0 & - & - & $f$ & HCHII \\
e11 & 19:23:45.58 & 14:30:26.2 & - & - & $b$ & HCHII \\
e11d & 19:23:45.69 & 14:30:29.1 & 3.6 & 0.09 & $E$ & HII \\
e12 & 19:23:42.86 & 14:30:30.4 & - & - & $E$ & - \\
e13 & 19:23:42.81 & 14:30:36.9 & - & - & $b,c,E$ & cCWB/HII \\
e14 & 19:23:42.61 & 14:30:42.1 & - & - & $b,c,E$ & cCWB/HII \\
e15 & 19:23:38.65 & 14:30:05.7 & - & - & $b,c,E$ & UCHII \\
e16 & 19:23:46.51 & 14:29:50.2 & - & - & $E$ & - \\
e17 & 19:23:45.85 & 14:29:50.2 & - & - & $E$ & - \\
e18 & 19:23:46.18 & 14:29:46.9 & - & - & $E$ & - \\
e18d & 19:23:46.18 & 14:29:44.0 & 3.2 & 0.08 & $E$ & HII \\
e19 & 19:23:44.83 & 14:29:45.0 & - & - & $E$ & - \\
e20 & 19:23:42.86 & 14:30:27.6 & - & - & $E$ & - \\
e21 & 19:23:42.83 & 14:30:27.8 & - & - & $E$ & - \\
e22 & 19:23:42.78 & 14:30:27.6 & - & - & $E$ & - \\
e23 & 19:23:43.06 & 14:30:34.9 & - & - & $E$ & - \\
\hline
\end{tabular}
\par
Objects with name e\#d are the diffuse counterparts to point sources.  The absolute positional accuracy is $\sim0.2\arcsec$.  Sources with no radius are unresolved, with upper limits of 0.3\arcsec (0.007 pc).Source names correspond to the labels in Figures \ref{fig:coverview_pointsrcs} and \ref{fig:coverview_diffuse}.\\
$a$: $\nu^2$ dependence \\
$b$: $\nu^1$ dependence \\
$c$: $\nu<5$ GHz excess or negative index \\
$d$: $\nu>15$ GHz flat \\
$f$: $\nu$-independent flat \\
$E$: extended (photometry not trustworthy) \\
$n$: near bright, extended emission (may affect photometry) \\
$v$: Likely variable \\
$w$: Too weak for SED classification \\
\newline \\
HCHII: Hypercompact \hii region \\
UCHII: Ultracompact \hii region \\
HII: Extended \hii region or part of a larger \hii region \\
cCWB: candidate colliding-wind binaries \\

\end{table*}

We include in this catalog any pointlike sources (at $\sim0.2-0.4\arcsec$
resolution) with emission in two bands, Ku and C (14 and 5 GHz; Table
\ref{tab:contsrcs}).  To identify point sources, we used uniformly-weighted
images.  These images exclude much of the extended emission, which makes point
source detection more straightforward.  There is some reason for concern about
this approach, since it can make local peaks of the diffuse emission appear to
be pointlike.  However, for many of the faint and low-significance peaks, we
have found clear near-infrared associations (Section \ref{sec:associations})
and have detected the sources in multiple bands, suggesting that these point
sources are real.

The point source photometry data are in Table \ref{tab:contsrcs}.  The Epoch column
describes the data source: Epoch 1 comes from \citet{Mehringer1994a}, Epoch 2
comes from 12A-274, 12B-365, or 13A-064 (mid 2012-early
2013), and Epoch 3 (C-band only) comes from
13A-064 in 2014. For the multi-configuration combined images, we list the date
of the highest-resolution observations.  Full details of the data provenance
are given in Table \ref{tab:obs_meta}.

There are two intensity columns in Table \ref{tab:contsrcs}.  The first shows
the peak intensity within a circular aperture with a radius equal to the beam
major FWHM.  The second column, `Peak - Background', shows the peak intensity
minus the minimum intensity in a box $6\times6$ beam FWHMs.  This
background-subtracted intensity is meant to account for negative bowling that
affects parts of the images; there are cases in which a bright point source is
seen on a deeply negative background such that the measured peak intensity is
near zero.  When these two values agree, they are reliable, but when they
differ significantly, they are probably affected by image reconstruction
artifacts and the background-subtracted version is preferred.

The RMS column gives the error measured in a signal-free region of the image.
It is approximately the same as that shown in Table \ref{tab:observations},
though slightly different regions are used in each case.

\begin{table*}[htp]
\caption{Continuum Point Sources (excerpt)}
\begin{tabular}{ccccccc}
\label{tab:contsrcs}
Object & Epoch & Obs. Date & Peak $S_{\nu}$ & Peak - Background & RMS & Frequency \\
$\mathrm{}$ & $\mathrm{}$ & $\mathrm{}$ & $\mathrm{mJy\,beam^{-1}}$ & $\mathrm{mJy\,beam^{-1}}$ & $\mathrm{mJy\,beam^{-1}}$ & $\mathrm{GHz}$ \\
\hline
d2 & 2 & 2012-10-16 & 5.5 & 7.7 & 0.2 & 2.5 \\
d2 & 1 & 1993-03-16 & 10 & 10 & 0.5 & 22.5 \\
d3-diffuse & 1 & 1992-10-25 & 0.3 & 0.3 & 0.2 & 4.9 \\
d3-diffuse & 2 & 2012-06-21 & - & - & 1 & 33.0 \\
d4e & 2 & 2013-03-02 & 1 & 1.3 & 0.1 & 12.6 \\
d4w & 2 & 2012-10-16 & 0.51 & 0.79 & 0.05 & 4.9 \\
d4w & 2 & 2012-06-21 & 1 & 4 & 1 & 27.0 \\
d5 & 2 & 2012-10-16 & -0.03 & 0.11 & 0.04 & 5.9 \\
d6 & 2 & 2012-10-16 & 1 & 1.4 & 0.2 & 2.5 \\
d6 & 1 & 1993-03-16 & 0.6 & 1.4 & 0.5 & 22.5 \\
d7 & 3 & 2014-04-19 & 0.24 & 0.38 & 0.08 & 4.9 \\
d7 & 2 & 2012-06-21 & 1 & 1 & 1 & 33.0 \\
e1 & 2 & 2013-03-02 & 23 & 23 & 0.1 & 12.6 \\
e10 & 2 & 2012-10-16 & 1.7 & 2 & 0.05 & 4.9 \\
e10 & 2 & 2012-06-21 & 2 & 5 & 1 & 27.0 \\
e11 & 2 & 2012-10-16 & 0.29 & 0.38 & 0.04 & 5.9 \\
e12 & 2 & 2012-10-16 & -0 & 0.9 & 0.2 & 2.5 \\
e12 & 1 & 1993-03-16 & 0.2 & 1.2 & 0.5 & 22.5 \\
e13 & 2 & 2012-10-16 & 0.76 & 0.97 & 0.05 & 4.9 \\
e13 & 2 & 2012-06-21 & 1 & 3 & 1 & 33.0 \\
e14 & 2 & 2013-03-02 & 1.6 & 1.2 & 0.1 & 12.6 \\
e15 & 3 & 2014-04-19 & 0.35 & 0.35 & 0.08 & 4.9 \\
e15 & 2 & 2012-06-21 & 5 & 7 & 1 & 27.0 \\
e16 & 2 & 2012-10-16 & 0.05 & 0.14 & 0.04 & 5.9 \\
e17 & 2 & 2012-10-16 & 0.1 & 0.5 & 0.2 & 2.5 \\
e17 & 1 & 1993-03-16 & - & - & 0.5 & 22.5 \\
e18 & 3 & 2014-04-19 & 0.51 & 0.37 & 0.08 & 4.9 \\
e18 & 2 & 2012-06-21 & -0 & 1 & 1 & 33.0 \\
e19 & 2 & 2013-03-02 & 1.3 & 1 & 0.1 & 12.6 \\
e2 & 1 & 1992-10-25 & 14 & 14 & 0.2 & 4.9 \\
e2 & 2 & 2012-06-21 & 1.6\ee{2} & 1.7\ee{2} & 1 & 27.0 \\
e20 & 3 & 2014-04-19 & 0.26 & 0.38 & 0.05 & 5.9 \\
e21 & 2 & 2012-10-16 & 0.5 & 1.4 & 0.2 & 2.5 \\
e21 & 1 & 1993-03-16 & 0.4 & 1.1 & 0.5 & 22.5 \\
e22 & 2 & 2012-10-16 & 0.12 & 0.33 & 0.05 & 4.9 \\
e22 & 2 & 2012-06-21 & 0 & 2 & 1 & 33.0 \\
e23 & 2 & 2013-03-02 & 0.5 & 0.4 & 0.1 & 12.6 \\
e3 & 1 & 1992-10-25 & 8 & 7.9 & 0.2 & 4.9 \\
e3 & 2 & 2012-06-21 & 8 & 11 & 1 & 27.0 \\
e4 & 3 & 2014-04-19 & 4.3 & 4.6 & 0.05 & 5.9 \\
e5 & 2 & 2012-10-16 & 0.1 & 4.1 & 0.2 & 2.5 \\
e5 & 1 & 1993-03-16 & 15 & 17 & 0.5 & 22.5 \\
e6 & 2 & 2012-10-16 & 2.5 & 2.6 & 0.05 & 4.9 \\
e6 & 2 & 2012-06-21 & 1 & 2 & 1 & 33.0 \\
e7 & 2 & 2013-03-02 & 0 & 0.2 & 0.1 & 12.6 \\
e8n & 1 & 1992-10-25 & 1.5 & 1.5 & 0.2 & 4.9 \\
e8n & 2 & 2012-06-21 & 4 & 6 & 1 & 27.0 \\
e8s & 3 & 2014-04-19 & 1.1 & 1.3 & 0.05 & 5.9 \\
e9 & 2 & 2012-10-16 & 2.6 & 3 & 0.2 & 2.5 \\
e9 & 1 & 1993-03-16 & 0.3 & 2.4 & 0.5 & 22.5 \\
\hline
\end{tabular}
\par
An excerpt from the point source catalog.  For the full catalog, see Table \ref{tab:contsrcs_full}
\end{table*}

\subsubsection{Variability}
\label{sec:variability}
The multi-epoch data demonstrate that at least one of these sources is
variable.  The most convincing case for variability is in the (double) source
d4.  In the \citet{Mehringer1994a} data, there is no
hint of emission at this location, with a 3-$\sigma$ upper limit of 0.6 mJy.
At the same position and frequency in 2014, there is a $1.03 \pm 0.06$ mJy
source at the position of d4e.

\subsubsection{Multi-wavelength Source Associations}
\label{sec:associations}
We compare our extracted source catalog with the MOXC Chandra X-ray catalog
\citep{Townsley2014a} and the UKIDSS Galactic Plane Survey
\citep[UGPS;][]{Lucas2008a} to identify sources that are detected at multiple
wavelengths.  The associations are listed in Table
\ref{tab:associations}.  We discuss some
individual source associations in more detail in this section.
The absolute astrometric offsets in the Chandra catalogs are $<0.1\arcsec$
\citep{Townsley2014a}, and typical absolute astrometric errors in JVLA
data\footnote{\url{https://science.nrao.edu/facilities/vla/docs/manuals/oss/performance/positional-accuracy}}
are $\sim10\%$ of the beam, or in our case $\lesssim0.02-0.04\arcsec$.

Both lobes of the variable source d4 are
0.3\arcsec from MOXC 192339.62+143130.3 and d6 is 0.3\arcsec from MOXC
192341.23+143111.8 (0.3\arcsec is about a full-width resolution element in the
radio data, so this offset
is highly significant).    These
separations may indicate that either the \hii regions or the MOXC point sources
are associated with diffuse material, e.g. outflows,
near stars rather than the stars themselves.  
Candidate sources e20 and e22 are $<0.1$\arcsec from
MOXC 192342.86+143027.5 and 192342.77+143027.5, respectively.
There are a handful of X-ray
sources within W51 IRS2 that could be associated with faint radio sources that
we are unable to detect against the bright diffuse background emission, but
these X-ray sources could also be associated with
outflowing material (e.g., the Lacy jet; see Section \ref{sec:lacyjet}).  The
lack of correlation between X-ray and \hchii regions indicates that the
embedded OB stars within these \hchii regions are at most weak X-ray emitters.  

We compared our point source catalog to the UKIDSS K-band and UWISH2 \hh images
to search for infrared associations with our detected radio sources
\citep{Lucas2008a,Froebrich2011a}.  Most of the W51e sources have no
association, as is expected given the high extinction in this region
($A_V>100$, based on mm emission).  We discuss the exceptions here.  Diffuse
source e7 to the south is associated with UGPS
J192344.79+142911.2 \citep{Lucas2008a}.  Three sources, e20, e21, and
e22, have NIR K-band counterparts, though the UGPS cataloged positions do not
align with the evident NIR K-band sources in the archive images. We nonetheless
consider these sources
to have infrared counterparts; likely adaptive optics imaging will be required
to obtain reliable flux density measurements for these due to both point source
and extended emission confusion in the region.   Similarly, e5 exhibits some
NIR K-band emission, though because it is diffuse, this emission is
not cataloged in the UGPS.

The source d7 is associated with a luminous infrared source, which is
seen both in UKIDSS \citep{Lucas2008a} and NACO \citep{Figueredo2008a} images.
However, the reported position of the near-infrared K-band
\citet{Goldader1994a} source RS15 is
0.7\arcsec away, even though in the UKIDSS images there is no detected emission
at this position.  Surprisingly, the associated MOXC Chandra source
\citep{Townsley2014a} is more closely aligned with the offset RS15 position
than the true source location in the NACO and UKIDSS images.  High proper motion
is unlikely to be the explanation for this discrepancy, since the source
would have to be moving at $v\sim10^3$ \kms to have shifted 0.7\arcsec in 30
years.  Despite this confusion, we regard the NIR, X-ray, and radio features to
have a common
origin.

\subsection{\formaldehyde \twotwo emission}
\label{sec:twotwoemission}
We detect \formaldehyde \twotwo emission around W51e2, W51e8, in a region
between e1, e8, and e10, and in W51 North.  We report the tentative detection
of an extended structure between e2 and e1, though this structure is weak
and diffuse.  The fitted
emission line parameters are listed in Table \ref{tab:emission22}.  The
emission is shown in Figures \ref{fig:w51northcore},
\ref{fig:w51mainemilabels}, and \ref{fig:w51mainemicontours} and discussed in
Section \ref{sec:h2coemission}.

\begin{table*}[htp]
\caption{\formaldehyde \twotwo emission line parameters}
\begin{tabular}{ccccccccc}
\label{tab:emission22}
Object Name & Amplitude & $E$(Amplitude) & $V_{LSR}$ & $E(V_{LSR})$ & $\sigma_V$ & $E(\sigma_V)$ & $r_{eff}$ & Detection Status \\
 & $\mathrm{mJy}$ &  & $\mathrm{km\,s^{-1}}$ &  &  &  &  &  \\
\hline
NorthCore & 0.652 & 0.018 & 58.877 & 0.074 & 2.375 & 0.074 & 1.8 & - \\
e2-e8 bridge & 0.369 & 0.022 & 56.554 & 0.07 & 1.0 & 0.07 & 1.4 & - \\
e2\_a & 0.641 & 0.026 & 57.412 & 0.1 & 2.167 & 0.1 & 0.9 & - \\
e2\_b & 1.085 & 0.035 & 56.083 & 0.097 & 2.605 & 0.097 & 0.9 & - \\
e2\_c & 0.612 & 0.026 & 56.01 & 0.19 & 3.95 & 0.19 & 0.9 & - \\
e8mol & 1.551 & 0.061 & 60.55 & 0.13 & 2.85 & 0.13 & 0.5 & - \\
e8mol\_ext & 1.045 & 0.024 & 60.16 & 0.089 & 3.439 & 0.089 & 1.0 & weak \\
e10mol & 0.999 & 0.039 & 58.01 & 0.21 & 4.69 & 0.21 & 0.5 & - \\
e10mol\_ext & 0.813 & 0.018 & 58.507 & 0.094 & 3.593 & 0.094 & 1.1 & weak \\
\hline
\end{tabular}
\par
Columns with $E$ denote the errors on the measured parameters.  $\sigma_{V}$ is the 1-dimensional gaussian velocity dispersion.  $r_{eff}$ is the effective aperture radius.
\end{table*}

\subsection{Line-of-sight velocities}
\label{sec:LOSvelo}
We have detected H77$\alpha$ emission from many of the ultracompact and
hypercompact \hii regions within W51.  We report their line-of-sight velocities
as measured from gaussian profile fits to their extracted spectra.

The H77$\alpha$ emission line parameters are listed in Table \ref{tab:h77a}, and
the \para \twotwo absorption line parameters are in Table \ref{tab:absorption22}.

The \formaldehyde absorption lines provide velocities of foreground molecular
gas.  The e5 and e6 sources are close to one another (separation 3.1\arcsec,
or projected distance 0.08 pc), and both exhibit \formaldehyde absorption at
62-63 \kms.  Source e6 is detected in H77$\alpha$ at 68 \kms, suggesting that
the true velocity of both e5 and e6 are redshifted from the \formaldehyde
absorption lines. 

\begin{table*}[htp]
\caption{\formaldehyde \twotwo absorption line parameters}
\begin{tabular}{ccccccccc}
\label{tab:absorption22}
Object Name & Amplitude & $E$(Amplitude) & $V_{LSR}$ & $E(V_{LSR})$ & $\sigma_V$ & $E(\sigma_V)$ & $r_{eff}$ & Detection Status \\
 & $\mathrm{mJy}$ &  & $\mathrm{km\,s^{-1}}$ &  &  &  & $\mathrm{{}^{\prime\prime}}$ &  \\
\hline
e1 & -2.922 & 0.034 & 62.531 & 0.078 & 5.81 & 0.078 & 0.6 & ambig \\
e2 & -21.186 & 0.082 & 56.8728 & 0.009 & 2.0231 & 0.009 & 0.6 & - \\
e3 & -2.57 & 0.1 & 64.26 & 0.11 & 2.35 & 0.11 & 0.4 & - \\
e5 & -1.944 & 0.094 & 62.739 & 0.032 & 0.576 & 0.032 & 0.6 & - \\
e6 & -0.598 & 0.015 & 63.771 & 0.061 & 2.149 & 0.061 & 1.8 & - \\
e9 & -0.477 & 0.078 & 55.4 & 0.22 & 1.15 & 0.22 & 0.5 & - \\
e10 & -0.63 & 0.1 & 66.74 & 0.21 & 1.1 & 0.21 & 0.5 & - \\
\hline
\end{tabular}
\par
Columns with $E$ denote the errors on the measured parameters.  $\sigma_{V}$ is the 1-dimensional gaussian velocity dispersion.  $r_{eff}$ is the effective aperture radius.
\end{table*}

\begin{table*}[htp]
\caption{H$77\alpha$ emission line parameters}
\begin{tabular}{cccccccc}
\label{tab:h77a}
Object Name & Amplitude & $E$(Amplitude) & $V_{LSR}$ & $E(V_{LSR})$ & $\sigma_V$ & $E(\sigma_V)$ & Detection Status \\
 & $\mathrm{mJy}$ & $\mathrm{mJy}$ & $\mathrm{km\,s^{-1}}$ & $\mathrm{km\,s^{-1}}$ & $\mathrm{km\,s^{-1}}$ & $\mathrm{km\,s^{-1}}$ &  \\
\hline
e1 & 3.0 & 0.2 & 54.87 & 0.85 & 10.9 & 0.85 & - \\
e2 & 0.6 & 0.13 & 56.0 & 3.8 & 15.3 & 3.8 & - \\
e3 & 1.48 & 0.33 & 59.8 & 2.7 & 10.5 & 2.7 & - \\
e4 & 0.56 & 0.39 & 57.2 & 4.4 & 5.4 & 4.4 & - \\
e5 & 0.25 & 0.18 & 52.6 & 4.7 & 5.8 & 4.7 & weak \\
e6 & 0.183 & 0.051 & 68.6 & 4.9 & 15.4 & 4.9 & - \\
e9 & 0.15 & 0.18 & 66.8 & 7.9 & 5.5 & 7.9 & weak \\
e10 & 0.27 & 0.2 & 51.2 & 9.8 & 11.6 & 9.8 & weak \\
\hline
\end{tabular}

\end{table*}

\subsection{Diffuse continuum emission features}
\label{sec:diffuseemission}
There are a few new notable continuum emission features detected in our data
that were not detected in previous shallower data.

Near source e11, there is a bow-shaped feature (Figure \ref{fig:e11bow}).
There are no known associated sources at other wavelengths, though in the
Spitzer GLIMPSE
\citep{Benjamin2003a} images, there is some diffuse emission at this location.
The bow-like structure points away from e11, suggesting that e11 produces the
bow-shaped nebula.

\Figure{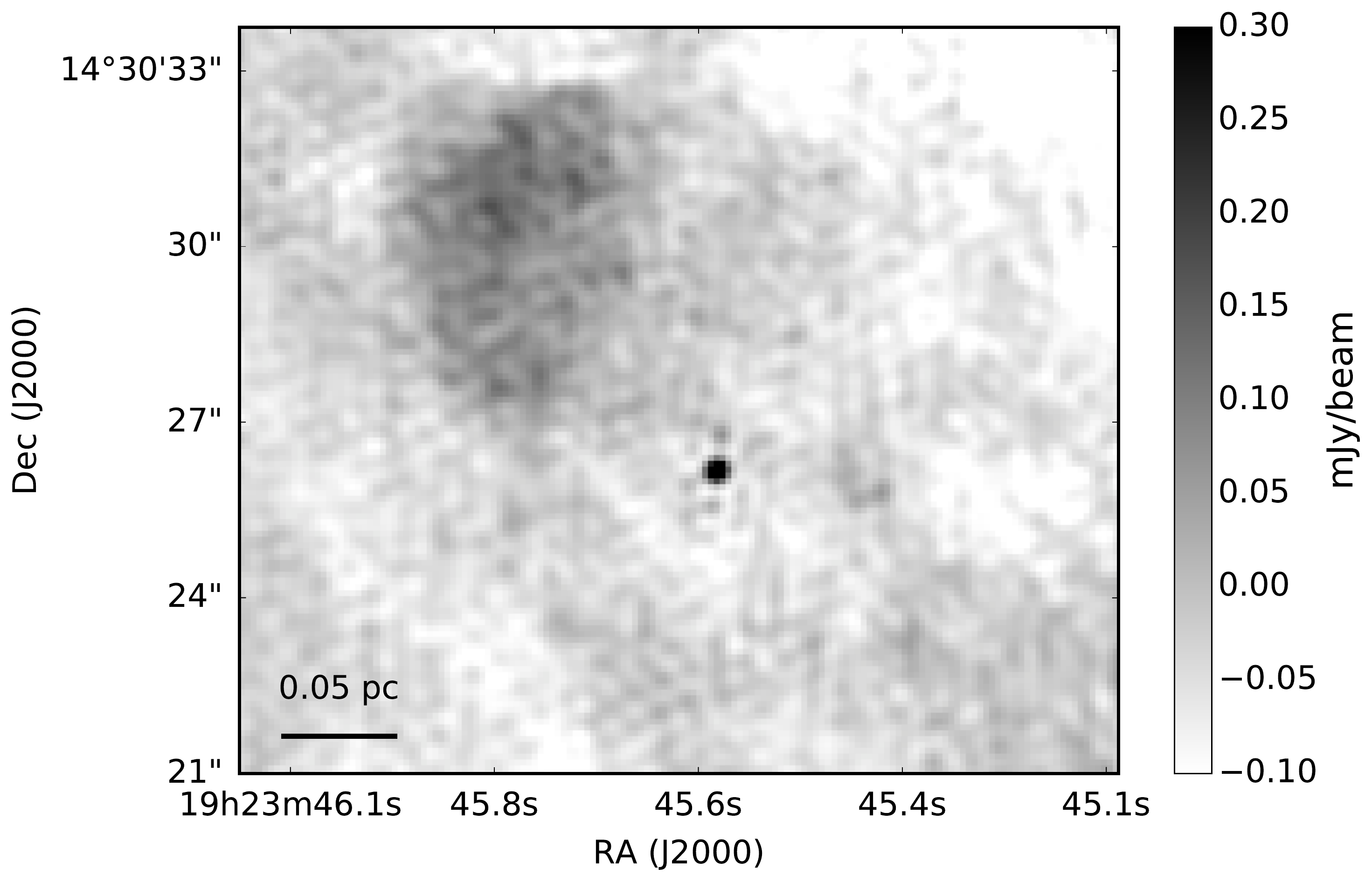}
{A bow-shaped feature toward the northeast of the source e11 in the Ku-band
(14.5 GHz)
continuum image.  The feature resembles the bow shocks of Herbig-Haro objects,
but is more likely to be an \hii region given its smooth structure.
The synthesized beam size is
$\theta_{FWHM}\approx0.33\arcsec$.
}
{fig:e11bow}{0.5}{6.5in}

The source d3 is an \hii region with radius $r\sim1.9$\arcsec.  It is
associated with a Spitzer source and the 2MASS source 2MASX J19233591+1431288.
The sources d3, e1, e6, and e7 are compact but resolved (0.02-0.07 pc) and
round; they are therefore classified as ultracompact \hii regions.

The diffuse emission associated with the W51 Main shell and W51 IRS1 traces a broad arc
($r\sim42\arcsec$) that has been observed in many previous data sets.  The new
deeper, higher-resolution data represent a substantial improvement in image
quality.  While in previous observations a relatively smooth and clumpy
structure was observed, the new images reveal a network of wispy, sharp-edged,
filamentary structures.  Figure \ref{fig:w51irs1} shows W51 IRS1, the
region between the W51e and
W51 IRS2 clusters.  While this area contains few clear individual sources, it
accounts for the majority of the radio luminosity of the W51 Main region and
somewhere between one quarter and one half the bolometric luminosity of the
whole W51A complex.

The filamentary structures in W51 IRS1 are unresolved along the short
axis, with aspect ratios $>25$.   These resemble the ``stringlike ionized
features'' noted by \citet[][see
\url{http://images.nrao.edu/402}]{Yusef-Zadeh1990a} in Orion, which are
associated primarily with the Orion Bar PDR and the nearby fainter PDR to the
northeast.  The similarity suggests that these features are PDRs, highlighting
the sharp interaction points between the \hii region and the surrounding
molecular cloud.

The two most prominent of these sharp features are on opposite sides of the W51
Peak (Figure \ref{fig:w51irs1}, blue arrows).  The long vertical filament
on the left (Filament A) approximately faces W51e, while the S-shaped filament
on the right (Filament B) faces outward, away from the e1/e2 group and toward a
region that contains no known OB stars.  The presence of multiple PDR features
raises questions about the ionizing source(s).  In Orion, the Trapezium is 0.2
pc in projection from the Bar PDR.  Filament A is 0.6 pc from the W51e cluster,
which contains enough O-stars to illuminate the filaments, but there may not be
a clear line-of-sight from those stars, which appear to be deeply embedded in
molecular material, to the filament.  These features therefore indicate the
presence of additional OB stars that are obscured along our line-of-sight but
that are not deeply embedded.

\FigureTwo
{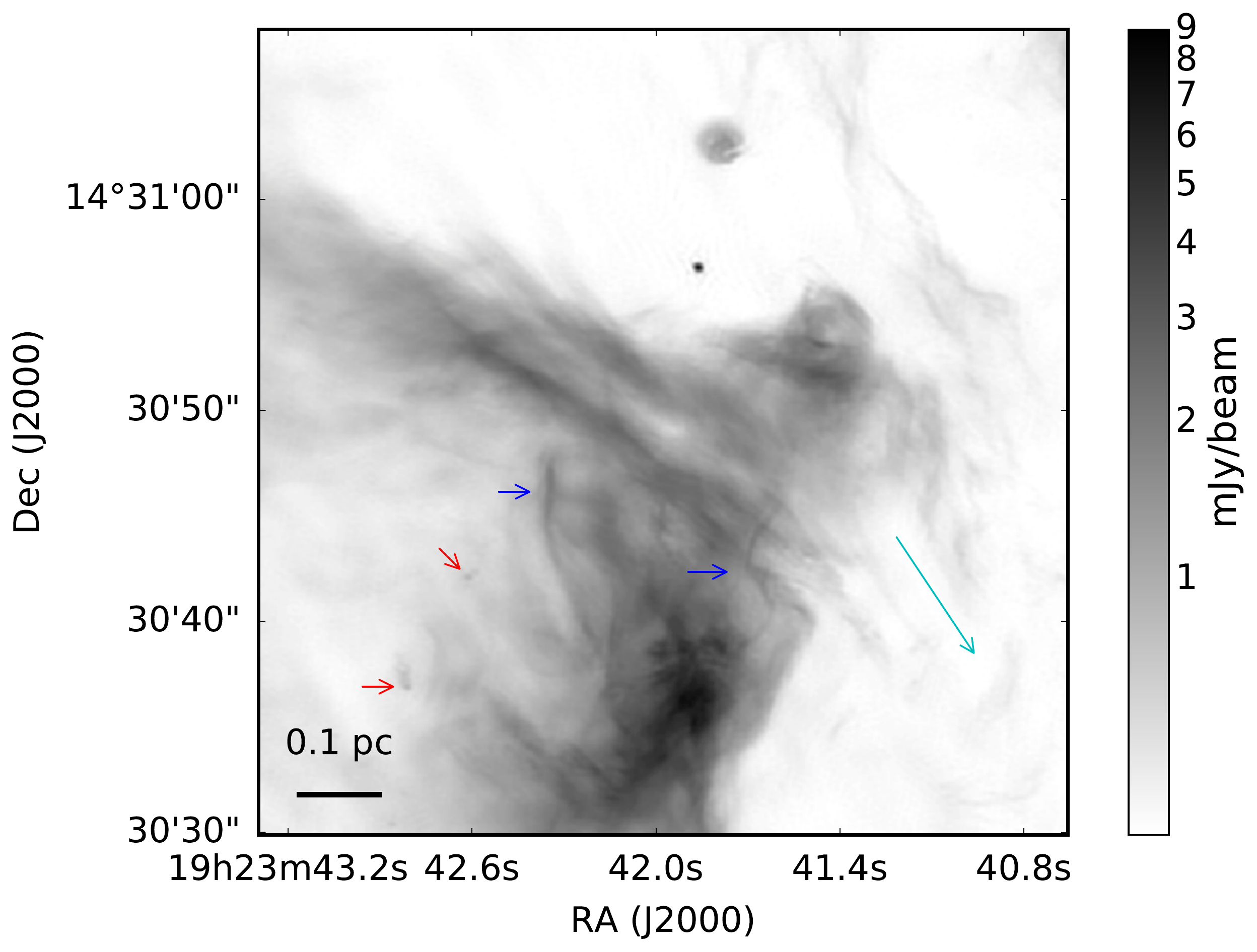}
{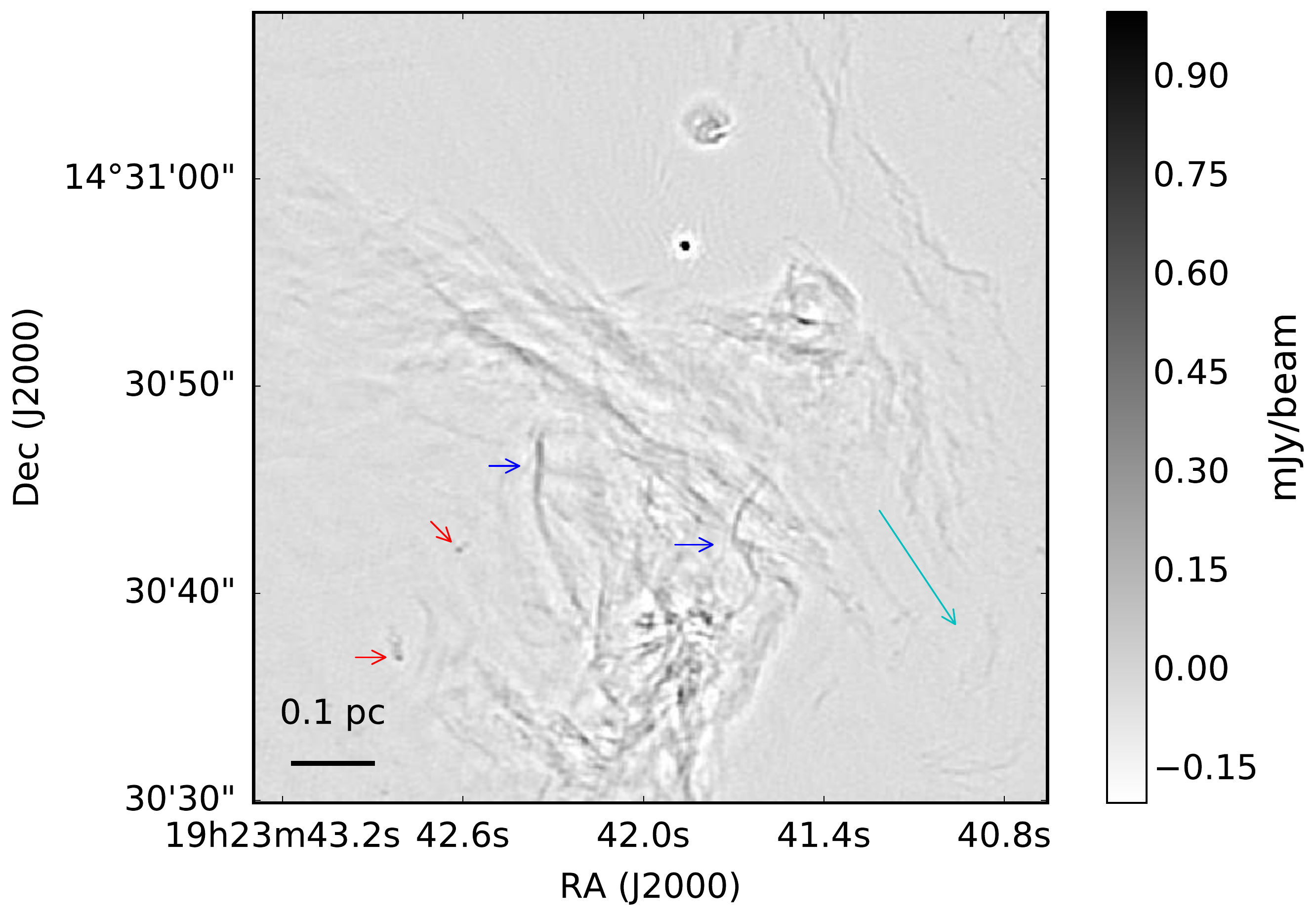}
{({\it a}) A C-band (4.9 GHz) image of W51 IRS1, the location of peak
intensity on the W51 Main shell.
This region accounts for more than half of the radio luminosity of W51.  The
colorbar has been selected to emphasize the pointlike sources even though it
saturates some of the diffuse emission.  The synthesized beam size
is $\theta_{FWHM}\approx0.3\arcsec$.
({\it b}) An unsharp-masked version of the image with a 0.3\arcsec smoothing
kernel used for the large-angular-scale flux removal. 
There are
two pointlike sources in field marked by red arrows, e14 at center-left and the
cometary e13 toward the lower left.
The filamentary features mentioned in Section \ref{sec:diffuseemission} are
identified by blue arrows.  The cyan arrow points along a possible ionized flow,
traced by elongated features parallel to the arrow and a rounded feature to the
southwest.  A counterpart to the northeast is barely visible.
}
{fig:w51irs1}{1}{3.5in}

Additionally, the W51 IRS1 ridge is detected in H77$\alpha$.  Filament
B peaks in the range 40-50 \kms.  Filament A is the sharp edge feature
spanning $v_{LSR}\sim38-48$ \kms.  From Filament A, there is continuous
H77$\alpha$ emission toward the northeast from 40 to 80 \kms that exhibits a
gradient from blue to red from southeast to northwest.  

The H77$\alpha$ lines reveal two circular features (bubbles) that overlap along
the line of sight but are distinct in velocity (Figure
\ref{fig:rrl_chan_main}).  A smaller bubble at 40 \kms shows a velocity
gradient orthogonal to the gradient seen next to filament A, increasing in
velocity from southwest to northeast; this gradient may trace the edges of an
expanding spherical shell.  The large bubble is centered at about 65 \kms and
shows no clear gradients (Figure \ref{fig:rrl_chan_main}, for
labels see Figure \ref{fig:coverview_diffuse}).

\Figure{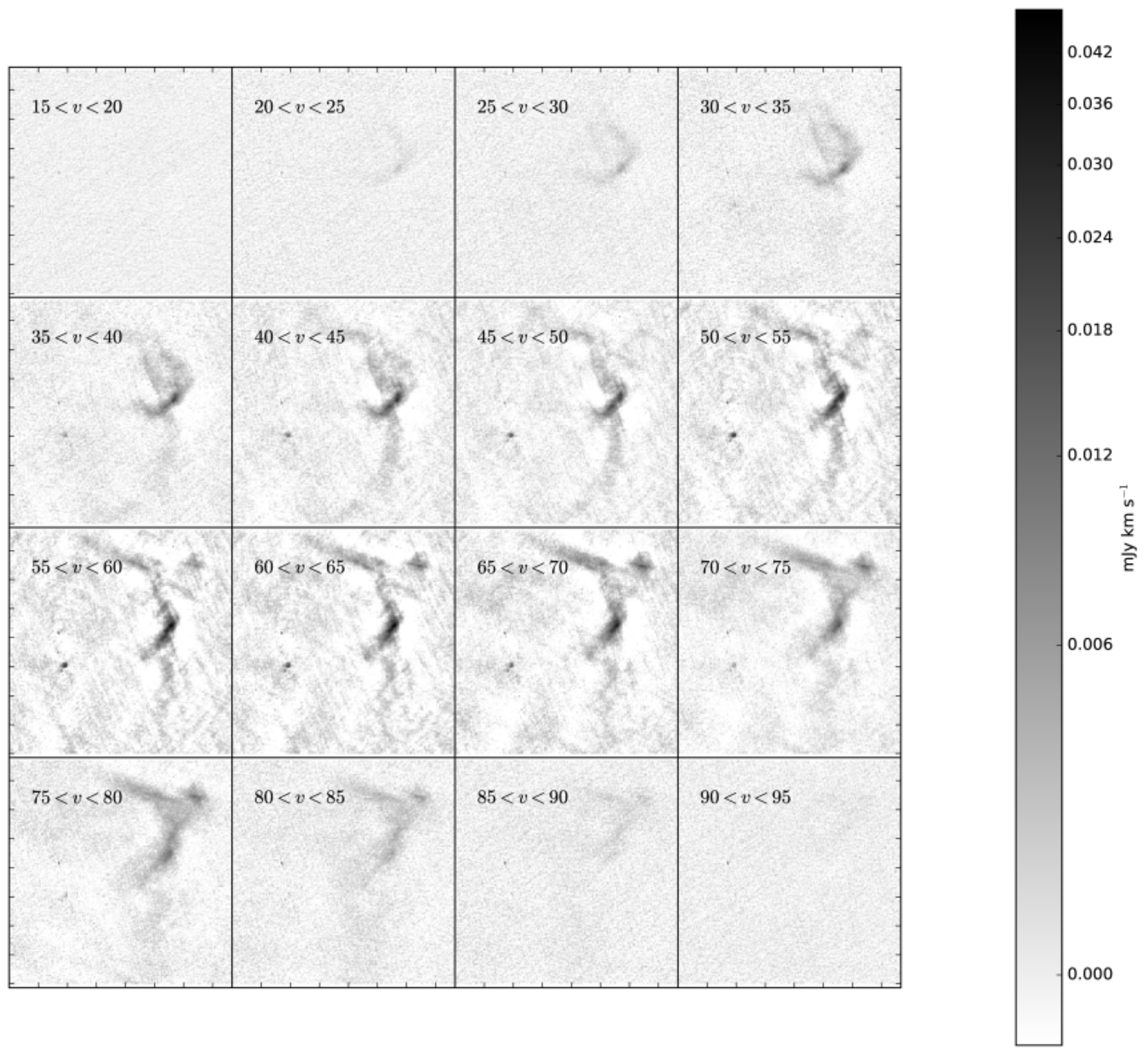}
{Channel maps of the H77$\alpha$ radio recombination line integrated over 5 \kms windows
for W51 Main and IRS1.  There are small ring features notable at low and high
velocities; these are identified in Figure \ref{fig:coverview_diffuse}.  These
images demonstrate that there are two distinct but possibly interacting bubbles
along the line of sight in W51 Main.
The synthesized beam size is $\theta_{FWHM}\approx0.42\arcsec$.
}
{fig:rrl_chan_main}{1}{7in}

The W51 IRS1 \hii region appears connected to the W51 IRS2
bubble in the continuum (Figure \ref{fig:w51irs1} shows the
apparently-connected region, Figure \ref{fig:w51irs2} shows the IRS2 bubble),
but the RRL emission peaks at around 75 \kms (Figure \ref{fig:rrl_chan_main}).
Along this
region, there are weak `striations' at PA 126 degrees, parallel to the cyan
arrow in Figure \ref{fig:w51irs1} and orthogonal to the direction toward
IRS2.  These striations  may trace the outer edge of the expanding IRS2
region.  There is also an arc to the southwest of this region that appears to
be a bow shock from gas outflowing along the same axis; the cyan arrow points
to this arc in Figure \ref{fig:w51irs1}.

The IRS2 region peaks in velocity around 62 \kms at the center, but exhibits a
gradient from its center to its surroundings, with a more extended component
peaking around 47 \kms.  There is a shell around IRS2 with maximum projected
radius 21\arcsec (0.5 pc; Figures \ref{fig:w51irs2} and
\ref{fig:rrl_chan_irs2}).  A prominent sharp edge feature (blue arrow in Figure
\ref{fig:w51irs2}) extends to the northeast from the central cluster; it may
trace the outflow of hot material from the cluster.

Parts of the shell close to IRS2 are detected in radio recombination lines
(H77$\alpha$), but the more distant regions marked in Figure \ref{fig:w51irs2}
with red arrows are not (they are too faint).  The shell to the northwest peaks
at $v_{lsr}=37.5$ \kms, and a gradient is observed from that shell to the peak
velocity of IRS2 at $v_{lsr} = 62.6$ \kms.  The velocity structure is
inconsistent with spherical outflow but is consistent with a conical flow
structure.  This flow structure likely
indicates that the cluster has evacuated a large cavity approximately toward
the observer, consistent with the low observed extinction toward
IRS2.  

\FigureTwo
{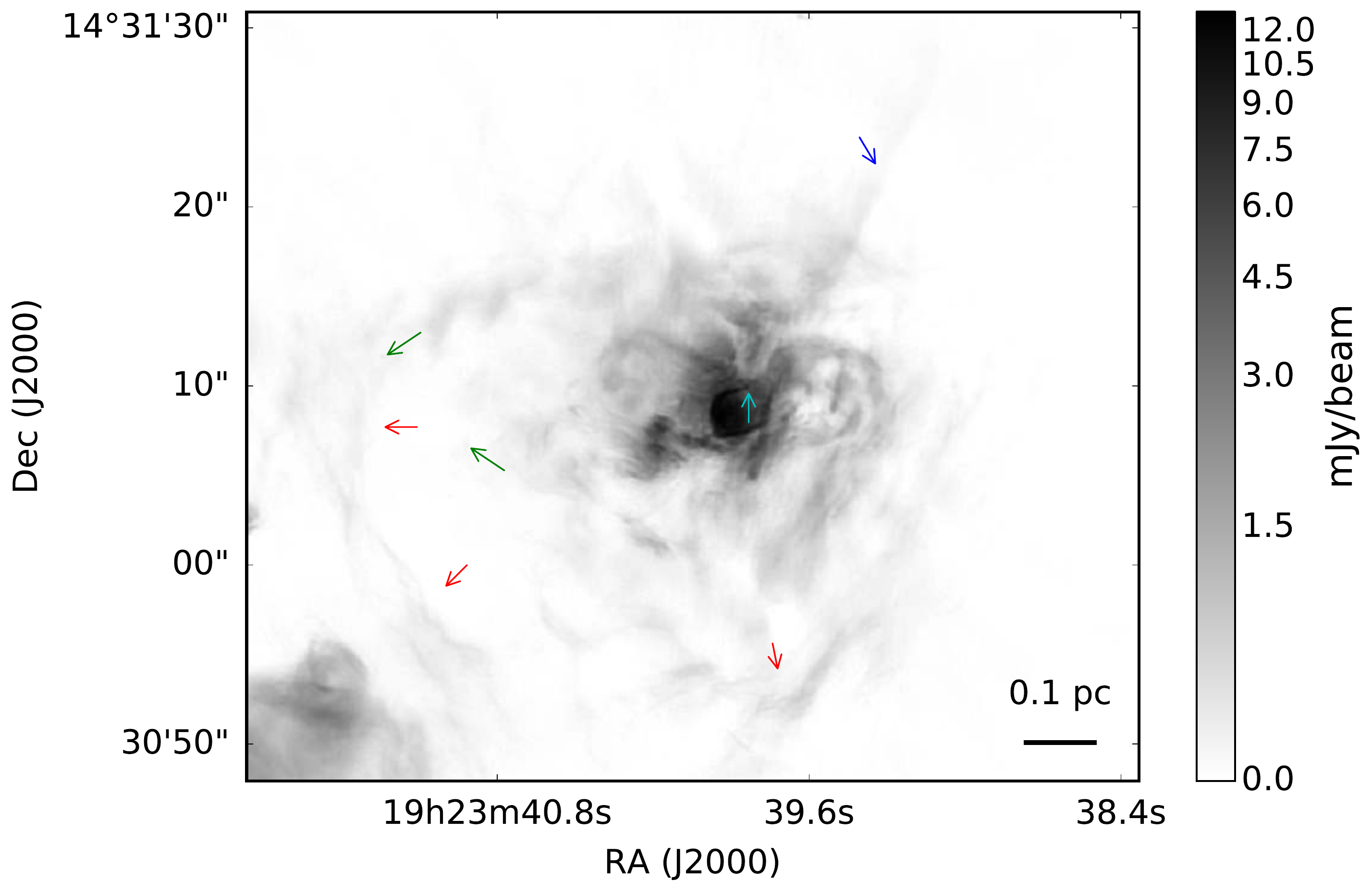}
{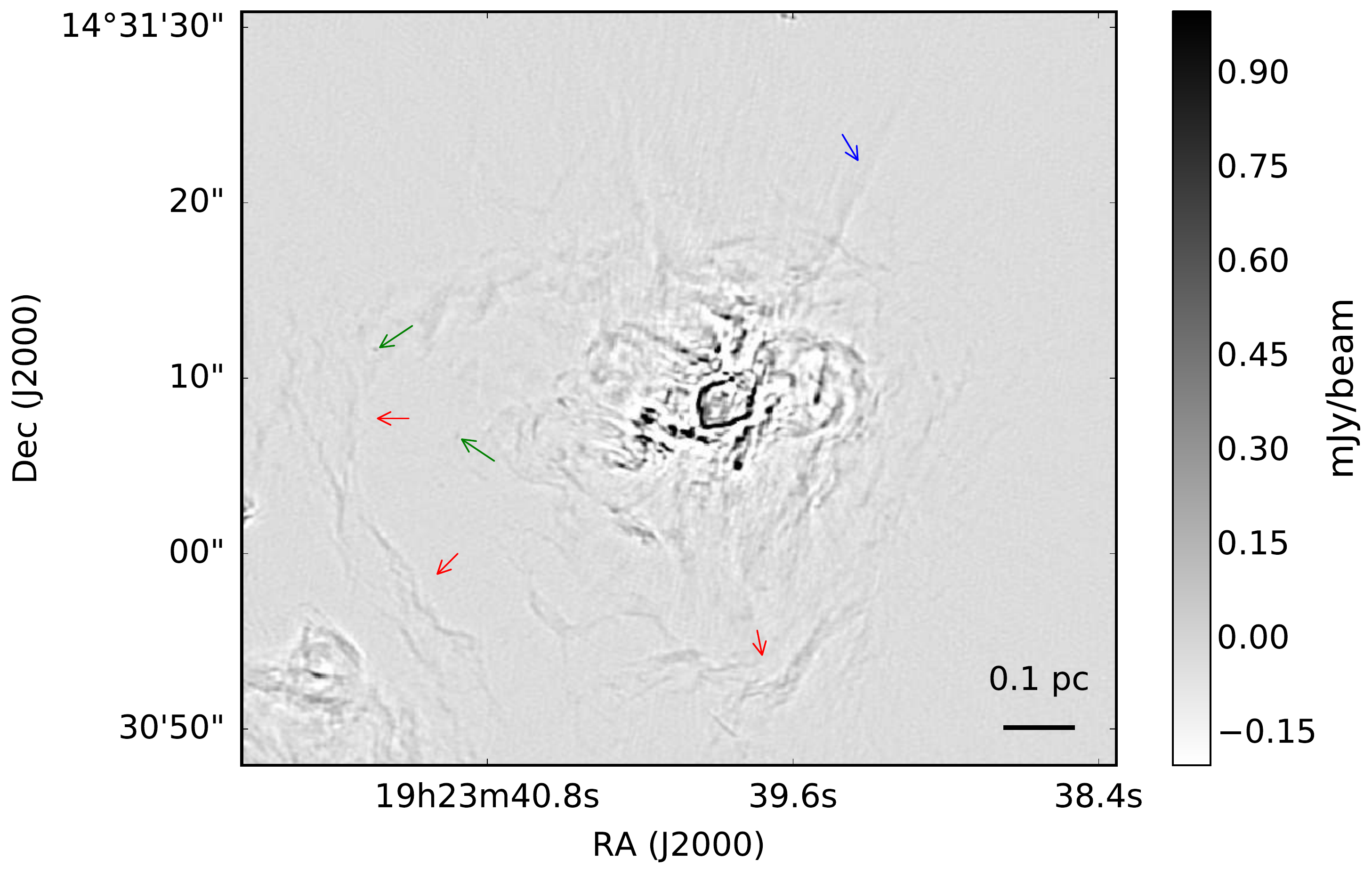}
{(\textit{a}) The W51 IRS2 region in C-band continuum and (\textit{b}) unsharp
masked with a 0.3\arcsec kernel.  Concentric ring features surround the IRS2
region.  A V-shaped feature to the north (cyan arrow) may highlight the edge of
gas flows out of the cluster, extending to the long linear feature pointing
northwest (blue arrow).  An
apparent shell is observed centered on the IRS2 cluster, with edge
features to the south and east (red arrows).  The compact sources d6 and d7,
which are candidate colliding-wind binaries (Section \ref{sec:contnature}),
are identified with green arrows.
}
{fig:w51irs2}{1}{3.5in}

\Figure{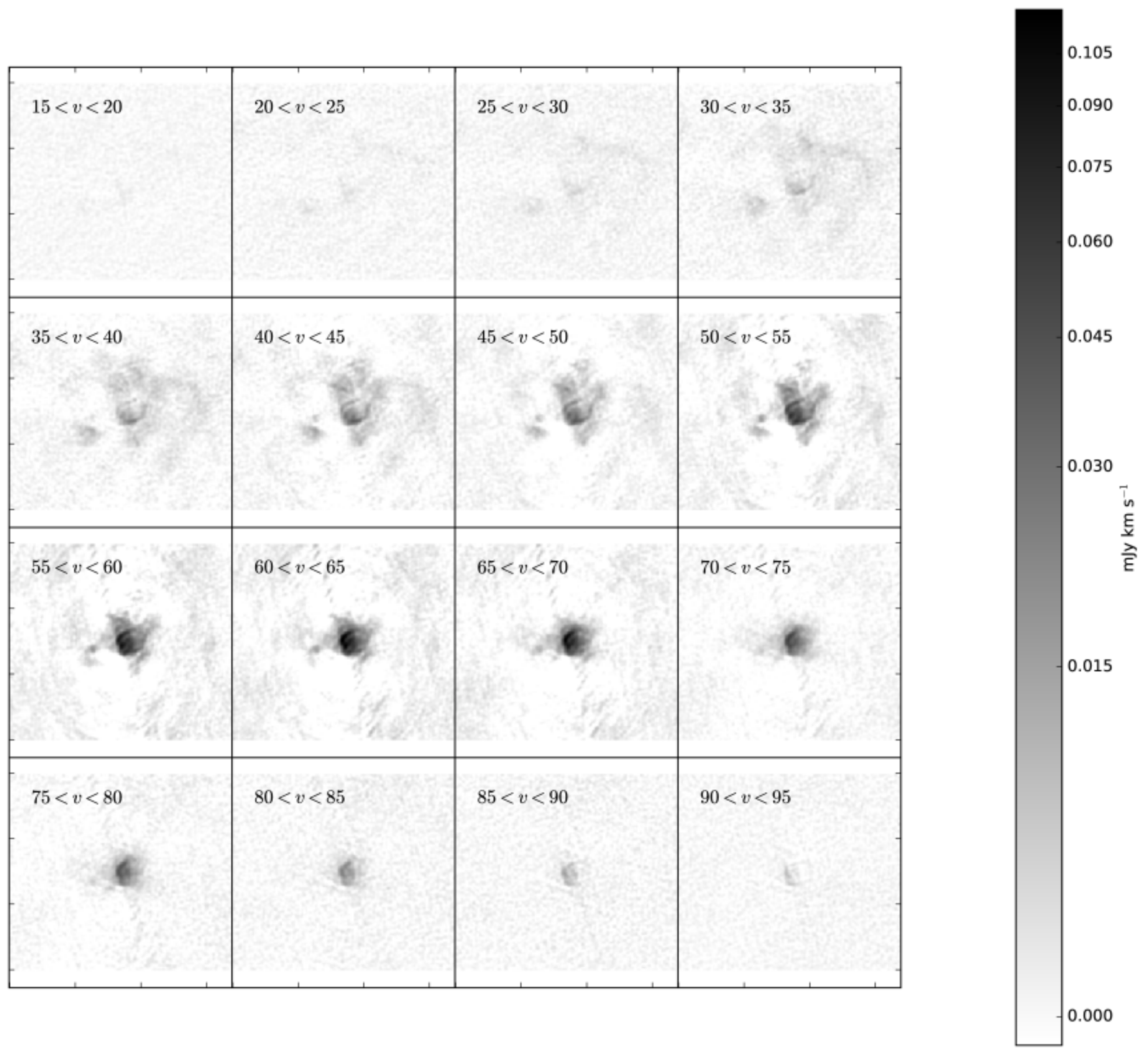}
{Channel maps of the H77$\alpha$ line integrated over 5 \kms windows
for the IRS2 region.  Around 30-35 \kms, the edges of the IRS2 region become
visible, illustrating the blueshifted cavity discussed in Section
\ref{sec:diffuseemission}.  These images demonstrate the presence of outflowing
material and are used in Section \ref{sec:diffuseemission} to estimate the
ionized gas mass loss rate from IRS2.
The synthesized beam size is $\theta_{FWHM}\approx0.42\arcsec$.
}
{fig:rrl_chan_irs2}{1}{7in}

\subsection{The Lacy IRS2 ionized outflow}
\label{sec:lacyjet}
\citet{Lacy2007a} reported the detection of very high velocity ionized gas
in the mid-infrared [Ne II] 12.8\um and S IV 10.5\um lines.  They observed the
gas at a velocity blueshifted about 100 \kms from the IRS2 ionized and molecular
gas velocity.  We have detected the same feature in the H77$\alpha$ RRL.
The RRL shows the same position-velocity structure as the infrared ionized
features.  No redshifted counterflow is detected (Figure \ref{fig:lacyjetslice}).

One minor but notable feature of the H77$\alpha$ flow is that the emission
occurs close to the rest velocity of the He77$\alpha$ line.  If
\citet{Lacy2007a} had not already unambiguously identified the high velocity of
this feature in two infrared lines, we would have assumed the detected emission
to be slightly redshifted Helium, since He77$\alpha$ is detected at $v\sim60$
\kms in IRS2, rather than correctly identifying it as H77$\alpha$.

\FigureTwo
{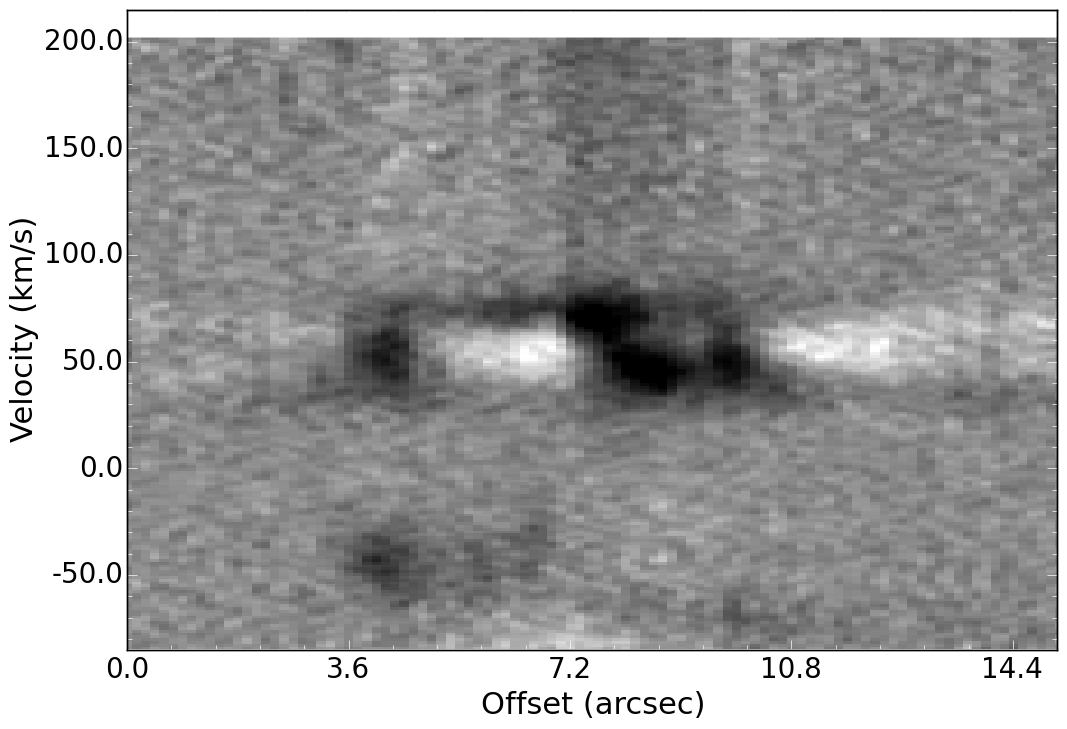}
{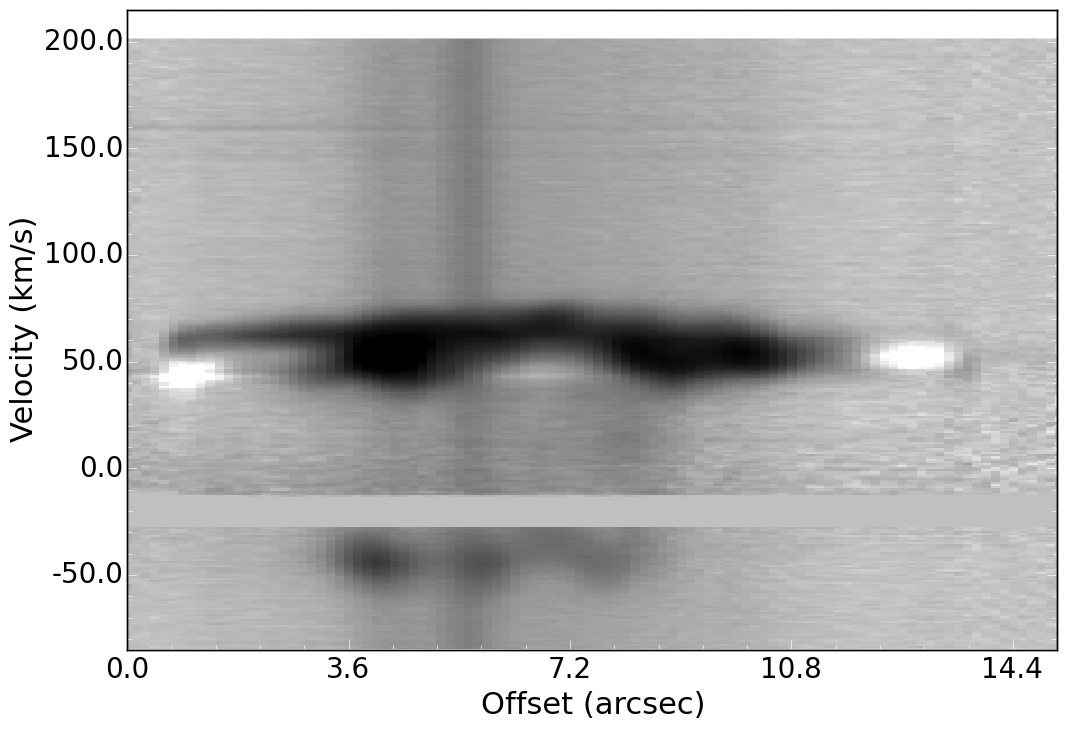}
{Position-velocity slices through (a) the H77$\alpha$ cube and (b) the [Ne II]
cube from \citet{Lacy2007a} tracking the approximate path of the Lacy jet.  Emission is black,
absorption (which in both images is due to reduction artifacts rather than true
physical absorption) is white.  The vertical streaks are continuum sources.
The
abscissa shows the offset position along the slice from J2000 19:23:40.464
+14:31:07.9536 to 19:23:39.4176 +14:31:04.7172, position angle 258\arcdeg.
The blueshifted lobe is evident at -50 \kms in both cubes, and neither shows a
redshifted counterpart.  The H77$\alpha$ data reveal that there is truly no
counterpart, ruling out the possibility that it was simply too extincted to be
observed in the infrared.}
{fig:lacyjetslice}{1}{3.5in}

\FigureTwo
{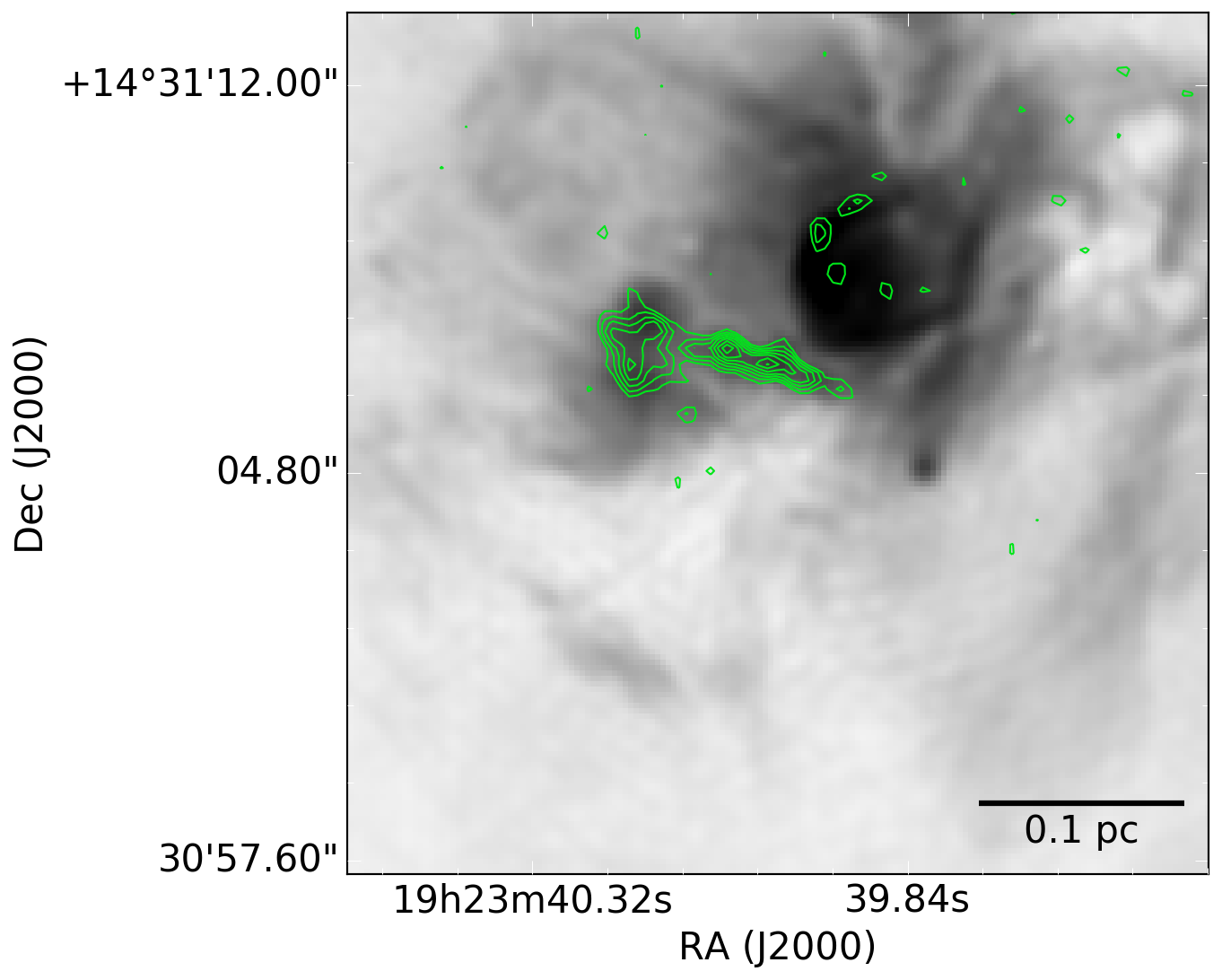} 
{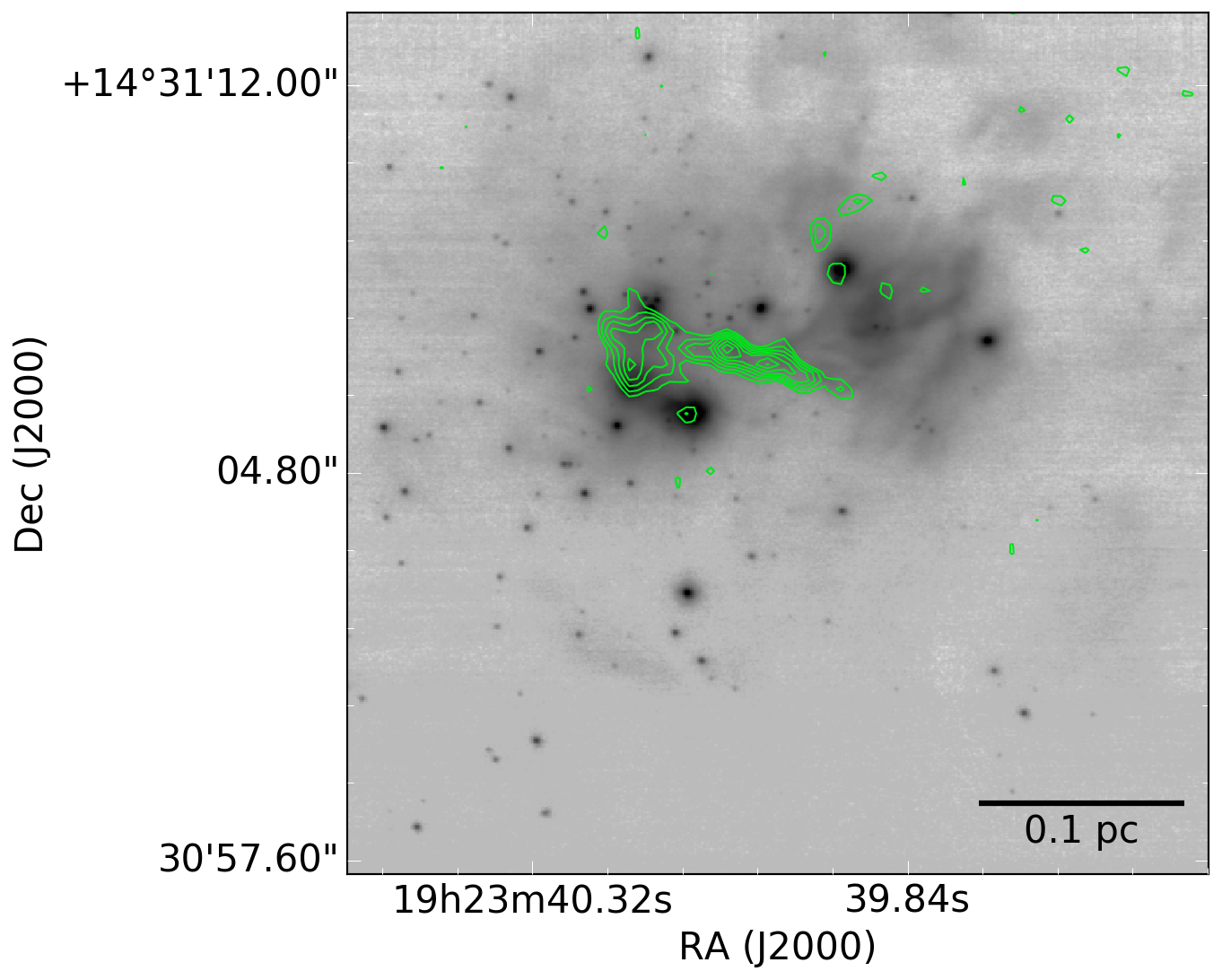} 
{The `Lacy Jet' as seen in H77$\alpha$ contours overlaid on (a) the 15 GHz
continuum image (synthesized beam $\theta_{FWHM}\approx0.33\arcsec$) and (b)
the NACO K-band image \citep[resolution $\approx0.2\arcsec$][]{Barbosa2008a}.
The H77$\alpha$ contours are at 0.1, 0.2, 0.3, 0.4, 0.5, 0.6, and 0.7 Jy \kms,
integrated over the velocity range -60 to -16 \kms.}
{fig:lacyjetoverlay}{1}{3.5in}

\section{Analysis}
\label{sec:analysis}
We use the data to examine the following properties of W51:
\begin{itemize}
    \item Section \ref{sec:stellarmass}: Determination of the total stellar
        mass in W51 based on its infrared luminosity and a brief examination
        of the probable locations of the massive stars.
    \item Section \ref{sec:contnature}: Determination of the object type
        (e.g., stellar wind, photoionized HII region) associated with each
        detected continuum source.
    \item Section \ref{sec:faintw51main}: A brief discussion of the compact
        sources near W51 Main and their role in ionizing the W51 Main bubbles.
    \item Section \ref{sec:faintw51irs2}: A brief discussion of the IRS2 faint
        point sources.
    \item Section \ref{sec:h2coemission}: Examination of the \ortho
        \twotwo-detected hot cores, including upper limits on 4.829 GHz \ortho
        \oneone emission.
    \item Section \ref{sec:d4}: Discussion of the d4 variable emission
        feature, concluding that it arises from a fast outflow interacting with
        the dense local ISM.
    \item Section \ref{sec:vdisp}: Measurements of the velocity dispersion
        within the e1 and e2 clusters based on their RRL and \formaldehyde
        emission velocities, and a derivation of the gas-to-stellar mass ratio
        on this basis.
    \item Section \ref{sec:irs2outflow}: Constraints on the mass loss rate of the
        relatively low-extinction IRS2 region using RRL emission to constrain
        the mass and velocity of escaping material.
\end{itemize}

\subsection{The FIR-deduced stellar mass}
\label{sec:stellarmass}

We have re-measured the infrared luminosity of the W51 protoclusters using Herschel
Hi-Gal data \citep{Molinari2010a,Traficante2011a,Molinari2016a}, fitting an SED from the 70
to 500 \um with a single blackbody component.
While a single-component fit provides a poor
measurement of the dust temperature - multiple temperature components are
evident \citep{Sievers1991a} - it provides a good approximation to the total
infrared luminosity, which is dominated by a single warm ($\sim60$ K)
component.  The infrared luminosity is about $L\sim2\ee{7}$ \lsun within a 2 pc
radius\footnote{In many bands, W51 Main and W51 IRS2 are
saturated, so we have interpolated from neighboring pixels to estimate the flux
density.
Our luminosity estimate is therefore a lower limit.}, which includes
both the W51 IRS2 and W51 Main protoclusters.  The luminosity estimate does not
include the mid-infrared luminosity, which may provide an additional
$\sim25-50\%$ based on
the IRAS 12 and 25 \um measurements.  Our measured total luminosity is somewhat
larger than the previous estimates based on IRAS and KAO measurements
\citep{Harvey1986a,Sievers1991a}.
A luminosity $L=2.0\pm0.5\ee{7}$ \lsun implies
a stellar mass $M_{cl} = 6.7 \pm 2.3\ee{3}$ \msun, with a corresponding number of
O-stars (greater than 20 \msun) $N_{\rm O} = 19 \pm 6$, assuming a
\citet{Kroupa2001a} IMF at the zero-age main sequence using \citet{Vacca1996a}
stellar parameters.  This estimate provides a lower limit on the present day
O-star population, since some of the short-wavelength radiation (i.e., the NIR
radiation) is able to escape rather than being reprocessed into the far
infrared, which we have used to infer the luminosity.

Of these expected $\sim20$ O-stars, six have been detected in the near infrared
and spectroscopically confirmed.  \citet{Figueredo2008a} found 4 exposed O-type
stars and \citet{Barbosa2008a} found an additional two with strong infrared
excess.  None of these coincide with ultracompact or hypercompact \hii regions,
but all are in the bright and diffuse IRS2 \hii region.

\citet{Mehringer1994a} found an additional eight ultracompact and hypercompact
\hii regions, all of which appear to be B0 or earlier stars based on their
radio-derived ionizing photon luminosity.  The other $\sim$10-20 O-stars
expected to have \emph{already} formed given the observed total luminosity may
be the hypercompact \hii regions we have detected.  However, it is difficult to
explain the total luminosity of the region from hypercompact \hii regions
alone, since their emission is necessarily confined by optical depth to a small
region, and they contribute minimally to the overall radio luminosity.  The
thermal radiation from the OB-stars in the diffuse \hii regions may illuminate
a large portion of the cloud and contribute substantially to the infrared
luminosity.

It is possible that the O-stars providing most of the observed reprocessed
luminosity are near or within the ``shell'' structure of the W51 IRS1 region
(Figure \ref{fig:coverview_diffuse}, Section \ref{sec:diffuseemission}).  This
region dominates the optical and radio luminosity of W51 Main (though it
provides $\lesssim 1/4$ of the FIR luminosity) and is bright enough in
near-infrared and radio continuum emission to make detection of point sources
impossible, explaining why no O-stars have been previously confirmed.  There
are multiple overlapping \hii bubbles in the W51 Main shell at different
velocities (Figure \ref{fig:rrl_chan_main}), suggesting that there are
interacting and  expanding bubbles, which in turn implies that there is a
separation between the driving sources and the \hii region of at least the
bubble sizes, $\sim0.2-0.3$ pc.  This scenario suggests the existence of a
population of un-embedded O-stars spread between the W51 protoclusters.

\subsection{The nature of the continuum sources}
\label{sec:contnature}

Out of the 27 compact continuum sources reported, the majority that we could
identify are ultra- or hyper-compact \hii regions.  The ultracompact \hii
regions are identified as resolved sources, with $r>0.005$
pc\footnote{The
\citet{Kurtz2002a} definition of \hchii regions is $r_{\hii}<0.05$ pc.  We
detect a few sources that are on the borderline between \hchii
and \uchii regions, with $r\sim0.05$ pc, and therefore we have identified all
unresolved sources at the highest resolution (0.005 pc) as \hchii regions, and
all that are well-resolved are classified as \uchii regions.}.  The \hchii
regions are more
challenging to identify, as they are observed only as unresolved sources.
Normally, these could be identified from their SEDs, which should follow an
$\alpha=2$ Rayleigh-Jeans law up to some turnover point, above which they 
become optically thin and turn over to an $\alpha={-0.1}$ power law
\citep{Wilson2009a}.  However, the SEDs we observed do not exhibit such unambiguous
behavior for any cases except e2.  For the rest, we see $\alpha\approx1$ power laws
at low frequencies (e.g., e4), negative power-laws at low frequencies (e.g.,
e9), or entirely flat SEDs (e.g., e10).  SED classifications for other sources
are given in Table \ref{tab:positions}, and plots of the SEDs can be seen in
Appendix \ref{sec:SEDs}.

An SED with $\alpha\sim1$ or $\alpha\sim0$ can be explained by density
gradients in the \hii region \citep{Keto2008a,Galvan-Madrid2009a,Tanaka2016a}.
However, a minimum at $\sim6$ GHz cannot.  Additionally, an optically thick
\hii region that is just unresolved would have $S_{5 GHz} \sim 70$ mJy, so for
the sources discussed in this section with $S_{5 GHz} < 1$ mJy, the upper limit
on the source radius of an optically thick \hii region with electron
temperature $T_e=10^4$ K is $\lesssim160$ au.  If any of these sources are \hii
regions, then, they are among the most compact of \hchii regions in the galaxy,
and by extension must be extremely young or accretion-dominated
\citep{Keto2003a,Peters2010b}.  It is unlikely to find a large
population of very young O-stars at exactly coincident evolutionary stages
distributed across a parsec-scale star-forming region, especially when the clustered
groups of compact \hii regions exhibit a significant size distribution from
$r<0.005$ pc to $r\sim0.05$ pc.  Additionally, most of the detected sources
are not correlated with regions of high molecular gas density: they do not
have a significant reservoir from which to accrete.  We therefore discard the
hypothesis that the faint sources are all extreme \hchii regions.

Negative spectral indices at low frequencies are usually assumed to indicate
synchrotron emission \citep{Wilson2009a,Condon2007a}, but young and forming
stars and \hii regions are generally not strong synchrotron emitters.  However,
colliding-wind binaries (CWBs) often exhibit nonthermal SEDs, likely caused by
accelerated particles in the wind-wind collision zone \citep{De-Becker2013a}.
Stellar winds around massive stars may also exhibit $\alpha<2$ power-laws, but
they are not expected to deviate far from $\alpha\sim0.6-0.7$
\citep{Wright1975a,Panagia1975b,Reynolds1986a}.

Radio jets are sometimes observed to have negative spectral indices
\citep{Moscadelli2013a,Moscadelli2016a}, and they are often faint at distances
comparable to W51, making them plausible candidates for the observed faint
emission.  However, previous radio jets have been identified by their
association with \methanol or \water masers, and the faint continuum sources we
discuss here do not have any associated masers.  Absent any signs of ongoing
accretion, jets are not the most likely source of the observed emission, but
the current data cannot definitively rule them out.

CWBs typically have radio luminosities
$L_{rad}\sim10^{29}-10^{30}$ erg \pers \citep{De-Becker2013a}.   At the
distance of W51, this corresponds to 0.5-5 mJy at 5 GHz, so such binaries
should be detected.  By contrast, single star winds range from $L_{rad} \sim
10^{27}-10^{29}$ erg \pers, with only a few known in the high range
\citep{Bieging1989a}, so these are less likely to be detected.

Both CWBs and radio-bright stellar wind sources should be bright in the
optical and near-infrared, with stellar luminosities $10^5-10^6$ \lsun.
Much of the W51 protocluster region is obscured in the infrared by extinction
from the Galactic plane and the W51 GMC, but in the near-IR an unextincted O9
star would have
$m_K\sim10.5$
\citep{Pecaut2013a}\footnote{\url{http://www.pas.rochester.edu/~emamajek/EEM_dwarf_UBVIJHK_colors_Teff.txt}},
or $m_K\sim 13.0$ with Galactic disk line-of-sight extinction $A_K=2.6$
\citep{Goldader1994a}.  Some of these stars therefore ought to be detected in
the NIR; we report such detections in Table \ref{tab:associations}.

Based on these arguments, we have classified each detected compact source in
Table \ref{tab:positions}.  We report 12 candidate CWBs throughout the observed
region.  These sources constitute a large fraction of the un-embedded O-star
population discussed in Section \ref{sec:stellarmass}, and their presence confirms
that un-embedded, main-sequence O-stars reside in the same cloud as the still-forming
clusters.

\subsubsection{The faint radio continuum sources in W51 Main}
\label{sec:faintw51main}

The sources e20, e21, and e22 are likely members of a distributed O-star
population, rather than members of the clusters of compact \hii regions (Figure
\ref{fig:w51e20cluster}).  They were detected by \citet{Goldader1994a} in the
K-band at $m_K < 12$, making these the brightest NIR sources outside of W51
IRS2 (Table \ref{tab:associations} gives their associated NIR source names
from large surveys).  Their luminosities suggest they are O-type stars.  Since
they were also detected in X-rays \citep{Townsley2014a}, these are strong
candidate CWBs.  The high infrared K-band brightness and weak radio
continuum from these sources, in contrast with the bright radio continuum and
absence of infrared emission from e1/e2, supports this hypothesis.  
Assuming they are early O-stars, e20, e21, and e22 are capable of providing the
ionizing and infrared luminosity observed from the W51 IRS1 \hii region.

\FigureTwo
{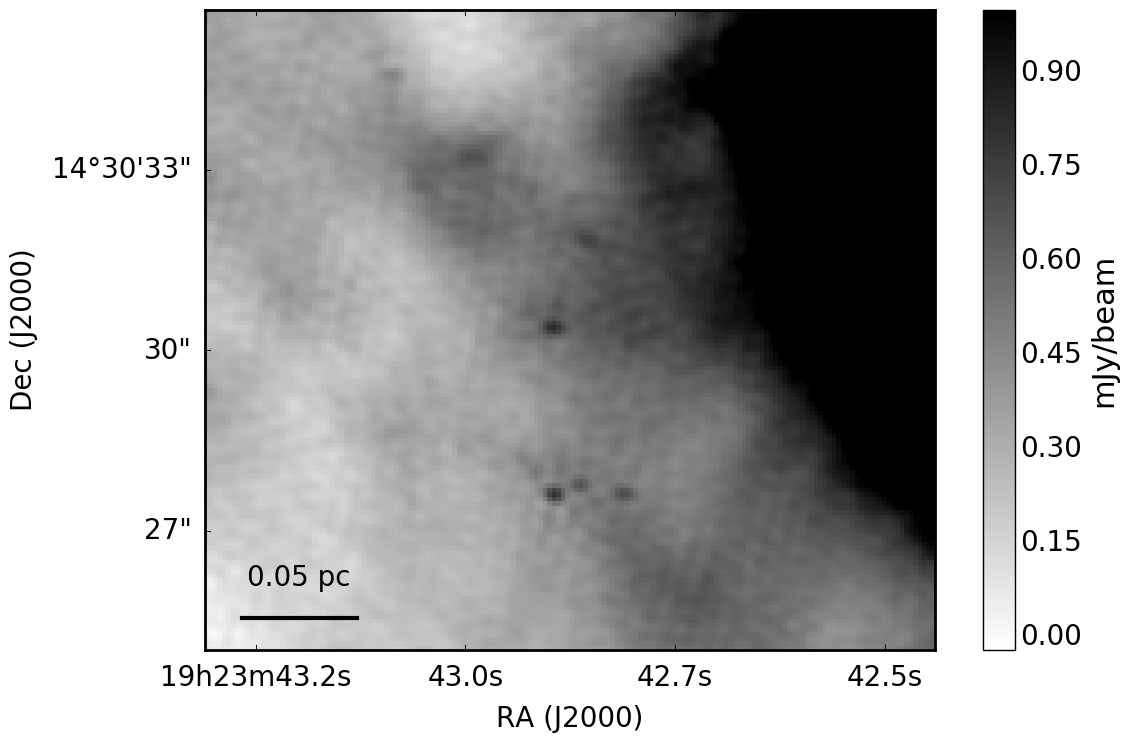}
{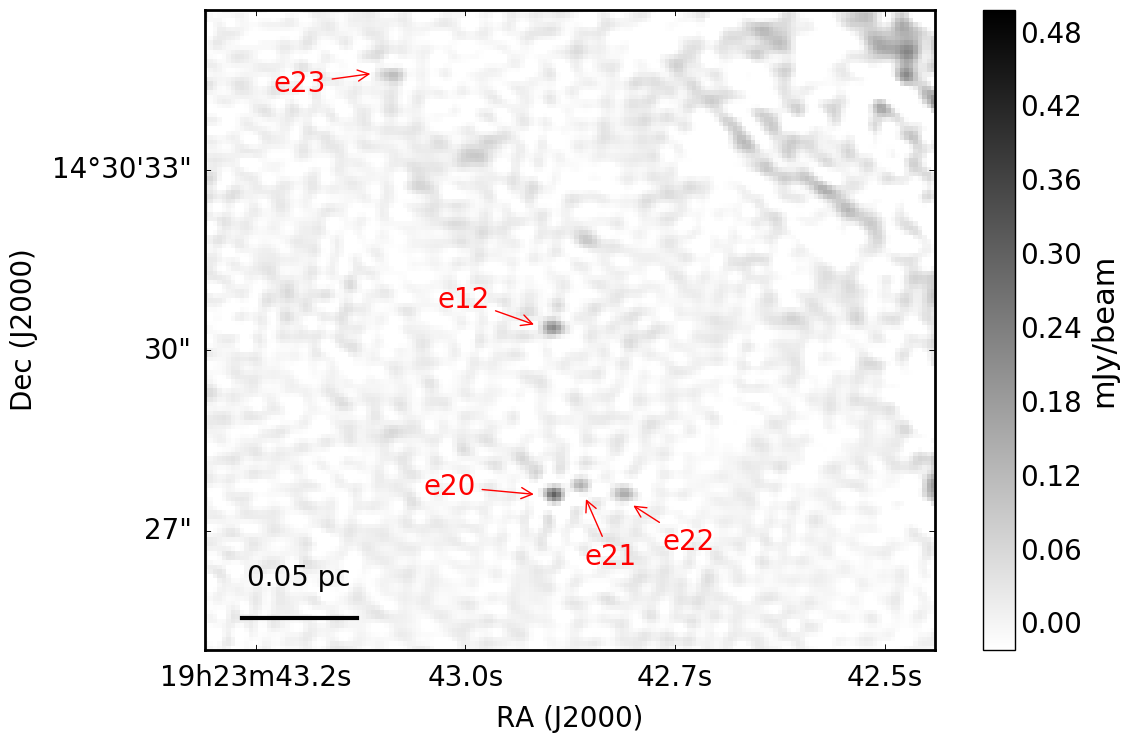}
{({\it a}) A C-band image of the W51e20/e21/e22 cluster.
({\it b}) An unsharp-masked version of the image with a 0.3\arcsec smoothing
kernel. 
Sources e20, e21, and e22 correspond to two detected infrared K-band sources from
\citet[][Table \ref{tab:associations}]{Goldader1994a} and have K-band
luminosities consistent with spectral type O4V or earlier.  The other point
sources seen in this field are not detected in the near infrared.
}
{fig:w51e20cluster}{1}{3.5in}

Some questions about the continuum sources remain unanswered: How close are the
e1/e2 and e20/e21/e22 clusters?  They are separated by 0.36 pc in projection,
but there is no direct information about the line-of-sight velocity of the e20
cluster.    The other sources in this general area that do not have infrared
counterparts, e12, e13, and e23, are also likely OB stars behind
infrared dark clouds that obscure them in the near infrared.  It is possible that
all of the W51 IRS1 and W51 Main compact sources are within a common $\sim
0.5$ pc sphere.

\Figure
{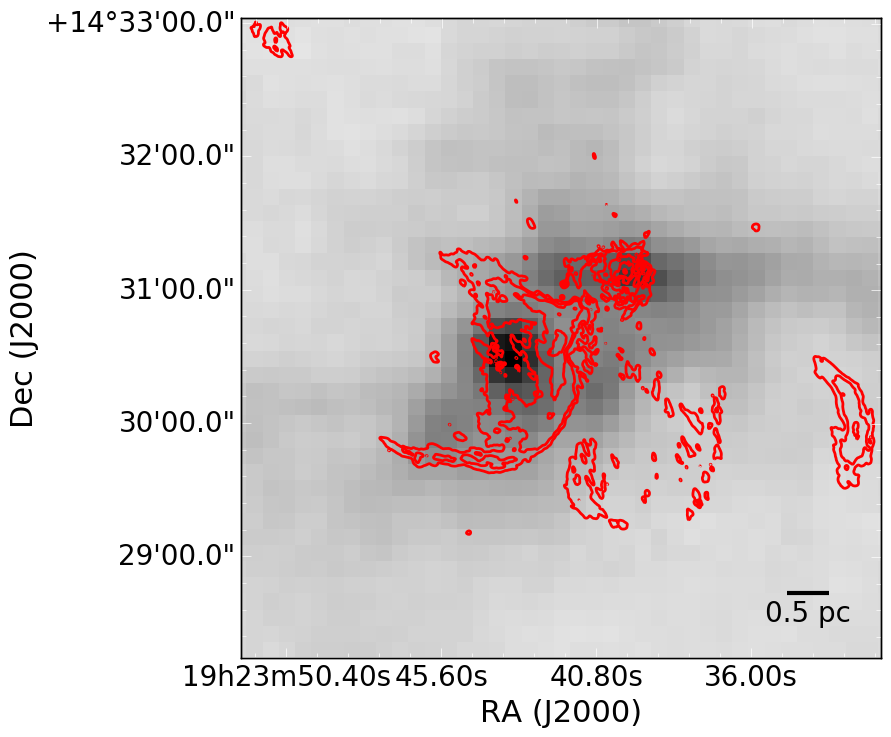}
{Contours of the C-band (4.9 GHz) continuum overlaid on an integrated intensity
map of \ceighteeno 3-2 from 45 to 65 \kms \citep{Parsons2012a}.  The contours
go from 0.1 to 10 mJy in 5 logarithmic steps.  The \ceighteeno peaks on the
e1/e2 region and has a clear minimum corresponding to the peak of the radio
continuum emission.
}
{fig:contonco}{1}{6.5in}

\subsubsection{The faint radio continuum sources in W51 IRS2}
\label{sec:faintw51irs2}

Of the compact radio sources identified near W51 IRS2, all but d2 have infrared
associations (Table \ref{tab:associations}).  Four of them also have X-ray
associations.  Except for d3, which is associated with a moderately
extended \hii region, these are candidate CWBs.

An alternate explanation for the SED, location, and variability of d4 is
discussed in Section \ref{sec:d4}.

\begin{table*}[htp]
\caption{Source Associations}
\begin{tabular}{llll}
\label{tab:associations}
Source Name & X-ray & NIR & Goldader 1994 \\
 &  &  &  \\
\hline
d3 & - & UGPSJ192335.88+143128.8 & - \\
d4e & CXOJ192339.6+143130 & UGPSJ192339.65+143130.9 & - \\
d4w & (same as d4e) & (same as d4e) & - \\
d5 & - & UGPSJ192341.77+143127.6 & - \\
d6 & CXOJ192341.1+143110 & UGPSJ192341.29+143111.8 & - \\
d7 & CXOU192340.96+143106.7 & UGPSJ192340.91+143106.7 & RS15 \\
e7 & - & UGPSJ192344.79+142911.2 & - \\
e14 & - & UGPSJ192342.60+143042.2 & - \\
e15 & - & UGPSJ192338.65+143005.8 & - \\
e20 & CXOU192342.86+143027.5 & UGPSJ192342.85+143027.7 & RS7 \\
e21 & (same as e20) & (same as e20) & (same as e20) \\
e22 & CXOU192342.77+143027.5 & UGPSJ192342.84+143027.5 & RS8 \\
\hline
\end{tabular}

\end{table*}

\subsection{\formaldehyde emission features}
\label{sec:h2coemission}

\formaldehyde is generally a good tracer of molecular gas, showing up at typical
abundances $\sim10^{-9}$ \hh wherever CO is detected \citep[e.g.,][]{Mangum1993a}.
The centimeter lines of \ortho are often used to measure gas density when
detected in absorption \citep[e.g.,][]{Ginsburg2011a,Zeiger2010a}.
\ortho \oneone and \twotwo are commonly observed in absorption
but rarely in emission \citep[e.g.,][]{Mangum1993a,Araya2007b}.  The \twotwo
line has been observed in emission in the starburst M82 \citep{Mangum2008a},
behind the Orion nebula
\citep{Evans1975a,Kutner1976a,Batrla1983a,Johnston1983a,Bastien1985a,Wilson1989a},
in $\rho$ Ophiucus B
\citep{Loren1980a,Loren1983a,Martin-Pintado1983a,Wadiak1985a}, and in DR 21
\citep{Wilson1982a,Johnston1984a}.  However, the \twotwo line has so far only
been observed in absorption at large distances ($d>2$ kpc) within the Galaxy.

We have detected three regions of \twotwo emission in the W51 region, all
corresponding to previously detected hot molecular cores
\citep{Zhang1997a,Shi2010a,Shi2010b,Goddi2015a,Goddi2016a}.   W51e2 is partially
surrounded by a `halo' of \formaldehyde \twotwo emission to its northeast; the
\hchii region itself shows only \twotwo
absorption because the continuum source is bright
(Figure \ref{fig:w51bridge22emispec} and \ref{fig:w51mainemicontours}).
The \formaldehyde emission corresponds to the cores W51e2e and W51e2nw and the diffuse
emission between them \citep{Goddi2016a,Shi2010a}.
The \hchii region W51e8 exhibits extended \twotwo emission, including a
somewhat diffuse region between W51e4 and W51e1.  Finally, in W51
IRS2, there is extended \twotwo
emission between W51d1 and W51d2, adjacent to the \ammonia masers
observed by \citet{Zhang1995a} and more recently \citet{Goddi2015a}, and
aligned with the \citet{Zapata2010a} W51 North Disk and thermal \ammonia emission (Figure
\ref{fig:w51northcore}).

\FigureTwo
{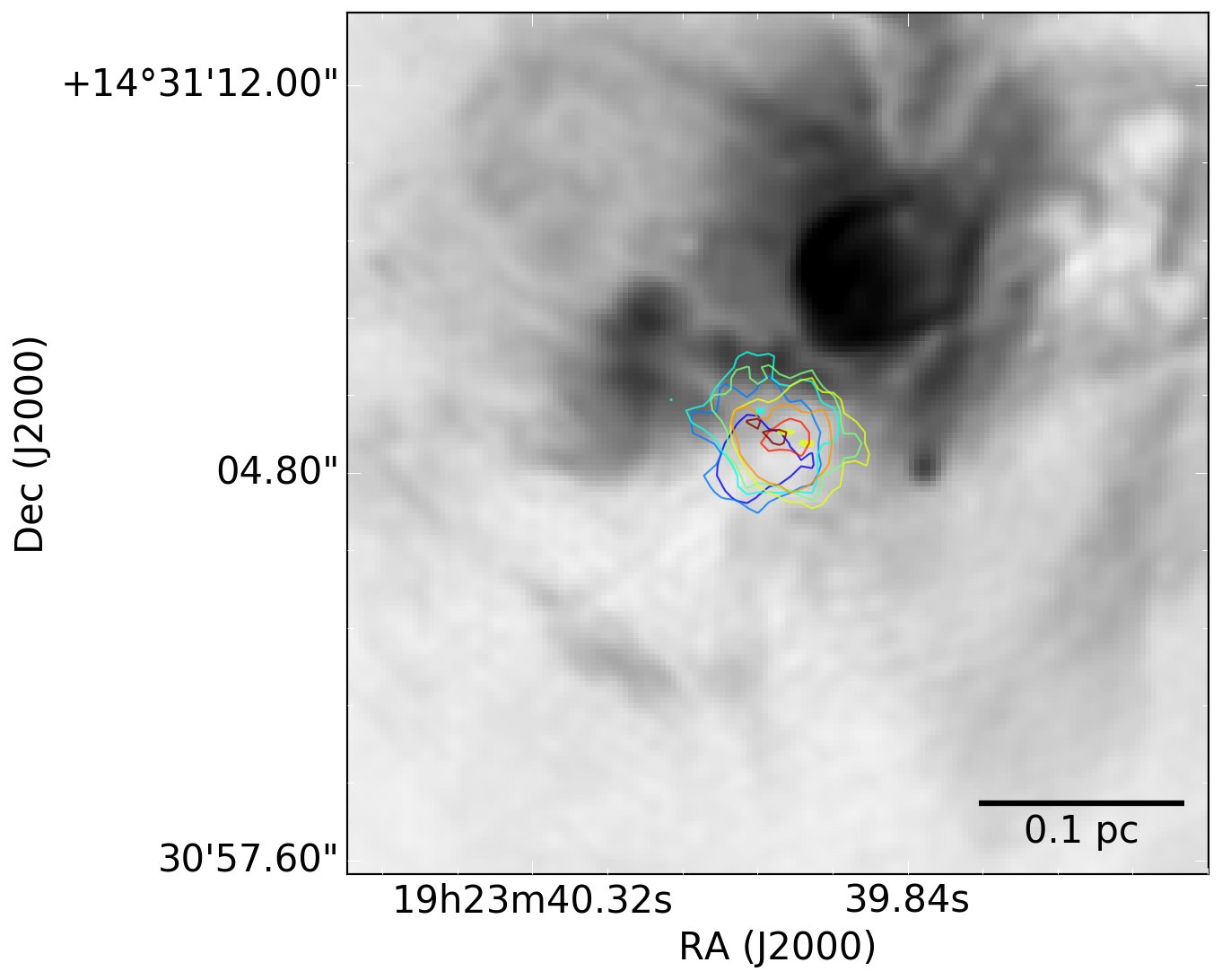}
{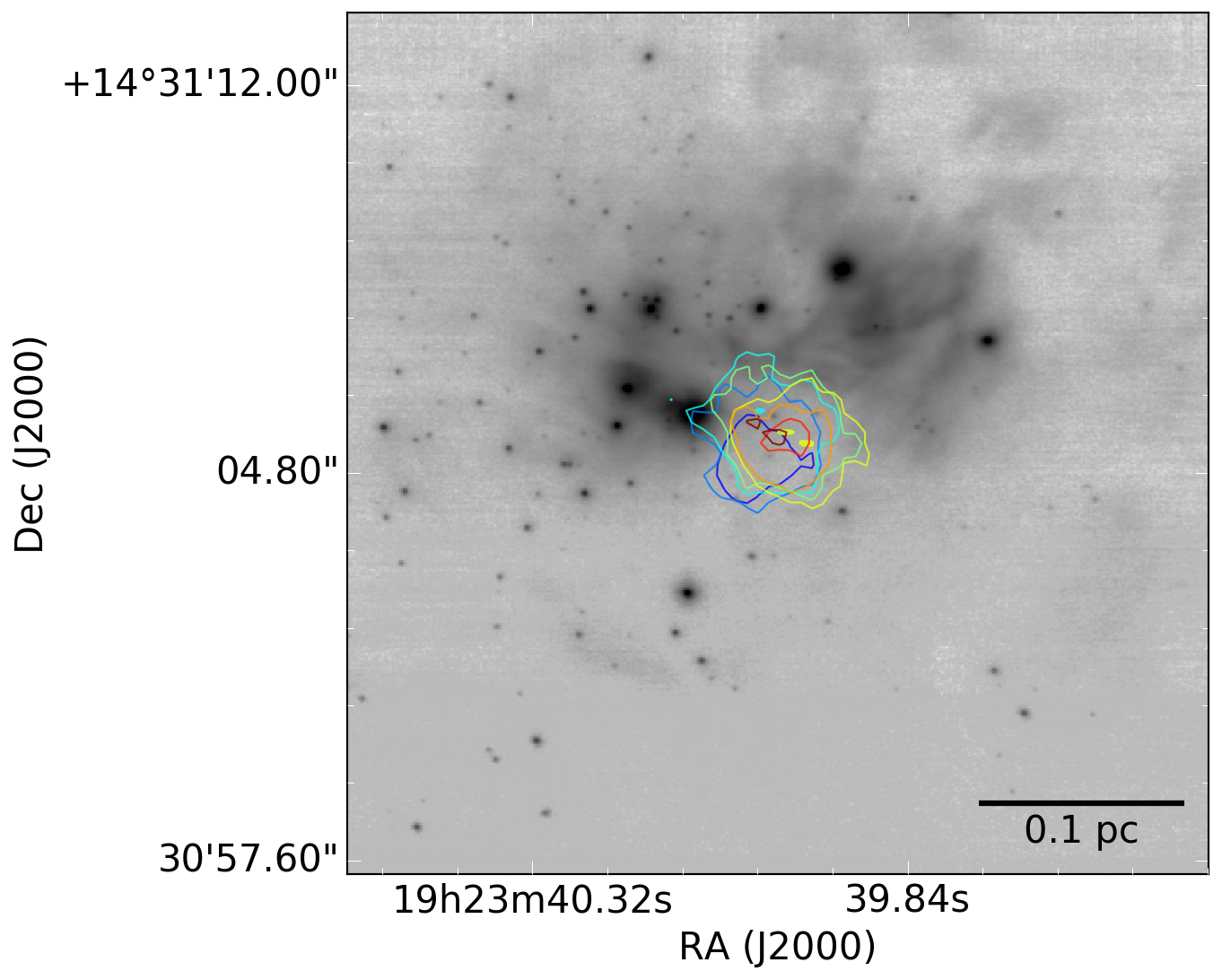}
{The W51 North core shown in \formaldehyde \twotwo emission in contours
overlaid on (a) the 14 GHz continuum and (b) the NACO K-band continuum image
\citep{Barbosa2008a}.  The contours are at 3 mJy/beam in each channel with
width 0.5 \kms at velocities 56 (blue) to 60 (red) \kms.
}
{fig:w51northcore}{1}{3.5in}

In all cases, the emission is extended and spread smoothly over multiple
velocity channels (Figures 
\ref{fig:w51bridge22emispec}, \ref{fig:w51e2emispec}, \ref{fig:w51e1emispec}).
It is therefore not maser emission.

None of these regions are detected in \oneone emission in our data.  These
nondetections are likely because our brightness sensitivity at C-band (4.9 GHz)
is poor.  The upper limit is $T_B<300$ K (Table \ref{tab:cubes}), ruling
out the presence of strong masers.  Additionally, the foreground cloud is a
much stronger absorber in the \oneone than in the \twotwo line, so any emission
is likely to be obscured by foreground absorption.

\Figure
{{f22}.png}
{Spectrum of the W51 North core in \ortho \twotwo.  The spectrum
is the average over a $2\arcsec\times1.5\arcsec$ elliptical aperture extracted
from the Briggs-weighted cube (about 14$\times$ the beam area).  The grey
shading shows the $1\sigma$
errorbars.
}
{fig:w51bridge22emispec}{0.5}{3.5in}

\Figure
{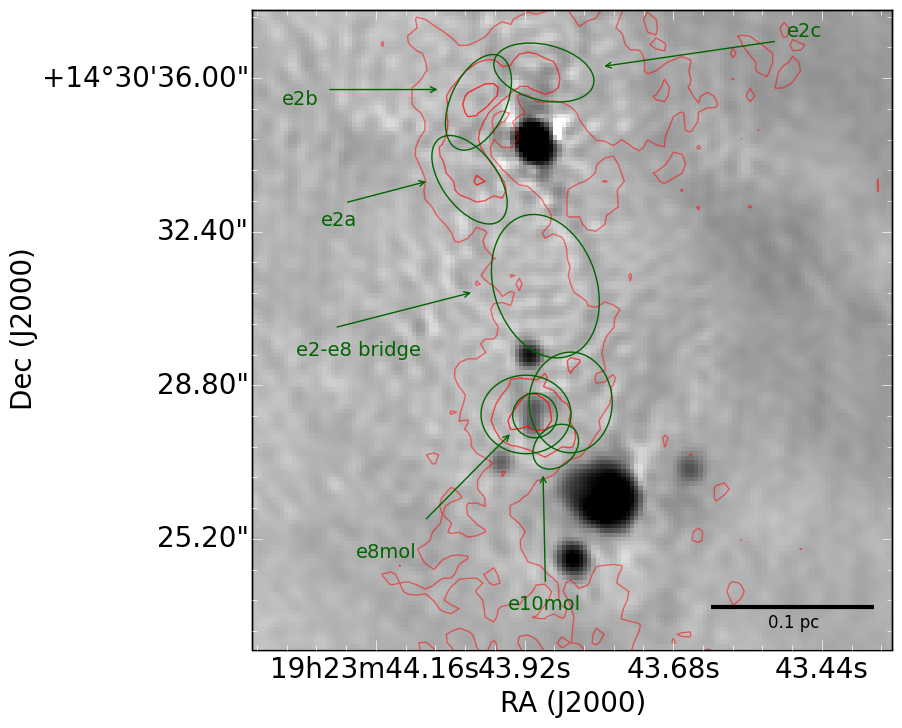}
{ Peak intensity contours (red) of the naturally weighted \formaldehyde \twotwo
emission ($\theta_{FWHM}\approx1\arcsec$) from 55 to 62 \kms in the e1/e2
region superposed on the Ku-band continuum image
($\theta_{FWHM}\approx0.33\arcsec$).  The contours are at 2, 4, 6
mJy/beam.  The green circles show
the apertures used for the spectra shown in Figures
\ref{fig:w51e2emispec} and
\ref{fig:w51e1emispec}.  The larger apertures around e8mol and e10mol are
e8mol\_ext and e10mol\_ext.  The feature denoted `e2-e8 bridge' is a diffuse
feature that is seen in the naturally-weighted but not in the robust-weighted
images.  e2a and e2b in this figure are both associated with
the SMA and VLA source e2e \citep{Shi2010a,Goddi2016a}.  e2c is associated with
the source e2nw.
}{fig:w51mainemilabels}{1}{6.5in}

\FigureThreeAA
{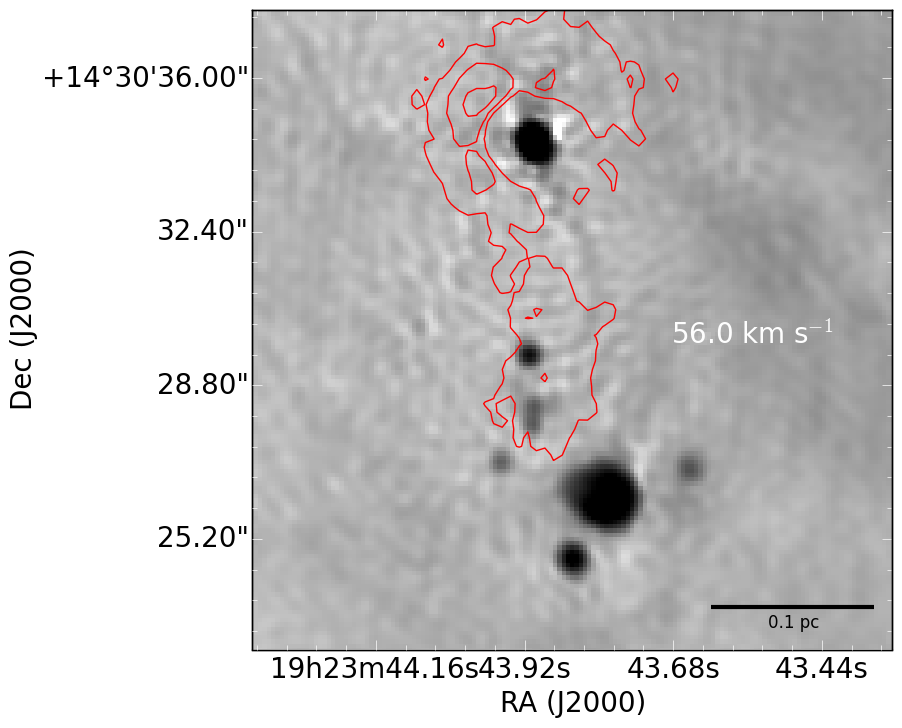}
{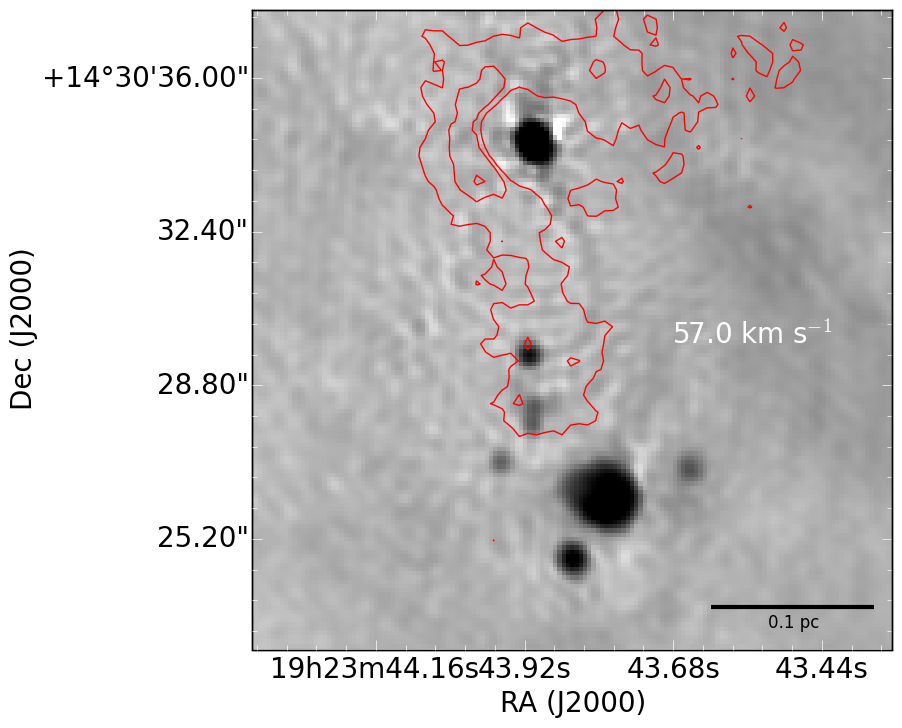}
{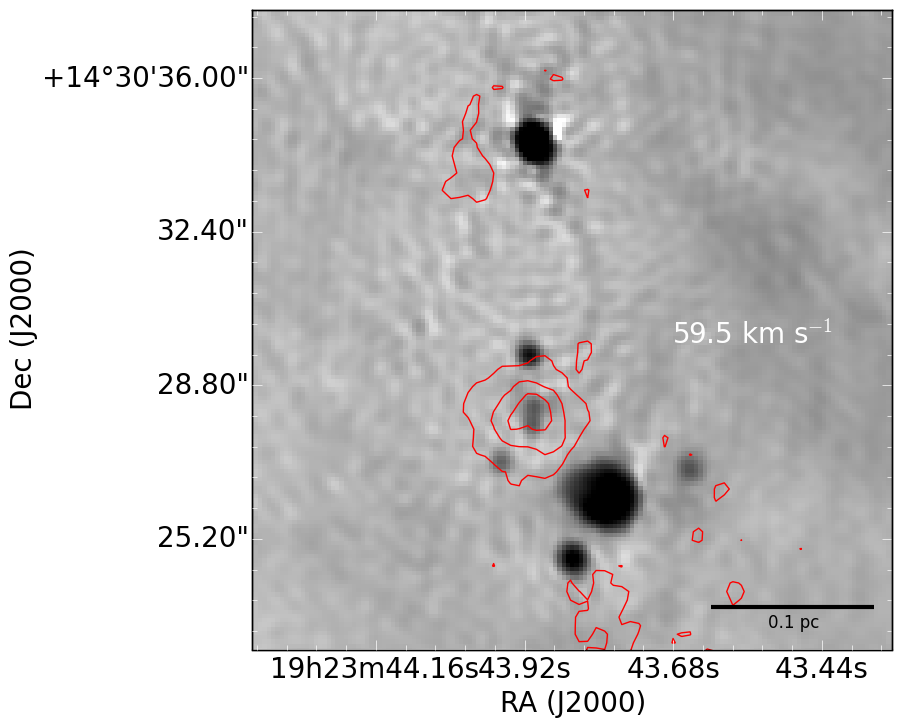}
{Contours of the \formaldehyde \twotwo naturally weighted emission in the W51
Main region at 3 velocities superposed on the Ku-band 14.5 GHz continuum map.
(a) shows the peak velocity of the e2 core, where the center of the core is
missed because it is
in absorption against the bright continuum peak, (b) shows the overlap
velocity, and
(c) shows the e8 core}
{fig:w51mainemicontours}{1}{2.1in}

\FigureThreeAA
{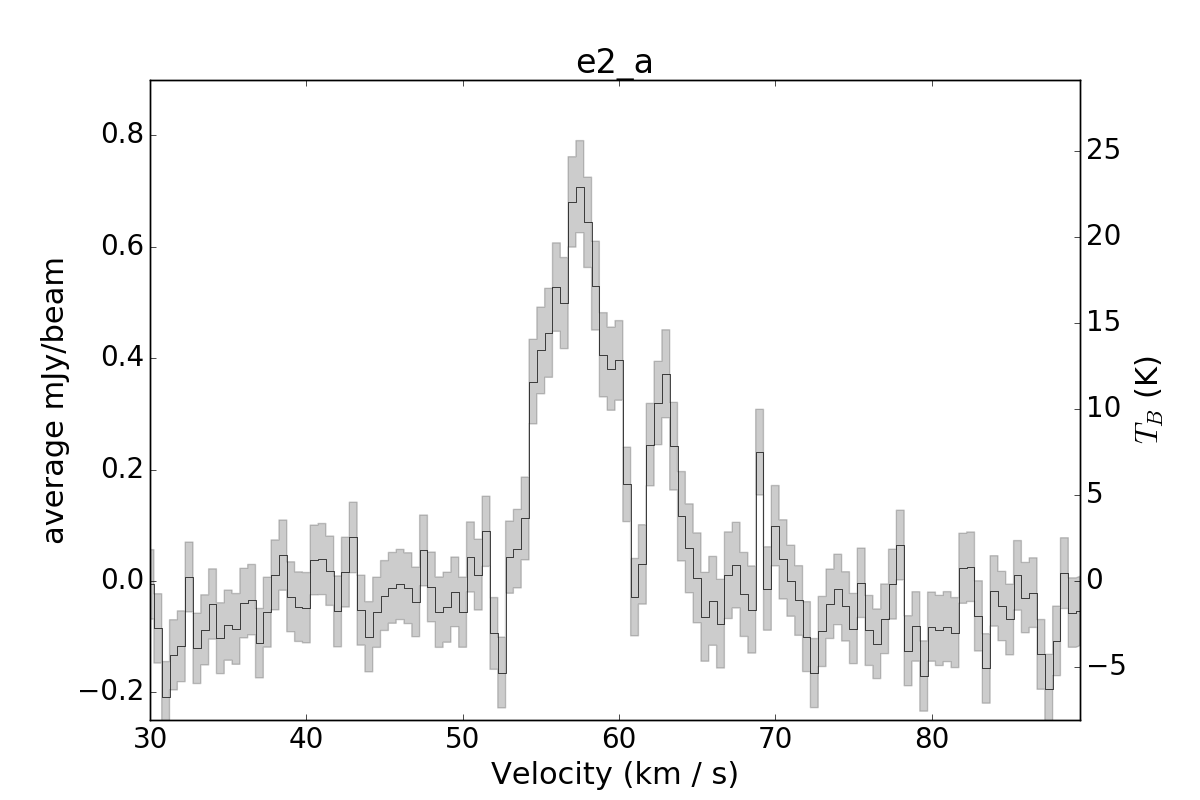}
{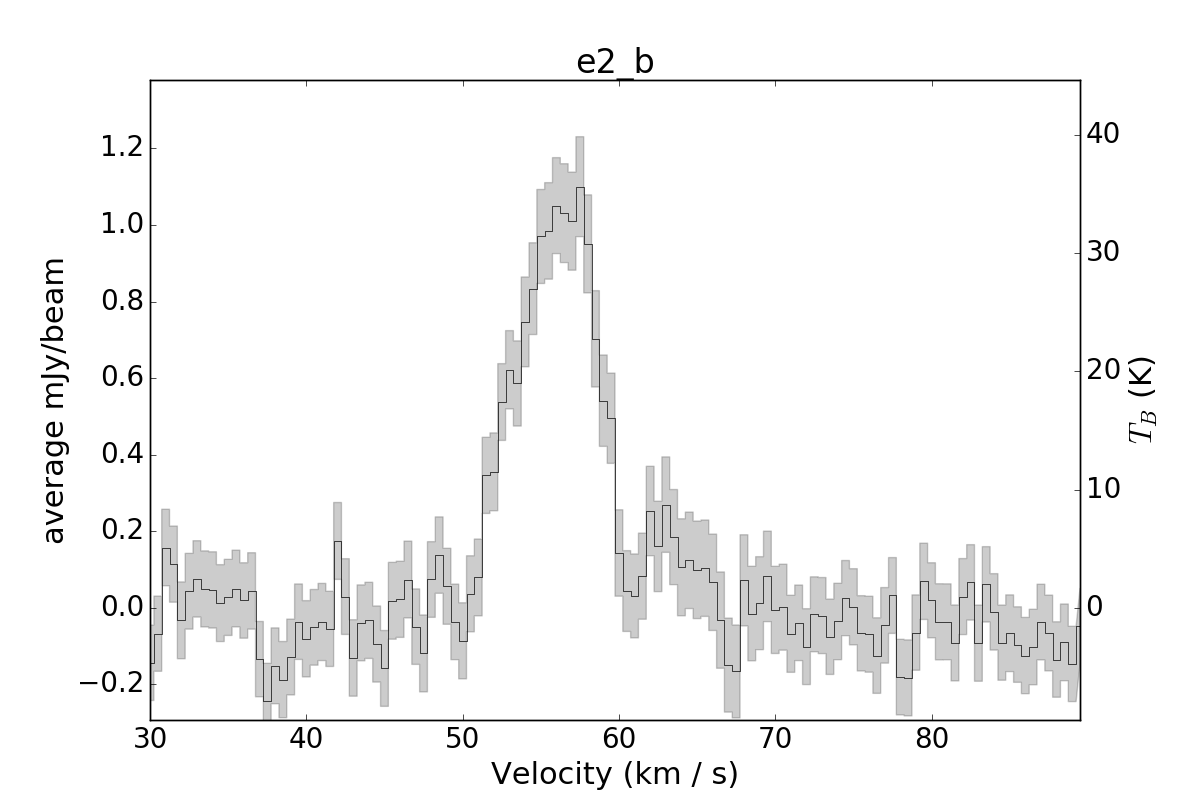}
{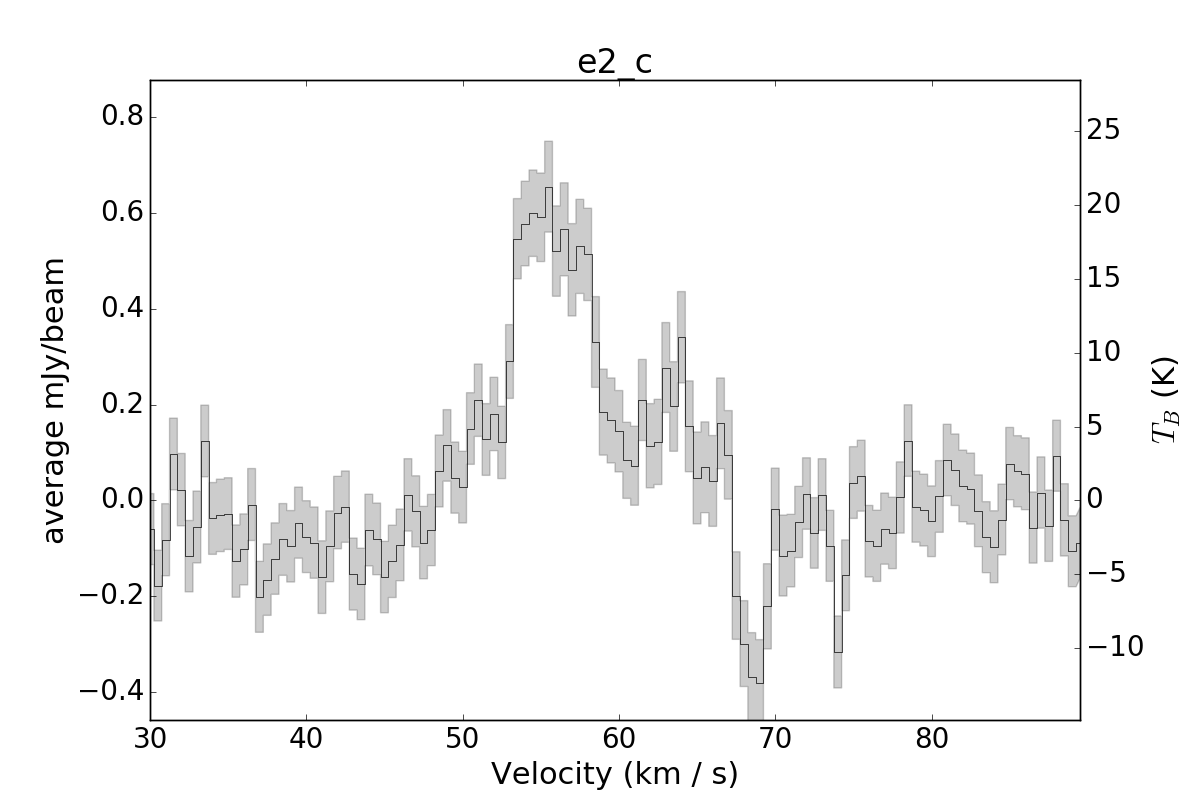}
{Spectra of the \twotwo emission around W51e2 at three positions (a)
north/northwest (b) northeast (c) southeast of the e2 \hchii region.  See Figure
\ref{fig:w51mainemilabels} to see where these were extracted.  The left axis
shows the average brightness over the aperture in mJy/beam, while the right axis
shows the brightness temperature.}
{fig:w51e2emispec}{1}{2.25in}

\FigureTwoAA
{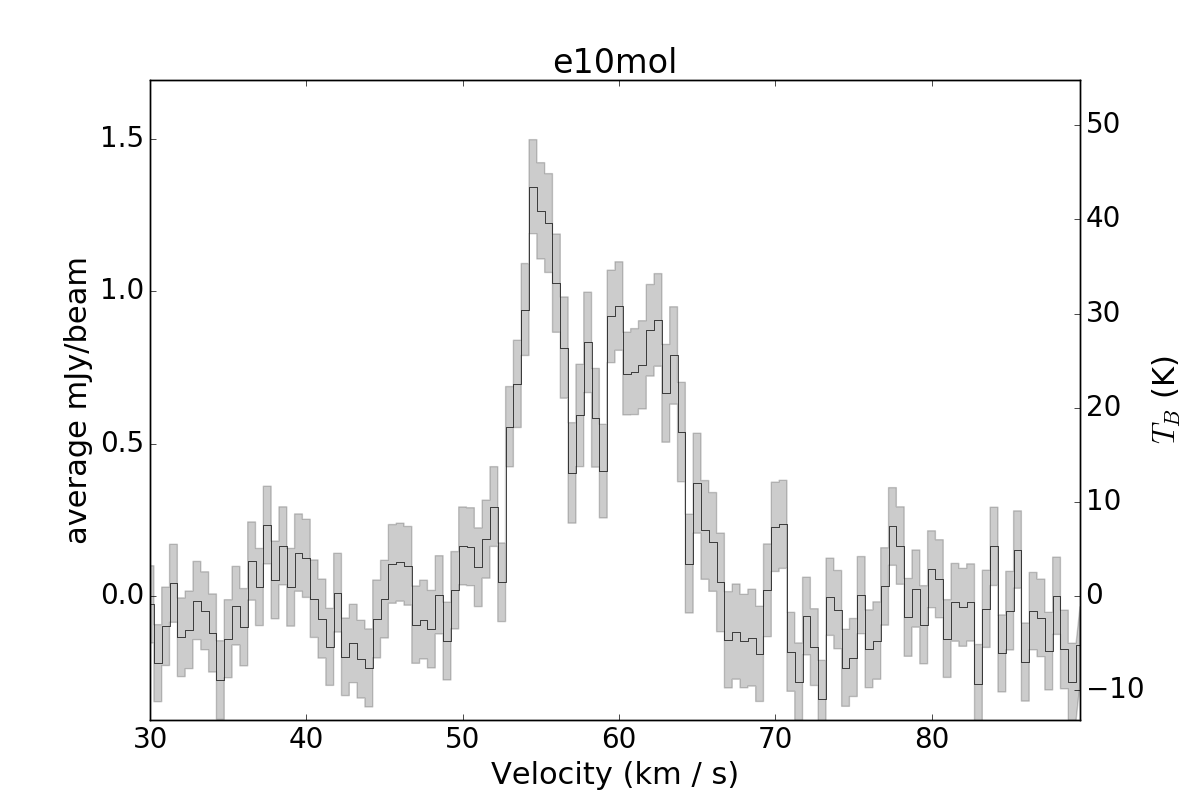}
{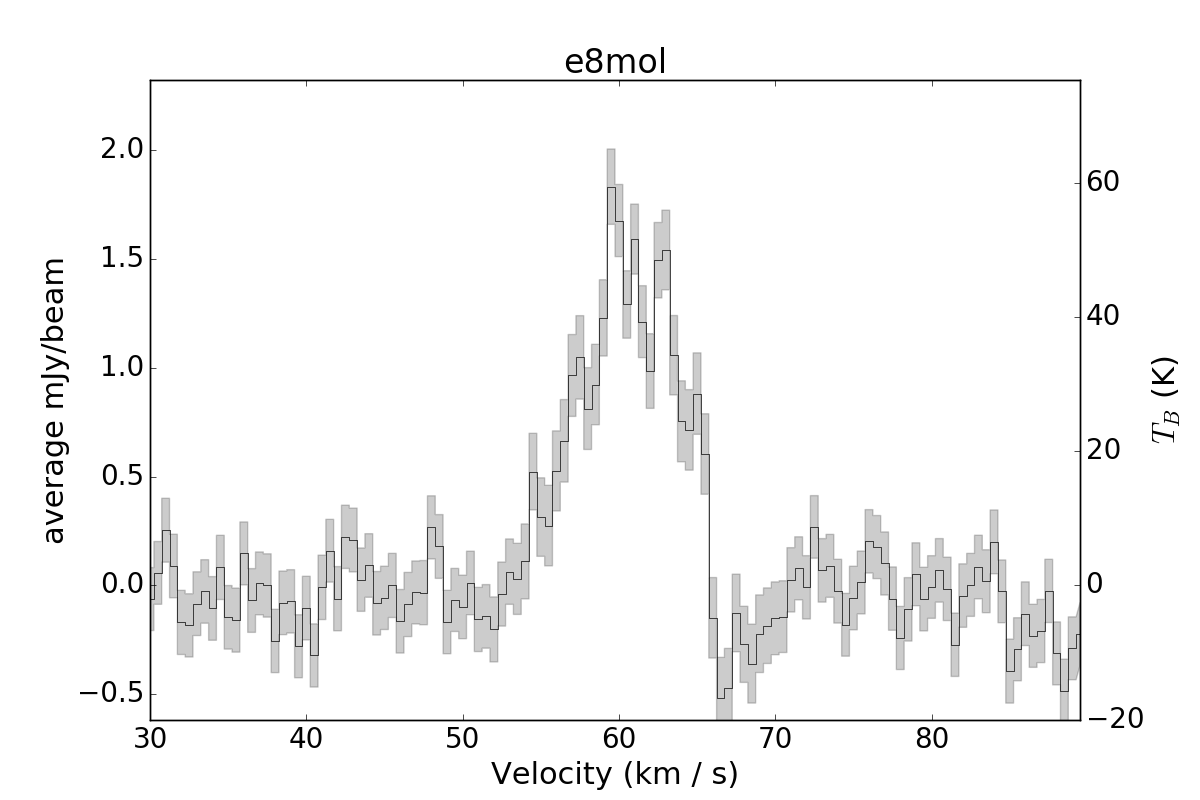}
{Spectra of the \twotwo emission around W51e1.  See Figure
\ref{fig:w51mainemilabels} to see where these were extracted.  The left axis
shows the average brightness over the aperture in mJy/beam, while the right axis
shows the brightness temperature.
}
{fig:w51e1emispec}{1}{3.5in}

The W51e2, W51e8, and North cores share some properties but are not identical.
All three exhibit peak \formaldehyde \twotwo brightness temperatures
$T\sim30-60$ K at 14.5 GHz.  W51e8 has a
brightness temperature and spatial extent similar to that of W51 North, though
it is also connected with the e10 molecular emission and is morphologically
more complex.  W51e8 contains a hypercompact \hii region, while W51 North does
not.  However, in W51 North, detection of a hypercompact \hii region may be prevented by
confusion with the nearby bright IRS2: it is therefore plausible that
W51 North contains a similar continuum source to W51e8 ($S_\nu \sim 1$ mJy).
Unlike W51e8 and North, W51e2 has a more extended morphology including
the cores W51e2e and W51e2nw, and it includes the bright \hchii region e2w
\citep{Shi2010a,Goddi2016a}.

We estimate the mass of the W51 North core in a few ways.  A reasonable lower limit
on the mass is given by assuming it has a volume density $n\gtrsim10^6$ \percc,
which is required to detect it in emission in \ortho \twotwo
\citep{Mangum1993a}.  The core radius, as measured by fitting a 2D gaussian to
its integrated intensity emission map, is
$\sigma=0.9\arcsec$ or 0.025 pc (FWHM=2.1\arcsec, 0.056 pc).  If we assume the core
is spherically symmetric, the resulting mass is $M\gtrsim14\msun$.
We compute the \ortho \twotwo column density using \citet{Mangum2015a} equation
100, with a measured integrated intensity of 55 mJy \kms or 65 K \kms given
the $2\times1.5$\arcsec source area.  The inferred column of \ortho,
assuming LTE, is $\sim5\ee{15} - 5\ee{16}$ \persc, depending on the assumed
kinetic temperature and excitation temperature (20-200 K).  This \formaldehyde column
implies an implausibly large mass ($M\sim5\ee{3}$ \msun) for typical assumed abundances
($X_{\formaldehyde}\sim10^{-9}$), so we conclude that the \formaldehyde abundance
must be 2-3 orders of magnitude greater than in the molecular cloud, with
$X_{\formaldehyde}\sim3\ee{-7}$ providing a mass consistent with the density
lower-limit based mass.  For a reasonable range of `core' masses, $20
\msun<M<110 \msun$, the implied abundance is $3\ee{-7} > X_{\formaldehyde} >
5\ee{-8}$.  The source e8 similarly requires high abundance to explain its
emission.  While these inferred abundances are high, they are consistent with
measurements of \formaldehyde abundance in the Orion Hot Core and Compact Ridge
\citep{Mangum1993b}.  The enhanced abundance is likely caused by radiation-driven
desorption of \formaldehyde from dust grains \citep{Shalabiea1994a,van-Dishoeck1998b}.

\subsection{The d4 variable source \& \hh region}
\label{sec:d4}
The source d4e exhibits $\sim$20-year timescale variability.  d4e has
brightened from $<0.6$ mJy ($3\sigma$ upper limit) to $>1.0\pm0.06$ mJy.  We
consider a few possibilities for the origin of this variable emission: a
protostellar jet,  a stellar wind, ionized accretion onto a star, or a knot in
a high-velocity outflow.  While we cannot definitively rule out any of these
scenarios, we favor the outflow-ISM interaction because it explains the
coincidence of d4e and d4w with extended NIR \hh and [Fe II] and X-ray
continuum emission.

Protostellar jets exhibit radio emission that can be variable.  Radio
variability has been observed in nearby star forming regions
\citep{Liu2014c,Forbrich2013a} and associated with variability in accretion or
jet related processes, but all sources seen in these regions are orders of
magnitude fainter than d4e, so a jet from a low-mass star is unlikely to be
responsible for the emission.  Radio jets from high-mass stars, e.g., the W75N
jet \citep{Carrasco-Gonzalez2015a}, are also too faint: the expanding jet in
W75N would have a flux density $S_{5 GHz} < 0.1$ mJy at the distance of W51.
Nonetheless, jets are highly variable and may interact with a surrounding
medium of dramatically different density in different regions, resulting in
wildly different luminosities.  We therefore cannot rule out the possibility
that the emission comes from the formation of a jet, and it is even plausible
that the d4 pair represents two opposing jets forming the base of a larger
outflow structure.

In keeping with the SED interpretation from Section \ref{sec:contnature}, it is
possible that the d4 sources are massive stars with strong stellar winds.  In
this interpretation, the overlap with the NIR \hh and [Fe II] outflow emission
\citep[Figure \ref{fig:d4h2}][]{Hodapp2002a} may either be pure
coincidence, or the winds of the massive stars may be interacting directly with
the outflow from IRS2.  While it is highly unlikely that an outflow would
impact these rare windy stars, the interaction would provide a natural
explanation for the large line widths observed in the near infrared and for the
presence of radio emission.  However, the weakness of the NIR K-band detection
($m_K \sim 15$, and the emission is extended) suggests no massive stars are
present.

The radio emission could come from ionized jets or magnetospheric emission
associated with accreting young stars.  However, the brightest T Tauri stars in the
Orion Nebula Cluster (ONC) have $L_{rad}<10^{28}$ erg \pers, or a flux density
$S_{5 GHz} < 0.1$ mJy \citep{Forbrich2013a,Zapata2004a}, so it is unlikely that
these are accreting low- or intermediate- mass stars.  

These sources could be very compact \hchii regions, in which case their
variability may be caused by variable accretion flow
\citep{Peters2010c,Galvan-Madrid2011d,de-Pree2014a}.  In this scenario, the
source d4e was fainter 20 years ago because the accretion rate was higher,
which meant that the \hchii region was smaller.

The interaction between protostellar outflows and the ISM can result in radio
and X-ray emission at great distances from the driving source \citep[e.g.,
HH80/81][]{Lopez-Santiago2013a,Masque2015a}.  There is a region of \hh emission,
MHO 2419, coincident with d4 \citep[Figure
\ref{fig:d4h2}][]{Hodapp2002a,Froebrich2011a}.  The \hh and [Fe II] emission
come from bow shocks based on morphological and velocity structure in the
\citet{Hodapp2002a} spectra, and these bow shocks point back to an origin in
IRS2.  The high velocity [Fe II] features ($\pm150$ \kms) are a strong
indication that this is a bow shock from a high-velocity outflow.  It is
therefore possible that the d4 sources are emission features within the bow
shock.  The variability arises because the shocks are short-lived transient
phenomena.  The flux density of d4e and d4w is consistent with that observed
in HH 80/81 \citep{Masque2015a} to within an order of magnitude, though d4
is more luminous.  Given all of the observed coincident features, d4 is best
explained as an outflow knot.  

\Figure{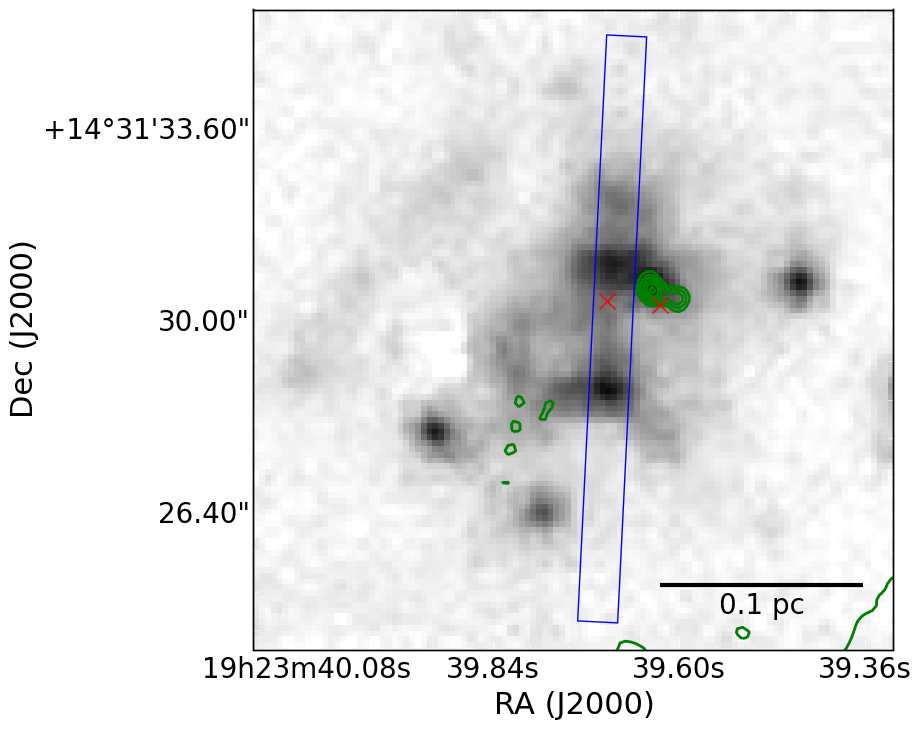}
{A UWISH2 \hh image with Ku-band contours of sources d4e and d4w overlaid at
0.15, 0.30, 0.45, 0.60, 0.75, and 0.90 mJy/beam.  The box shows the slit
position used in \citet{Hodapp2002a}, in which [Fe II] emission spanning
$>200$ \kms was detected.  The X's mark positions of MOXC X-ray
sources \citep{Townsley2014a}.  Both radio sources are considered
associated with the closer MOXC source in Table \ref{tab:associations}.
This figure is discussed in Section \ref{sec:d4}.
}
{fig:d4h2}{1}{3.5in}

\subsection{The velocity dispersion in the W51e cluster}
\label{sec:vdisp}
We detect H77$\alpha$ toward five of the eight hyper/ultra compact \hii regions
in the W51e1/e2 cluster; four of these five are firm detections, one
is weak.  We also determine a velocity from the \formaldehyde emission from
e8mol and more uncertainly from e10mol.  Although the molecular and ionized
emission trace different processes, we treat the line-of-sight velocities of
both tracers as though they are representative of the stellar velocity for this
analysis.  With such compact \hii regions, the ionized gas is unlikely to be
significantly shifted from the stellar velocity.  The core emission is likely
to be dominated by the densest component nearest the central accreting source.

The resulting 1D velocity dispersion is $\sigma=2.0$ \kms if we exclude the
uncertain sources e9 and e10mol; with e9 and e10mol included the dispersion
increases to 3.6 \kms.  This velocity dispersion is confined to $r<5.4\arcsec$
or $r<0.13$ pc, though the (proto)stars are not symmetrically distributed. 
The more compact and round e1 subcluster, which contains the
symmetrically-distributed sources e1, e3, e4, e9, e8mol and e10mol, has a
velocity dispersion $\sigma=2.0$ \kms in $r=2.9$ \arcsec (0.07 pc).

If we assume all six sources in the e1 subcluster represent OB stars
($M>8\msun$) and the stellar masses are distributed according to a standard
\citet{Kroupa2001a} mass function, the cluster should be star-dominated with a
mass $M_{cluster}=600\pm170$ \msun.  
The stellar virial mass in the e1 subcluster implied by its velocity dispersion 
and radius is $M_{vir,*}=670$ \msun, given that (1)
the stars are virialized, which
is appropriate if the stellar mass is dominant \citep{Kruijssen2012b}
and (2)
the stars follow a \citet{Plummer1911a}
profile \citep[cf.][though the inferred mass is insensitive to the
assumed radial profile]{Portegies-Zwart2010a}. 
The agreement between the luminous and virial masses hints that the e1
subcluster is stellar dominated, which can be checked independently by putting
constraints on the gas mass.  The
lower limit on the gas density from detecting \formaldehyde in emission is
$n>10^6$ \percc, which implies a gas mass $M_{gas} > 100$ \msun within the e1
subcluster.  Overall, the observations of e1 are consistent with it being a
stellar-dominated but nonetheless gas-rich ($M_{gas}/M_{*} > 0.15$) system.

Given the mass of W51 Main and its $\sim10$ \kms escape velocity, all of these
(proto)stars are bound to the overall cluster gas.  In the innermost region,
where the stellar mass is dominant, the stars are also likely to be bound to
themselves.

\FigureTwo
{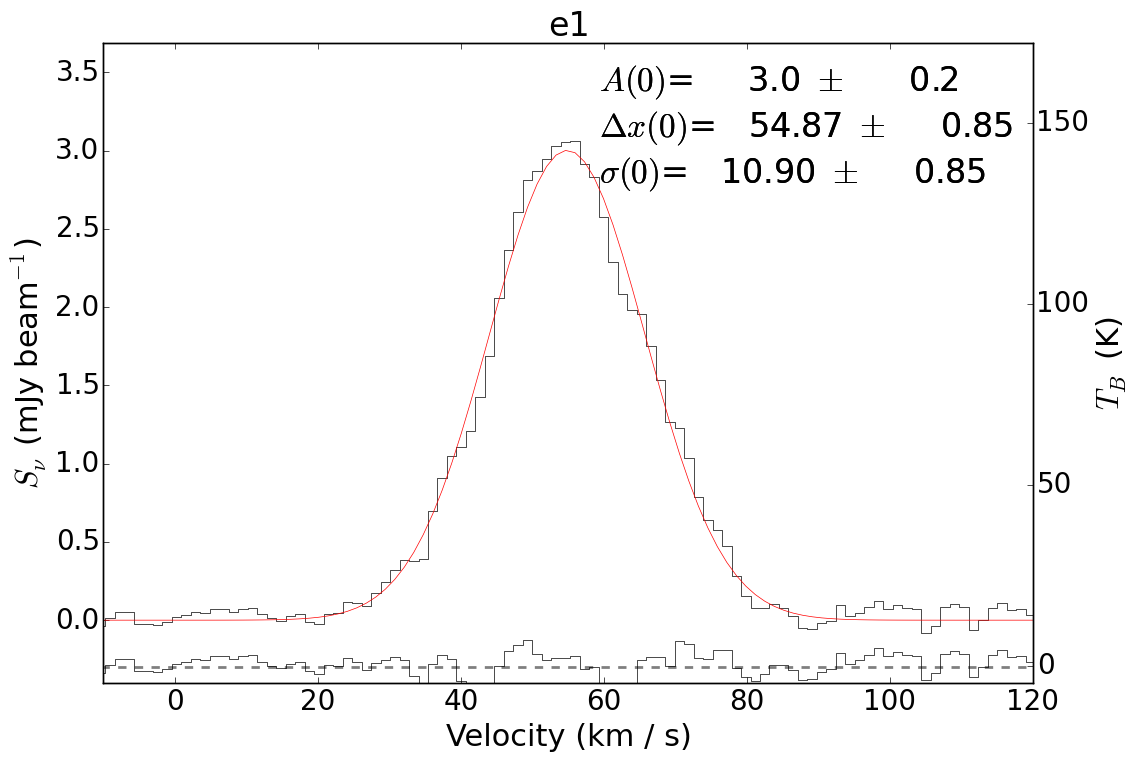}
{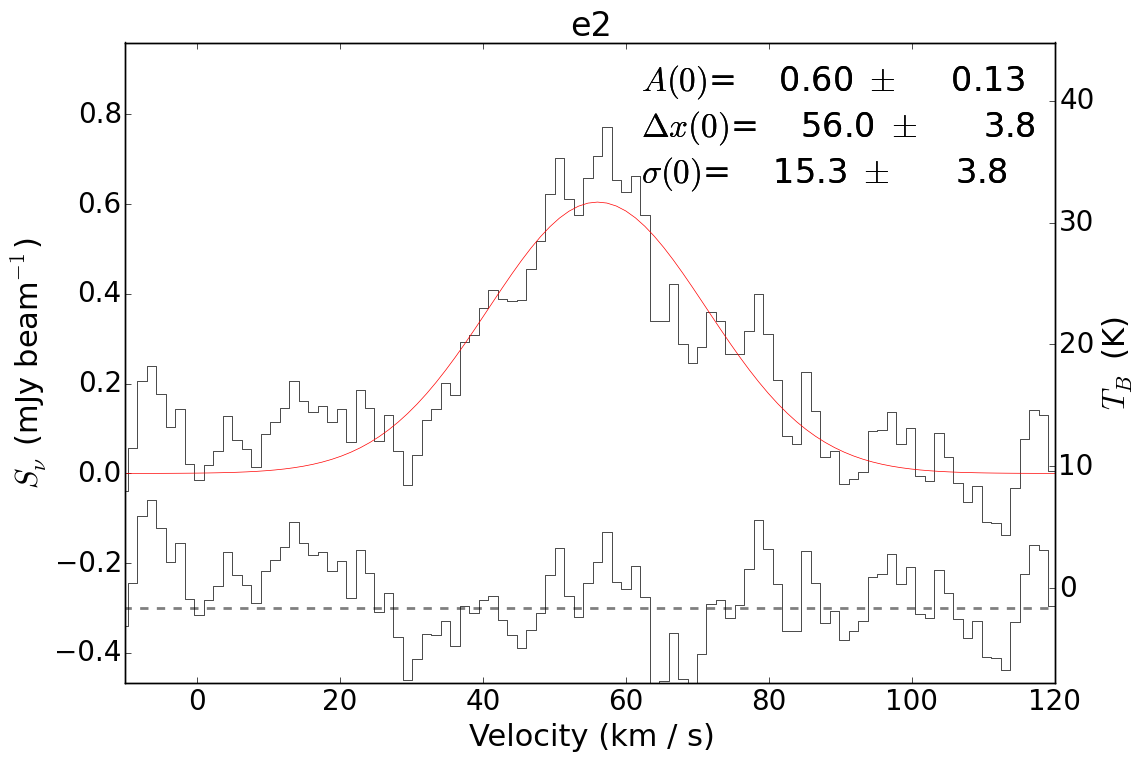}
{H77$\alpha$ spectra of W51e1 and W51e2.  These velocities are used in
conjunction with the \formaldehyde emission fits to determine the velocity
dispersion in Section \ref{sec:vdisp}.
The red curve shows the fit.  The lower spectrum shows the fit residual, with
the dashed line indicating the zero level.  
}
{fig:h77afits}{1}{3in}

\subsection{Constraints on the bulk outflow from IRS2}
\label{sec:irs2outflow}
In Section \ref{sec:diffuseemission}, we noted that the velocity structure of
the RRL emission in W51 IRS2 is consistent with outflowing ionized gas.  We use
the RRL emission to constrain the ionized gas mass loss rate from this cluster.
The outflowing ionized gas we observe is unbound from the system, moving at
$\sim25$ \kms relative to the cluster rest velocity, which is greater than the
escape velocity $v_{esc}\approx8$ \kms assuming $M_{cluster}=10^4$ \msun and
$r_{\mathrm{protocluster~gas}}=1.5$ pc.  The remaining ionized gas is bound to the system, with
thermal line width $c_s\approx6.7$ \kms assuming $T_e=7.5\ee{4}$ K \citep{Ginsburg2015a}.

We use Equation 14.28 of \citet{Wilson2009a} to determine the emission measure:
\begin{equation}
    EM = 5.2\ee{-4} 
    \left(\frac{T_B}{\mathrm{K}}\right)
    \left(\frac{T_e}{\mathrm{K}}\right)^{3/2} \left(\frac{\Delta
    \nu}{\mathrm{kHz}}\right) \mathrm{cm}^{-6}\mathrm{~pc}
\end{equation}
To estimate an upper limit on the outflow mass, we use the peak brightness in the shell at 37.5
\kms of $S_\nu=1.7$ mJy beam$^{-1}$ in a 10 \kms band ($T_B=60$ K;
 see Figure
\ref{fig:rrl_chan_main}, panel 4) and assume it fills the entire shell with
$r=0.25$ pc.  The resulting emission measure is $EM\approx10^7$ cm$^{-6}$~pc,
which for a shell with width 0.03 pc yields a density $n_e\approx2.4\ee{4}$
\percc and a mass $M_{shell}=3.5$ \msun.  For a more conservative upper limit,
we assume the entire surface area (a circle, rather than a shell) has the same
emission measure, resulting in $M_{outflow}=12$ \msun.  The shell has a velocity
difference from the peak of the \hii region emission $\Delta v=25$ \kms, which
at its current radius translates to a time $t = r / \Delta v = 7.5$ kyr.
The implied \emph{upper limit} mass loss rate is $\dot{M} = 1.6\ee{3}$ \msun
Myr$^{-1}$, which gives an evaporation timescale for this $M>10^4$ \msun
clump $t>6$ Myr.  The free-fall time averaged over the full volume of the protocluster gas is $t_{ff} < 0.5$ Myr
assuming $n=10^4$ \percc within $R=1.5$ pc, so the photoevaporation timescale is
at least an order of magnitude longer.  However, the molecular gas is centrally
condensed (most of the volume has no molecular gas) and under pressure from an
HII region, so the local gas density is likely to be much higher and the
collapse timescale therefore shorter.
Given the conservative assumptions adopted above, the true photoevaporative
mass loss is likely an order of magnitude lower, suggesting that this mass loss
mechanism is negligible at present.

\section{Discussion}
\label{sec:discussion}

\subsection{The massive stars and the clusters}
\label{sec:mstarscluster}
There are three distinct massive star populations identified in W51: W51 IRS2,
W51e1/e2, and the new W51e20 group.  The W51 IRS2 cluster is closer to
exhausting its available gas reservoir than the W51e1/e2 cluster, since it is
visible in the near-infrared.  The presence of many stars with no ongoing
accretion, along with the large volumes that are not occupied by dense gas,
suggest the IRS2 cluster is older than W51e1/e2. The e1/e2 protocluster shows
no signs of gas ejection, has no holes through which short-wavelength radiation
is escaping, and contains no demonstrably non-accreting stars.  
Because they have no evident surrounding dense gas or suggestions of ongoing
accretion, the W51e20 stars are likely the oldest in the W51A system.

However, W51e20 and the related W51e candidate CWBs may constitute an
unclustered population.  The Chandra MOXC catalog \citep{Townsley2014a} reveals
that the most evident cluster is W51 IRS2, and while there is a mild
overdensity associated with the W51 Main region, it is more amorphous than
centrally concentrated (Figure \ref{fig:moxconc}).  While it is possible that
the e20 group will merge with the e1/e2 cluster, the later evolutionary stage
of the e20 group (they are not embedded or accreting) suggests that they formed
separately and in smaller groupings.

\Figure{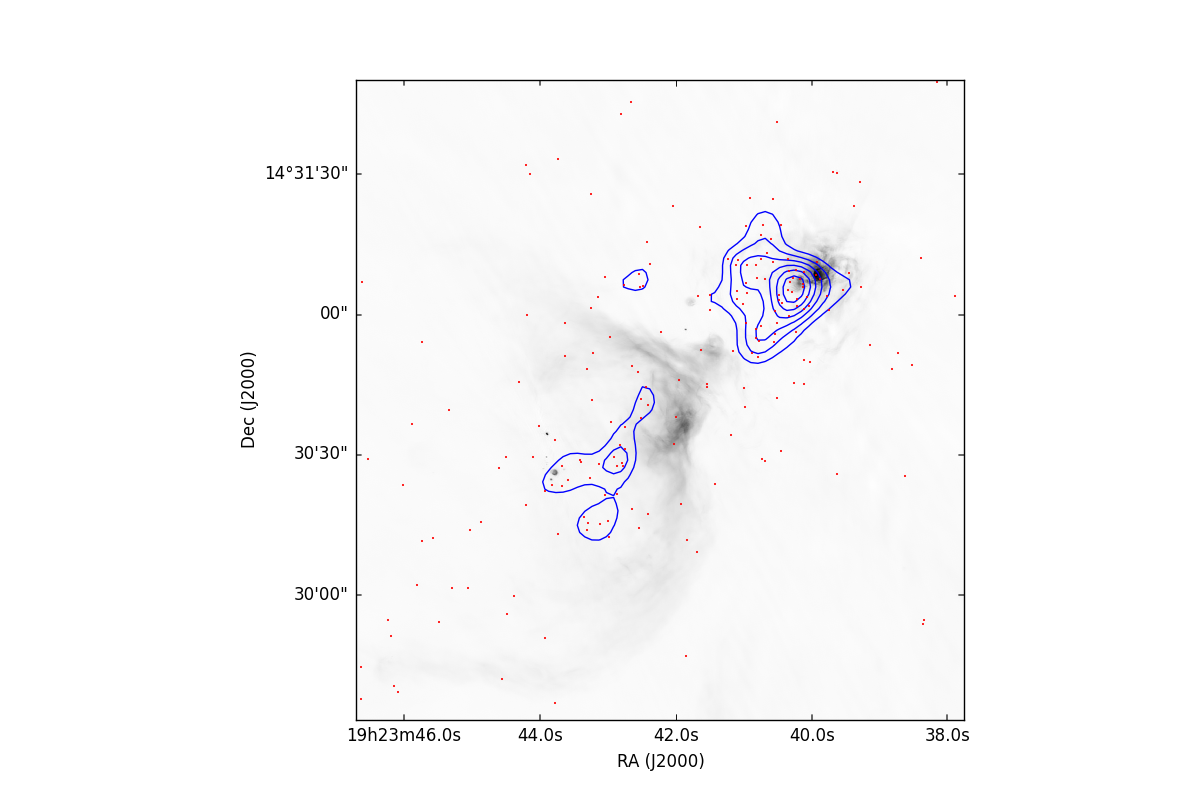}
{Contours \& individual points showing the X-ray source density from the 
MOXC \citep{Townsley2014a} catalog overlaid on the C-band image of W51.
See Section \ref{sec:mstarscluster}.}
{fig:moxconc}{1}{6.5in}

The W51d CWB candidates, by contrast, are each isolated, and they are
distributed around W51 IRS2 (Figures \ref{fig:w51irs2},
\ref{fig:coverview_pointsrcs}).  These stars could therefore be runaways from
the IRS2 cluster.  Source d5 is the furthest of these in projection, so with a
distance of 0.9 pc, its ejection time is $t = 0.08 \left({v_*}/{10
~\kms}\right)^{-1}$ Myr.  The runaway scenario provides an alternate
explanation for the observed SEDs (Appendix \ref{sec:SEDs}): synchrotron
emission from the wind-ISM bow shock might generate the detected emission.
This scenario can be readily verified if high ($>10~\kms$) line-of-sight
velocities to the near-infrared-visible stars are measured.

\subsection{The future evolution of the W51 clusters and molecular cloud}
The W51 clusters are capable of reaching a high star formation efficiency,
since the escape velocity from them is greater than the sound speed in ionized gas
\citep{Matzner2002a,Ginsburg2012a,Bressert2012a,Dale2014a} and
the \hii regions evidently leak optical photons, reducing the effectiveness of
radiative and wind feedback.  If they really are
forming one or more
young massive clusters, the gas within $\sim 1$ pc must be evacuated or
exhausted to achieve a final appearance like that of NGC 3603 or Trumpler 14,
our Galaxy's prototypical young massive clusters.

While the feedback from massive stars throughout W51 is intense, it is not
halting star formation.  There is evidence for ongoing star formation in both
W51e1/e2 and IRS2 in the form of molecular cores (Section
\ref{sec:h2coemission}), ongoing accretion onto massive young stellar objects
(MYSOs) like IRS2E \citep{Figueredo2008a}, and the numerous \hchii regions
that are likely to be accreting.  In e1/e2, there is no sign that feedback
has evacuated \emph{any} gas yet, and the \hii regions appear to be surrounded
by molecular gas.

In IRS2, accretion is ongoing despite much of its volume being evacuated of
molecular gas.  Ionizing photons are escaping the inner cluster and ionized gas
is being driven out of the inner cluster where the stars are forming.  However,
the timescale for complete gas evacuation in IRS2 at the presently observed
mass loss rate is ${>}10\times$ longer than the gas free-fall time on the
$\sim1$ parsec scale of the clump, and the star-forming gas is much more
concentrated than this parsec-scale average (Section \ref{sec:irs2outflow}).
Feedback is therefore progressing more slowly than star formation.  The
inefficiency of feedback in removing the star-forming gas supports the argument
that the most massive clumps in the galaxy  end their formation by exhausting
their parent reservoir rather than by disrupting their parent cloud
\citep{Kruijssen2012b,Matzner2015a}.

The coexistence of distributed and clustered populations over a fairly compact
region affects how their feedback couples to the local cloud.  Because these
populations are nearly coeval and spatially coincident, the feedback from the
OB star population onto the gas within $\lesssim5$ pc is not spherically symmetric as
if it were injected from a single cluster.  The distributed population will
render the $\sim5$ pc-scale cloud more holey (porous) than would be
implied by a smooth or even a very turbulent molecular cloud.  These holes
provide networks of low-density gaps through which radiation can escape,
reducing the overall impact of radiative feedback on the cloud.  Radiative
escape helps explain why feedback has been inefficient at halting further star
formation.  We conclude that feedback only affects the low-density gas, while
the high-density, star-forming gas is largely unaffected.

\subsubsection{Merging \& age spread}
\label{sec:merging}
Given the high binding energy of the clusters and the proximity of many other
stars, it is likely that some of the observed subclusters will merge.  Merger
or collapse of clusters is a requirement for the formation of massive
clusters in the Galactic center and disk \citep{Walker2015a,Walker2016a}, so it is reasonable to
expect similar dynamic evolution here as well.

If any of the observed subclusters merge (e.g., e1/e2 with e20-e22 or with
IRS2), the
resulting cluster will have an age spread greater than the present
age spread.  If we assume the e20 cluster is responsible for driving the
nearest part of the W51 Main shell (W51 IRS1) at a separation of 0.35 parsecs with a
D-type ionization front at $v\sim2$ \kms, we derive a lower limit on its age
$t_{e20} > 0.17$ Myr.  This lower limit is comparable to the gas free fall time
in the cluster and is still consistent with the most accurate stellar age
spread estimates available from photometry of young clusters
\citep[e.g.,][]{Kudryavtseva2012a}.

If instead e20-e22 are part of a distributed population that will not merge
with e1/e2, the W51 protoclusters are embedded within an OB association.  The
OB association will dissolve over a few Myr while the clusters will survive.
However, before this happens, the OB association is likely to shut off
accretion onto the protoclusters.  The radiative feedback from the OB
association can readily ionize the lower-density molecular gas in the outskirts
of the clusters.  In this way, the cluster's final mass would be externally
feedback regulated rather than, or perhaps in addition to, being
self-regulated.

\subsubsection{Final mass}
The present stellar mass as inferred from the infrared luminosity (Section
\ref{sec:stellarmass}) is a lower limit on the final mass of the clusters,
i.e., the mass after the remaining gas has either been converted to stars or
ejected from the system.  The final mass of the combined clusters will be at
least $\gtrsim7\ee{3} + 5\ee{4} \eta$ \msun, where $5\ee{4}$ \msun is the
approximate dense gas mass within $\sim2.5$ pc and $\eta$ is a star formation
efficiency, likely to be at least 10\% in the dense gas, so the stellar
$M_{final} > 1.2\ee{4}$ \msun.  This is a conservative lower limit, as it may
miss a substantial population of stars whose luminosity is not reprocessed into
the infrared and it uses conservatively low estimates for the gas mass and the
star formation efficiency.

\subsubsection{Cloud-scale feedback}
Since the free-fall time is short within these clusters ($t_{ff} < 0.5$ Myr out
to 2.5 pc containing 5\ee{4} \msun of gas, i.e. the region containing both
clusters), supernovae are not likely
to affect their formation.  However, once the clusters have exhausted gas
locally, they
will remain in the cloud long enough for the stars to evolve and die.
Exhaustion and porosity also provide a relatively free line-of-sight from the
clusters to the rest
of the cloud.  The binding energy of a $10^6$ \msun, $r=50$ pc 
molecular cloud is only about $10^{51}$ ergs, so a single supernova perfectly
coupled to the gas could unbind it all \citep[though typical energy delivered
to molecular clouds is lower, $\sim0.01-0.25$; see references in][appendix
B]{Kruijssen2012a}.  The $\sim20$ O-stars presently formed,
plus at least that many expected to form, could provide enough energy
to unbind the whole W51 GMC in a few Myr.  However, the location of the clusters
on the outskirts of the GMC imply a substantially lower coupling factor,
suggesting instead that supernovae will only partially disrupt the cloud.

\subsubsection{Evidence against triggering}
The star formation ongoing within this cloud shows no signs of having been
triggered by expanding \hii regions, though there are multiple generations
separated by temporal and spatial scales that might otherwise suggest a typical
expanding-\hii-region-driven triggering
scenario.  The main regions of ongoing star formation at present are W51e1/e2
and IRS2.  However, the most evident region of feedback is the W51 Main Shell,
especially W51 IRS1 
(Figure \ref{fig:coverview_diffuse}).  Throughout the W51 Main Shell, there is no sign
of any excess of newly formed protostars (Figure \ref{fig:moxconc}) and there
is a clear deficit of molecular gas in W51 IRS1 as traced by \ceighteeno \citep[Figure
\ref{fig:contonco};][]{Parsons2012a}.  Where there is the
strongest evidence of interacting and colliding \hii region shells (Figure
\ref{fig:rrl_chan_main} as discussed in Section \ref{sec:diffuseemission}),
no sign of ongoing star formation is observed.

\section{Conclusions}
\label{sec:conclusion}
We present deep, high-resolution radio continuum images in many low-frequency
bands and H$77\alpha$ and \ortho \twotwo spectral line cubes  from the Jansky
Very Large Array with angular resolution $0.3$\arcsec to $0.5$\arcsec.   These
observations have revealed many intriguing features of the W51 protocluster
that together suggest that thermal and radiative feedback from massive stars is
ineffective at halting star formation in massive protoclusters, similar to what
has been concluded in other high-mass protoclusters \citep[e.g.]{Ginsburg2012a,
Galvan-Madrid2013a}.
These observations support theoretical models that require ineffective feedback
as a necessary step in the formation of gravitationally bound stellar clusters.

\begin{itemize}
    \item Using the H$77\alpha$ emission and \formaldehyde emission and
        absorption line information, we have measured the velocity dispersion
        of the W51 e1 protocluster $\sigma_v=2.0$ \kms.  Although this cluster
        consists of hypercompact and ultracompact \hii regions, indicating that
        it is extremely gas-rich, the velocity dispersion implies a stellar
        virial mass that is consistent with an extrapolated IMF stellar mass,
        suggesting that the innermost cluster ($r<0.07$ pc) is already
        dynamically star-dominated.

\item The SEDs of the point sources detected throughout the W51A region have
peculiar shapes.  Most of them exhibit an excess at low frequencies ($<5$ GHz)
that renders their overall SED shape inconsistent with either an optically
thick or thin free-free spectrum.  These sources are likely to be colliding
wind binaries, exhibiting a combination of free-free and synchrotron emission.

\item We detected radio recombination line (H77$\alpha$) emission throughout
    W51.  The RRLs trace a number of overlapping bubbles toward the W51 IRS1
    emission peak and show that the IRS2 region has excavated a cavity along
    our line of sight.  

\item The \citet{Lacy2007a} high-velocity ionized jet is detected in
    H77$\alpha$ at velocities consistent with the [Ne II] observations.
    The jet remains mysterious as there is still no direct information about
    what drives it.  

\item We have detected the three massive cores from \citet{Zhang1997a} in
\formaldehyde \twotwo emission.  For W51 North, we evaluated the mass and
derived a lower limit $M>15 \msun$, though it is likely to be significantly
higher.  The \formaldehyde abundance in these cores is enhanced compared to the
surrounding molecular cloud by at least 100$\times$, with $X_{\formaldehyde}
\sim 10^{-7}$, similar to the Orion BN/KL hot core \citep{Mangum1993b}.  In the W51e8 molecular
core, we have found a compact continuum source, indicating unambiguously that
these cores are forming massive stars.

\item Because of its association with high-mass protostellar cores,
    the \formaldehyde \twotwo 14.488 GHz transition is a good tracer of
    early-stage massive star formation.  Formaldehyde is apparently enhanced in
    abundance by 2-3
orders of magnitude over ISM values in high-mass protostellar cores, allowing
it to be detected in emission at long wavelengths.  This transition will be a
powerful tool
for studying the earliest stagest of high-mass star formation throughout the
Galaxy with the Next Generation VLA or a Square Kilometer Array.  Ongoing JVLA surveys (e.g.,
KuGARS, PI Thompson, \url{http://library.nrao.edu/proposals/catalog/10267}) may
detect a significant additional population of proto-O-stars.

\item The W51e1/e2 and IRS2 protoclusters appear to be high-mass clusters
    forming within a larger OB association.  The coexistence of the distributed
    and clustered OB-star populations suggests that \emph{external} feedback
    may be partly responsible for halting protocluster feeding.

\item The weak outflow from IRS2 and ongoing accretion in IRS2 and e1/e2
    suggest that gas exhaustion is a more important mechanism than expulsion in
    halting cluster growth.
    Recent theoretical models predict gas exhaustion to be a necessary step in
    the formation of gravitationally bound stellar clusters
    \citep[e.g.,][]{Kruijssen2012a} and our results provide evidence of this
    process in action.

\end{itemize}

These results show that even the low-pressure interstellar medium of the Milky
Way can reach densities high enough for gravitational collapse to form clusters
on $<$Myr timescales, rendering stellar feedback ineffective.  Future
observational studies aiming to understand the formation of massive stellar
clusters should target regions with properties similar to W51, where gas
exhaustion may be prevalent and the conditions therefore enable gravitationally
bound clusters to form.

\textbf{Acknowledgements}:

The National Radio Astronomy Observatory is a facility of the National Science
Foundation operated under cooperative agreement by Associated Universities,
Inc.
JMDK gratefully acknowledges financial support in the form of a Gliese
Fellowship and an Emmy Noether Research Group from the Deutsche
Forschungsgemeinschaft (DFG), grant number KR4801/1-1.

\textbf{Code Packages Used}:

\begin{itemize}
    \item pyradex \url{https://github.com/adamginsburg/pyradex}
    \item myRadex \url{https://github.com/fjdu/myRadex}
    \item pyspeckit \url{http://pyspeckit.bitbucket.org} \citet{Ginsburg2011c}
    \item aplpy \url{https://aplpy.github.io/}
    \item wcsaxes \url{http://wcsaxes.readthedocs.org}
    \item spectral cube \url{http://spectral-cube.readthedocs.org}
    \item pvextractor \url{http://pvextractor.readthedocs.org/}
    \item ds9 \url{http://ds9.si.edu}
\end{itemize}

\appendix
\section{Figures labeling point sources and diffuse regions}
\label{sec:appendix_labels}
We include additional versions of Figure \ref{fig:coverview} with labels for
the compact (Figure \ref{fig:coverview_pointsrcs}) and extended (Figure
\ref{fig:coverview_diffuse}) \hii regions.

\Figure{f36}
{The C-band image of the W51 region produced with a combination of JVLA
C and A array data using uniform weighting.  Annotated versions of this figure
identifying the regions discussed are available in the Appendix.}
{fig:coverview}{0.9}{6.5in}

\Figure{f37}
{The C-band image of the W51A region with diffuse \hii regions labeled.  The labels
are above the circles except where arrows are used.  The regions labeled
{W51 Main 65 kms RRL shell} and {W51 Main 40 kms RRL shell} are part of  W51
IRS1 \citep[e.g.,][]{Zhang1997a}; we have given them independent labels here
because we are now able to kinematically
distinguish them in RRL emission and because they are evidently part of (or
interacting with) a large shell centered near W51 Main.}
{fig:coverview_diffuse}{0.9}{6.5in}

\onecolumn
\begin{longtable}{ccccccc}
\caption{Continuum Point Sources}\\

\label{tab:contsrcs_full}
Object & Epoch & Obs. Date & Peak $S_{\nu}$ & Peak - Background & RMS & Frequency \\
$\mathrm{}$ & $\mathrm{}$ & $\mathrm{}$ & $\mathrm{mJy\,beam^{-1}}$ & $\mathrm{mJy\,beam^{-1}}$ & $\mathrm{mJy\,beam^{-1}}$ & $\mathrm{GHz}$ \\
\hline
d2 & 2 & 2012-10-16 & 5.5 & 7.7 & 0.2 & 2.5 \\
d2 & 2 & 2012-10-16 & 4.4 & 6.3 & 0.06 & 3.5 \\
d2 & 1 & 1992-10-25 & 10 & 8.7 & 0.2 & 4.9 \\
d2 & 2 & 2012-10-16 & 6 & 6.4 & 0.05 & 4.9 \\
d2 & 3 & 2014-04-19 & 6.5 & 6.2 & 0.08 & 4.9 \\
d2 & 2 & 2012-10-16 & 3.5 & 4.6 & 0.04 & 5.9 \\
d2 & 3 & 2014-04-19 & 4.8 & 4.8 & 0.05 & 5.9 \\
d2 & 1 & 1992-10-25 & 15 & 13 & 0.06 & 8.4 \\
d2 & 2 & 2013-03-02 & 18 & 17 & 0.1 & 12.6 \\
d2 & 2 & 2013-03-02 & 18 & 17 & 0.09 & 14.1 \\
d2 & 1 & 1993-03-16 & 10 & 10 & 0.5 & 22.5 \\
d2 & 2 & 2012-05-31 & 28 & 30 & 0.8 & 25.0 \\
d2 & 2 & 2012-06-21 & 26 & 28 & 1 & 27.0 \\
d2 & 2 & 2012-08-07 & 22 & 24 & 2 & 29.0 \\
d2 & 2 & 2012-06-21 & 26 & 27 & 1 & 33.0 \\
d2 & 2 & 2012-08-07 & 29 & 32 & 1 & 36.0 \\
d3-diffuse & 2 & 2012-10-16 & 0.4 & 0.4 & 0.2 & 2.5 \\
d3-diffuse & 2 & 2012-10-16 & 0.07 & 0.19 & 0.06 & 3.5 \\
d3-diffuse & 3 & 2014-04-19 & 0.19 & 0.17 & 0.08 & 4.9 \\
d3-diffuse & 2 & 2012-10-16 & 0.03 & 0.08 & 0.05 & 4.9 \\
d3-diffuse & 1 & 1992-10-25 & 0.3 & 0.3 & 0.2 & 4.9 \\
d3-diffuse & 2 & 2012-10-16 & 0.04 & 0.08 & 0.04 & 5.9 \\
d3-diffuse & 3 & 2014-04-19 & 0.03 & 0.1 & 0.05 & 5.9 \\
d3-diffuse & 1 & 1992-10-25 & 0.47 & 0.37 & 0.06 & 8.4 \\
d3-diffuse & 2 & 2013-03-02 & 0.4 & 0.4 & 0.1 & 12.6 \\
d3-diffuse & 2 & 2013-03-02 & 0.12 & 0.19 & 0.09 & 14.1 \\
d3-diffuse & 1 & 1993-03-16 & - & - & 0.5 & 22.5 \\
d3-diffuse & 2 & 2012-05-31 & - & - & 0.8 & 25.0 \\
d3-diffuse & 2 & 2012-06-21 & - & - & 1 & 27.0 \\
d3-diffuse & 2 & 2012-08-07 & - & - & 2 & 29.0 \\
d3-diffuse & 2 & 2012-06-21 & - & - & 1 & 33.0 \\
d3-diffuse & 2 & 2012-08-07 & - & - & 1 & 36.0 \\
d4e & 2 & 2012-10-16 & 1.1 & 1.3 & 0.2 & 2.5 \\
d4e & 2 & 2012-10-16 & 0.91 & 1.2 & 0.06 & 3.5 \\
d4e & 3 & 2014-04-19 & 0.85 & 1 & 0.08 & 4.9 \\
d4e & 1 & 1992-10-25 & -0.1 & 0.5 & 0.2 & 4.9 \\
d4e & 2 & 2012-10-16 & 0.79 & 1.1 & 0.05 & 4.9 \\
d4e & 3 & 2014-04-19 & 0.62 & 0.75 & 0.05 & 5.9 \\
d4e & 2 & 2012-10-16 & 0.7 & 0.91 & 0.04 & 5.9 \\
d4e & 1 & 1992-10-25 & 0.23 & 0.21 & 0.06 & 8.4 \\
d4e & 2 & 2013-03-02 & 1 & 1.3 & 0.1 & 12.6 \\
d4e & 2 & 2013-03-02 & 0.95 & 1.2 & 0.09 & 14.1 \\
d4e & 1 & 1993-03-16 & -0.4 & 0.7 & 0.5 & 22.5 \\
d4e & 2 & 2012-05-31 & 0.5 & 3 & 0.8 & 25.0 \\
d4e & 2 & 2012-06-21 & 1 & 3 & 1 & 27.0 \\
d4e & 2 & 2012-08-07 & 1 & 5 & 2 & 29.0 \\
d4e & 2 & 2012-06-21 & - & - & 1 & 33.0 \\
d4e & 2 & 2012-08-07 & - & - & 1 & 36.0 \\
d4w & 2 & 2012-10-16 & 1.1 & 1.3 & 0.2 & 2.5 \\
d4w & 2 & 2012-10-16 & 0.91 & 1.2 & 0.06 & 3.5 \\
d4w & 2 & 2012-10-16 & 0.51 & 0.79 & 0.05 & 4.9 \\
d4w & 1 & 1992-10-25 & 0.1 & 0.6 & 0.2 & 4.9 \\
d4w & 3 & 2014-04-19 & 0.6 & 0.77 & 0.08 & 4.9 \\
d4w & 2 & 2012-10-16 & 0.45 & 0.66 & 0.04 & 5.9 \\
d4w & 3 & 2014-04-19 & 0.4 & 0.53 & 0.05 & 5.9 \\
d4w & 1 & 1992-10-25 & 0.2 & 0.17 & 0.06 & 8.4 \\
d4w & 2 & 2013-03-02 & 0.9 & 1.2 & 0.1 & 12.6 \\
d4w & 2 & 2013-03-02 & 0.67 & 0.91 & 0.09 & 14.1 \\
d4w & 1 & 1993-03-16 & -0.3 & 1 & 0.5 & 22.5 \\
d4w & 2 & 2012-05-31 & 0 & 1.9 & 0.8 & 25.0 \\
d4w & 2 & 2012-06-21 & 1 & 4 & 1 & 27.0 \\
d4w & 2 & 2012-08-07 & 1 & 5 & 2 & 29.0 \\
d4w & 2 & 2012-06-21 & - & - & 1 & 33.0 \\
d4w & 2 & 2012-08-07 & - & - & 1 & 36.0 \\
d5 & 2 & 2012-10-16 & 0.3 & 0.4 & 0.2 & 2.5 \\
d5 & 2 & 2012-10-16 & 0.04 & 0.24 & 0.06 & 3.5 \\
d5 & 1 & 1992-10-25 & -0 & 0.5 & 0.2 & 4.9 \\
d5 & 3 & 2014-04-19 & 0.01 & 0.11 & 0.08 & 4.9 \\
d5 & 2 & 2012-10-16 & 0.03 & 0.09 & 0.05 & 4.9 \\
d5 & 3 & 2014-04-19 & 0.04 & 0.09 & 0.05 & 5.9 \\
d5 & 2 & 2012-10-16 & -0.03 & 0.11 & 0.04 & 5.9 \\
d5 & 1 & 1992-10-25 & 0.1 & 0.14 & 0.06 & 8.4 \\
d5 & 2 & 2013-03-02 & 0.1 & 0.5 & 0.1 & 12.6 \\
d5 & 2 & 2013-03-02 & 0.3 & 0.54 & 0.09 & 14.1 \\
d5 & 1 & 1993-03-16 & 0.7 & 1.5 & 0.5 & 22.5 \\
d5 & 2 & 2012-05-31 & 1.7 & 2.3 & 0.8 & 25.0 \\
d5 & 2 & 2012-06-21 & 1 & 3 & 1 & 27.0 \\
d5 & 2 & 2012-08-07 & 0 & 3 & 2 & 29.0 \\
d5 & 2 & 2012-06-21 & 1 & 2 & 1 & 33.0 \\
d5 & 2 & 2012-08-07 & 2 & 4 & 1 & 36.0 \\
d6 & 2 & 2012-10-16 & 1 & 1.4 & 0.2 & 2.5 \\
d6 & 2 & 2012-10-16 & 0.2 & 0.6 & 0.06 & 3.5 \\
d6 & 3 & 2014-04-19 & 0.42 & 0.47 & 0.08 & 4.9 \\
d6 & 1 & 1992-10-25 & 0.6 & 0.8 & 0.2 & 4.9 \\
d6 & 2 & 2012-10-16 & 0.34 & 0.54 & 0.05 & 4.9 \\
d6 & 2 & 2012-10-16 & 0.33 & 0.47 & 0.04 & 5.9 \\
d6 & 3 & 2014-04-19 & 0.38 & 0.43 & 0.05 & 5.9 \\
d6 & 1 & 1992-10-25 & 0.78 & 0.47 & 0.06 & 8.4 \\
d6 & 2 & 2013-03-02 & 0.3 & 0.7 & 0.1 & 12.6 \\
d6 & 2 & 2013-03-02 & 0.34 & 0.59 & 0.09 & 14.1 \\
d6 & 1 & 1993-03-16 & 0.6 & 1.4 & 0.5 & 22.5 \\
d6 & 2 & 2012-05-31 & 0.2 & 0.8 & 0.8 & 25.0 \\
d6 & 2 & 2012-06-21 & -0 & 2 & 1 & 27.0 \\
d6 & 2 & 2012-08-07 & -0 & 2 & 2 & 29.0 \\
d6 & 2 & 2012-06-21 & -1 & 1 & 1 & 33.0 \\
d6 & 2 & 2012-08-07 & 0 & 2 & 1 & 36.0 \\
d7 & 2 & 2012-10-16 & 0.4 & 0.9 & 0.2 & 2.5 \\
d7 & 2 & 2012-10-16 & 0.22 & 0.7 & 0.06 & 3.5 \\
d7 & 1 & 1992-10-25 & 0.7 & 1 & 0.2 & 4.9 \\
d7 & 2 & 2012-10-16 & 0.14 & 0.33 & 0.05 & 4.9 \\
d7 & 3 & 2014-04-19 & 0.24 & 0.38 & 0.08 & 4.9 \\
d7 & 2 & 2012-10-16 & 0.16 & 0.27 & 0.04 & 5.9 \\
d7 & 3 & 2014-04-19 & 0.21 & 0.29 & 0.05 & 5.9 \\
d7 & 1 & 1992-10-25 & 0.71 & 0.55 & 0.06 & 8.4 \\
d7 & 2 & 2013-03-02 & 0.3 & 0.6 & 0.1 & 12.6 \\
d7 & 2 & 2013-03-02 & 0.46 & 0.61 & 0.09 & 14.1 \\
d7 & 1 & 1993-03-16 & -0 & 1 & 0.5 & 22.5 \\
d7 & 2 & 2012-05-31 & 0.1 & 1.3 & 0.8 & 25.0 \\
d7 & 2 & 2012-06-21 & 0 & 1 & 1 & 27.0 \\
d7 & 2 & 2012-08-07 & 2 & 2 & 2 & 29.0 \\
d7 & 2 & 2012-06-21 & 1 & 1 & 1 & 33.0 \\
d7 & 2 & 2012-08-07 & 1 & 2 & 1 & 36.0 \\
e1 & 2 & 2012-10-16 & 11 & 11 & 0.2 & 2.5 \\
e1 & 2 & 2012-10-16 & 11 & 11 & 0.06 & 3.5 \\
e1 & 2 & 2012-10-16 & 10 & 11 & 0.05 & 4.9 \\
e1 & 3 & 2014-04-19 & 8.3 & 8.6 & 0.08 & 4.9 \\
e1 & 1 & 1992-10-25 & 16 & 16 & 0.2 & 4.9 \\
e1 & 3 & 2014-04-19 & 7.2 & 7.5 & 0.05 & 5.9 \\
e1 & 2 & 2012-10-16 & 8.2 & 8.2 & 0.04 & 5.9 \\
e1 & 1 & 1992-10-25 & 20 & 20 & 0.06 & 8.4 \\
e1 & 2 & 2013-03-02 & 23 & 23 & 0.1 & 12.6 \\
e1 & 2 & 2013-03-02 & 19 & 19 & 0.09 & 14.1 \\
e1 & 1 & 1993-03-16 & 12 & 14 & 0.5 & 22.5 \\
e1 & 2 & 2012-05-31 & 13 & 14 & 0.8 & 25.0 \\
e1 & 2 & 2012-06-21 & 8 & 11 & 1 & 27.0 \\
e1 & 2 & 2012-08-07 & 5 & 8 & 2 & 29.0 \\
e1 & 2 & 2012-06-21 & 7 & 9 & 1 & 33.0 \\
e1 & 2 & 2012-08-07 & 4 & 7 & 1 & 36.0 \\
e10 & 2 & 2012-10-16 & 1.8 & 1.8 & 0.2 & 2.5 \\
e10 & 2 & 2012-10-16 & 1.6 & 1.9 & 0.06 & 3.5 \\
e10 & 2 & 2012-10-16 & 1.7 & 2 & 0.05 & 4.9 \\
e10 & 1 & 1992-10-25 & 2.2 & 2.3 & 0.2 & 4.9 \\
e10 & 3 & 2014-04-19 & 1.5 & 1.7 & 0.08 & 4.9 \\
e10 & 3 & 2014-04-19 & 1.3 & 1.5 & 0.05 & 5.9 \\
e10 & 2 & 2012-10-16 & 1.5 & 1.7 & 0.04 & 5.9 \\
e10 & 1 & 1992-10-25 & 1.4 & 1.1 & 0.06 & 8.4 \\
e10 & 2 & 2013-03-02 & 2.1 & 2.4 & 0.1 & 12.6 \\
e10 & 2 & 2013-03-02 & 1.7 & 2 & 0.09 & 14.1 \\
e10 & 1 & 1993-03-16 & 0.8 & 2.4 & 0.5 & 22.5 \\
e10 & 2 & 2012-05-31 & 1.7 & 2.8 & 0.8 & 25.0 \\
e10 & 2 & 2012-06-21 & 2 & 5 & 1 & 27.0 \\
e10 & 2 & 2012-08-07 & 1 & 3 & 2 & 29.0 \\
e10 & 2 & 2012-06-21 & 3 & 7 & 1 & 33.0 \\
e10 & 2 & 2012-08-07 & 1 & 6 & 1 & 36.0 \\
e11 & 2 & 2012-10-16 & 0.1 & 0.4 & 0.2 & 2.5 \\
e11 & 2 & 2012-10-16 & 0.17 & 0.25 & 0.06 & 3.5 \\
e11 & 1 & 1992-10-25 & 0.5 & 0.8 & 0.2 & 4.9 \\
e11 & 3 & 2014-04-19 & 0.24 & 0.28 & 0.08 & 4.9 \\
e11 & 2 & 2012-10-16 & 0.25 & 0.35 & 0.05 & 4.9 \\
e11 & 3 & 2014-04-19 & 0.28 & 0.35 & 0.05 & 5.9 \\
e11 & 2 & 2012-10-16 & 0.29 & 0.38 & 0.04 & 5.9 \\
e11 & 1 & 1992-10-25 & 0.37 & 0.32 & 0.06 & 8.4 \\
e11 & 2 & 2013-03-02 & 0.5 & 0.8 & 0.1 & 12.6 \\
e11 & 2 & 2013-03-02 & 0.55 & 0.63 & 0.09 & 14.1 \\
e11 & 1 & 1993-03-16 & 0.3 & 0.9 & 0.5 & 22.5 \\
e11 & 2 & 2012-05-31 & 0.7 & 1.6 & 0.8 & 25.0 \\
e11 & 2 & 2012-06-21 & 3 & 3 & 1 & 27.0 \\
e11 & 2 & 2012-08-07 & 0 & 4 & 2 & 29.0 \\
e11 & 2 & 2012-06-21 & 0 & 2 & 1 & 33.0 \\
e11 & 2 & 2012-08-07 & 2 & 3 & 1 & 36.0 \\
e12 & 2 & 2012-10-16 & -0 & 0.9 & 0.2 & 2.5 \\
e12 & 2 & 2012-10-16 & 0.6 & 0.75 & 0.06 & 3.5 \\
e12 & 1 & 1992-10-25 & 2.1 & 1.3 & 0.2 & 4.9 \\
e12 & 2 & 2012-10-16 & 0.38 & 0.46 & 0.05 & 4.9 \\
e12 & 3 & 2014-04-19 & 0.03 & 0.47 & 0.08 & 4.9 \\
e12 & 2 & 2012-10-16 & 0.25 & 0.41 & 0.04 & 5.9 \\
e12 & 3 & 2014-04-19 & 0.42 & 0.48 & 0.05 & 5.9 \\
e12 & 1 & 1992-10-25 & 1.7 & 0.72 & 0.06 & 8.4 \\
e12 & 2 & 2013-03-02 & 1 & 0.7 & 0.1 & 12.6 \\
e12 & 2 & 2013-03-02 & 0.6 & 0.55 & 0.09 & 14.1 \\
e12 & 1 & 1993-03-16 & 0.2 & 1.2 & 0.5 & 22.5 \\
e12 & 2 & 2012-05-31 & 0.3 & 1.1 & 0.8 & 25.0 \\
e12 & 2 & 2012-06-21 & 1 & 2 & 1 & 27.0 \\
e12 & 2 & 2012-08-07 & -0 & 3 & 2 & 29.0 \\
e12 & 2 & 2012-06-21 & 1 & 2 & 1 & 33.0 \\
e12 & 2 & 2012-08-07 & 0 & 2 & 1 & 36.0 \\
e13 & 2 & 2012-10-16 & 0.8 & 2.4 & 0.2 & 2.5 \\
e13 & 2 & 2012-10-16 & 1.1 & 1.5 & 0.06 & 3.5 \\
e13 & 3 & 2014-04-19 & 0.57 & 0.92 & 0.08 & 4.9 \\
e13 & 1 & 1992-10-25 & 2.7 & 2 & 0.2 & 4.9 \\
e13 & 2 & 2012-10-16 & 0.76 & 0.97 & 0.05 & 4.9 \\
e13 & 3 & 2014-04-19 & 0.57 & 0.72 & 0.05 & 5.9 \\
e13 & 2 & 2012-10-16 & 0.6 & 0.81 & 0.04 & 5.9 \\
e13 & 1 & 1992-10-25 & 2.1 & 1.4 & 0.06 & 8.4 \\
e13 & 2 & 2013-03-02 & 1.9 & 1.8 & 0.1 & 12.6 \\
e13 & 2 & 2013-03-02 & 1.5 & 1.4 & 0.09 & 14.1 \\
e13 & 1 & 1993-03-16 & 0.6 & 1.5 & 0.5 & 22.5 \\
e13 & 2 & 2012-05-31 & 1 & 1.6 & 0.8 & 25.0 \\
e13 & 2 & 2012-06-21 & 1 & 2 & 1 & 27.0 \\
e13 & 2 & 2012-08-07 & 1 & 2 & 2 & 29.0 \\
e13 & 2 & 2012-06-21 & 1 & 3 & 1 & 33.0 \\
e13 & 2 & 2012-08-07 & 1 & 2 & 1 & 36.0 \\
e14 & 2 & 2012-10-16 & 0.9 & 1.6 & 0.2 & 2.5 \\
e14 & 2 & 2012-10-16 & 0.77 & 0.86 & 0.06 & 3.5 \\
e14 & 1 & 1992-10-25 & 2.6 & 1.6 & 0.2 & 4.9 \\
e14 & 3 & 2014-04-19 & 0.59 & 0.84 & 0.08 & 4.9 \\
e14 & 2 & 2012-10-16 & 0.43 & 0.84 & 0.05 & 4.9 \\
e14 & 3 & 2014-04-19 & 0.38 & 0.52 & 0.05 & 5.9 \\
e14 & 2 & 2012-10-16 & 0.33 & 0.58 & 0.04 & 5.9 \\
e14 & 1 & 1992-10-25 & 1.9 & 0.6 & 0.06 & 8.4 \\
e14 & 2 & 2013-03-02 & 1.6 & 1.2 & 0.1 & 12.6 \\
e14 & 2 & 2013-03-02 & 1 & 0.95 & 0.09 & 14.1 \\
e14 & 1 & 1993-03-16 & 0.8 & 1 & 0.5 & 22.5 \\
e14 & 2 & 2012-05-31 & 1.8 & 1.9 & 0.8 & 25.0 \\
e14 & 2 & 2012-06-21 & 2 & 2 & 1 & 27.0 \\
e14 & 2 & 2012-08-07 & 1 & 2 & 2 & 29.0 \\
e14 & 2 & 2012-06-21 & 1 & 2 & 1 & 33.0 \\
e14 & 2 & 2012-08-07 & 1 & 2 & 1 & 36.0 \\
e15 & 2 & 2012-10-16 & 1 & 1.1 & 0.2 & 2.5 \\
e15 & 2 & 2012-10-16 & 0.42 & 0.6 & 0.06 & 3.5 \\
e15 & 3 & 2014-04-19 & 0.35 & 0.35 & 0.08 & 4.9 \\
e15 & 1 & 1992-10-25 & 0.7 & 0.8 & 0.2 & 4.9 \\
e15 & 2 & 2012-10-16 & 0.31 & 0.36 & 0.05 & 4.9 \\
e15 & 2 & 2012-10-16 & 0.17 & 0.26 & 0.04 & 5.9 \\
e15 & 3 & 2014-04-19 & 0.25 & 0.31 & 0.05 & 5.9 \\
e15 & 1 & 1992-10-25 & 1 & 0.79 & 0.06 & 8.4 \\
e15 & 2 & 2013-03-02 & 0.4 & 0.5 & 0.1 & 12.6 \\
e15 & 2 & 2013-03-02 & 0.31 & 0.41 & 0.09 & 14.1 \\
e15 & 1 & 1993-03-16 & - & - & 0.5 & 22.5 \\
e15 & 2 & 2012-05-31 & 0.9 & 1.8 & 0.8 & 25.0 \\
e15 & 2 & 2012-06-21 & 5 & 7 & 1 & 27.0 \\
e15 & 2 & 2012-08-07 & 4 & 8 & 2 & 29.0 \\
e15 & 2 & 2012-06-21 & - & - & 1 & 33.0 \\
e15 & 2 & 2012-08-07 & - & - & 1 & 36.0 \\
e16 & 2 & 2012-10-16 & 0.3 & 0.4 & 0.2 & 2.5 \\
e16 & 2 & 2012-10-16 & 0.16 & 0.2 & 0.06 & 3.5 \\
e16 & 1 & 1992-10-25 & 0.4 & 0.5 & 0.2 & 4.9 \\
e16 & 3 & 2014-04-19 & 0.13 & 0.15 & 0.08 & 4.9 \\
e16 & 2 & 2012-10-16 & 0.05 & 0.15 & 0.05 & 4.9 \\
e16 & 3 & 2014-04-19 & 0.08 & 0.13 & 0.05 & 5.9 \\
e16 & 2 & 2012-10-16 & 0.05 & 0.14 & 0.04 & 5.9 \\
e16 & 1 & 1992-10-25 & 0.41 & 0.25 & 0.06 & 8.4 \\
e16 & 2 & 2013-03-02 & 0.2 & 0.3 & 0.1 & 12.6 \\
e16 & 2 & 2013-03-02 & 0.31 & 0.37 & 0.09 & 14.1 \\
e16 & 1 & 1993-03-16 & 0.6 & 1 & 0.5 & 22.5 \\
e16 & 2 & 2012-05-31 & - & - & 0.8 & 25.0 \\
e16 & 2 & 2012-06-21 & - & - & 1 & 27.0 \\
e16 & 2 & 2012-08-07 & - & - & 2 & 29.0 \\
e16 & 2 & 2012-06-21 & -0 & 2 & 1 & 33.0 \\
e16 & 2 & 2012-08-07 & -0 & 2 & 1 & 36.0 \\
e17 & 2 & 2012-10-16 & 0.1 & 0.5 & 0.2 & 2.5 \\
e17 & 2 & 2012-10-16 & 0.15 & 0.29 & 0.06 & 3.5 \\
e17 & 3 & 2014-04-19 & 0.19 & 0.21 & 0.08 & 4.9 \\
e17 & 1 & 1992-10-25 & 0.8 & 0.6 & 0.2 & 4.9 \\
e17 & 2 & 2012-10-16 & 0.07 & 0.22 & 0.05 & 4.9 \\
e17 & 2 & 2012-10-16 & 0.11 & 0.17 & 0.04 & 5.9 \\
e17 & 3 & 2014-04-19 & 0.13 & 0.15 & 0.05 & 5.9 \\
e17 & 1 & 1992-10-25 & 0.53 & 0.29 & 0.06 & 8.4 \\
e17 & 2 & 2013-03-02 & 0.1 & 0.3 & 0.1 & 12.6 \\
e17 & 2 & 2013-03-02 & 0.2 & 0.26 & 0.09 & 14.1 \\
e17 & 1 & 1993-03-16 & - & - & 0.5 & 22.5 \\
e17 & 2 & 2012-05-31 & - & - & 0.8 & 25.0 \\
e17 & 2 & 2012-06-21 & - & - & 1 & 27.0 \\
e17 & 2 & 2012-08-07 & - & - & 2 & 29.0 \\
e17 & 2 & 2012-06-21 & 0 & 2 & 1 & 33.0 \\
e17 & 2 & 2012-08-07 & -1 & 2 & 1 & 36.0 \\
e18 & 2 & 2012-10-16 & 0.8 & 0.8 & 0.2 & 2.5 \\
e18 & 2 & 2012-10-16 & 0.47 & 0.49 & 0.06 & 3.5 \\
e18 & 1 & 1992-10-25 & 1.2 & 0.8 & 0.2 & 4.9 \\
e18 & 2 & 2012-10-16 & 0.43 & 0.48 & 0.05 & 4.9 \\
e18 & 3 & 2014-04-19 & 0.51 & 0.37 & 0.08 & 4.9 \\
e18 & 2 & 2012-10-16 & 0.37 & 0.5 & 0.04 & 5.9 \\
e18 & 3 & 2014-04-19 & 0.52 & 0.48 & 0.05 & 5.9 \\
e18 & 1 & 1992-10-25 & 0.81 & 0.38 & 0.06 & 8.4 \\
e18 & 2 & 2013-03-02 & 1.1 & 1.2 & 0.1 & 12.6 \\
e18 & 2 & 2013-03-02 & 1.1 & 1.1 & 0.09 & 14.1 \\
e18 & 1 & 1993-03-16 & 1.5 & 1.8 & 0.5 & 22.5 \\
e18 & 2 & 2012-05-31 & - & - & 0.8 & 25.0 \\
e18 & 2 & 2012-06-21 & - & - & 1 & 27.0 \\
e18 & 2 & 2012-08-07 & - & - & 2 & 29.0 \\
e18 & 2 & 2012-06-21 & -0 & 1 & 1 & 33.0 \\
e18 & 2 & 2012-08-07 & 2 & 3 & 1 & 36.0 \\
e19 & 2 & 2012-10-16 & 1.9 & 2 & 0.2 & 2.5 \\
e19 & 2 & 2012-10-16 & 0.56 & 0.73 & 0.06 & 3.5 \\
e19 & 2 & 2012-10-16 & 0.33 & 0.52 & 0.05 & 4.9 \\
e19 & 3 & 2014-04-19 & 0.82 & 0.57 & 0.08 & 4.9 \\
e19 & 1 & 1992-10-25 & 1.9 & 1.1 & 0.2 & 4.9 \\
e19 & 3 & 2014-04-19 & 0.51 & 0.38 & 0.05 & 5.9 \\
e19 & 2 & 2012-10-16 & 0.41 & 0.48 & 0.04 & 5.9 \\
e19 & 1 & 1992-10-25 & 1.5 & 0.92 & 0.06 & 8.4 \\
e19 & 2 & 2013-03-02 & 1.3 & 1 & 0.1 & 12.6 \\
e19 & 2 & 2013-03-02 & 0.88 & 0.81 & 0.09 & 14.1 \\
e19 & 1 & 1993-03-16 & 0.3 & 1.1 & 0.5 & 22.5 \\
e19 & 2 & 2012-05-31 & -1.9 & 2.2 & 0.8 & 25.0 \\
e19 & 2 & 2012-06-21 & -0 & 6 & 1 & 27.0 \\
e19 & 2 & 2012-08-07 & 0 & 0 & 2 & 29.0 \\
e19 & 2 & 2012-06-21 & 2 & 2 & 1 & 33.0 \\
e19 & 2 & 2012-08-07 & 2 & 3 & 1 & 36.0 \\
e2 & 2 & 2012-10-16 & 6 & 6.4 & 0.2 & 2.5 \\
e2 & 2 & 2012-10-16 & 8.5 & 9.2 & 0.06 & 3.5 \\
e2 & 1 & 1992-10-25 & 14 & 14 & 0.2 & 4.9 \\
e2 & 2 & 2012-10-16 & 13 & 13 & 0.05 & 4.9 \\
e2 & 3 & 2014-04-19 & 11 & 11 & 0.08 & 4.9 \\
e2 & 2 & 2012-10-16 & 14 & 15 & 0.04 & 5.9 \\
e2 & 3 & 2014-04-19 & 14 & 15 & 0.05 & 5.9 \\
e2 & 1 & 1992-10-25 & 27 & 27 & 0.06 & 8.4 \\
e2 & 2 & 2013-03-02 & 98 & 98 & 0.1 & 12.6 \\
e2 & 2 & 2013-03-02 & 1\ee{2} & 1\ee{2} & 0.09 & 14.1 \\
e2 & 1 & 1993-03-16 & 1.6\ee{2} & 1.6\ee{2} & 0.5 & 22.5 \\
e2 & 2 & 2012-05-31 & 1.9\ee{2} & 1.9\ee{2} & 0.8 & 25.0 \\
e2 & 2 & 2012-06-21 & 1.6\ee{2} & 1.7\ee{2} & 1 & 27.0 \\
e2 & 2 & 2012-08-07 & 1.1\ee{2} & 1.2\ee{2} & 2 & 29.0 \\
e2 & 2 & 2012-06-21 & 1.5\ee{2} & 1.5\ee{2} & 1 & 33.0 \\
e2 & 2 & 2012-08-07 & 89 & 92 & 1 & 36.0 \\
e20 & 2 & 2012-10-16 & 0.5 & 1.3 & 0.2 & 2.5 \\
e20 & 2 & 2012-10-16 & 0.52 & 0.74 & 0.06 & 3.5 \\
e20 & 2 & 2012-10-16 & 0.4 & 0.6 & 0.05 & 4.9 \\
e20 & 1 & 1992-10-25 & 1.9 & 1.1 & 0.2 & 4.9 \\
e20 & 3 & 2014-04-19 & -0.09 & 0.49 & 0.08 & 4.9 \\
e20 & 2 & 2012-10-16 & 0.29 & 0.45 & 0.04 & 5.9 \\
e20 & 3 & 2014-04-19 & 0.26 & 0.38 & 0.05 & 5.9 \\
e20 & 1 & 1992-10-25 & 1.3 & 0.34 & 0.06 & 8.4 \\
e20 & 2 & 2013-03-02 & 0.8 & 0.6 & 0.1 & 12.6 \\
e20 & 2 & 2013-03-02 & 0.46 & 0.42 & 0.09 & 14.1 \\
e20 & 1 & 1993-03-16 & 0 & 0.9 & 0.5 & 22.5 \\
e20 & 2 & 2012-05-31 & 0.3 & 1 & 0.8 & 25.0 \\
e20 & 2 & 2012-06-21 & -0 & 2 & 1 & 27.0 \\
e20 & 2 & 2012-08-07 & -2 & 2 & 2 & 29.0 \\
e20 & 2 & 2012-06-21 & -1 & 2 & 1 & 33.0 \\
e20 & 2 & 2012-08-07 & 1 & 2 & 1 & 36.0 \\
e21 & 2 & 2012-10-16 & 0.5 & 1.4 & 0.2 & 2.5 \\
e21 & 2 & 2012-10-16 & 0.52 & 0.76 & 0.06 & 3.5 \\
e21 & 1 & 1992-10-25 & 1.9 & 1 & 0.2 & 4.9 \\
e21 & 2 & 2012-10-16 & 0.25 & 0.46 & 0.05 & 4.9 \\
e21 & 3 & 2014-04-19 & -0.17 & 0.4 & 0.08 & 4.9 \\
e21 & 2 & 2012-10-16 & 0.11 & 0.28 & 0.04 & 5.9 \\
e21 & 3 & 2014-04-19 & 0.12 & 0.24 & 0.05 & 5.9 \\
e21 & 1 & 1992-10-25 & 1.4 & 0.43 & 0.06 & 8.4 \\
e21 & 2 & 2013-03-02 & 0.8 & 0.6 & 0.1 & 12.6 \\
e21 & 2 & 2013-03-02 & 0.41 & 0.38 & 0.09 & 14.1 \\
e21 & 1 & 1993-03-16 & 0.4 & 1.1 & 0.5 & 22.5 \\
e21 & 2 & 2012-05-31 & 0 & 1.3 & 0.8 & 25.0 \\
e21 & 2 & 2012-06-21 & -0 & 2 & 1 & 27.0 \\
e21 & 2 & 2012-08-07 & -2 & 3 & 2 & 29.0 \\
e21 & 2 & 2012-06-21 & -0 & 2 & 1 & 33.0 \\
e21 & 2 & 2012-08-07 & 2 & 3 & 1 & 36.0 \\
e22 & 2 & 2012-10-16 & 0.2 & 1.1 & 0.2 & 2.5 \\
e22 & 2 & 2012-10-16 & 0.1 & 0.33 & 0.06 & 3.5 \\
e22 & 3 & 2014-04-19 & -0.27 & 0.31 & 0.08 & 4.9 \\
e22 & 1 & 1992-10-25 & 2.1 & 1.2 & 0.2 & 4.9 \\
e22 & 2 & 2012-10-16 & 0.12 & 0.33 & 0.05 & 4.9 \\
e22 & 3 & 2014-04-19 & 0.16 & 0.25 & 0.05 & 5.9 \\
e22 & 2 & 2012-10-16 & 0.12 & 0.28 & 0.04 & 5.9 \\
e22 & 1 & 1992-10-25 & 1.5 & 0.46 & 0.06 & 8.4 \\
e22 & 2 & 2013-03-02 & 0.5 & 0.3 & 0.1 & 12.6 \\
e22 & 2 & 2013-03-02 & 0.32 & 0.24 & 0.09 & 14.1 \\
e22 & 1 & 1993-03-16 & 0.4 & 1 & 0.5 & 22.5 \\
e22 & 2 & 2012-05-31 & 0.5 & 2 & 0.8 & 25.0 \\
e22 & 2 & 2012-06-21 & -0 & 1 & 1 & 27.0 \\
e22 & 2 & 2012-08-07 & 0 & 3 & 2 & 29.0 \\
e22 & 2 & 2012-06-21 & 0 & 2 & 1 & 33.0 \\
e22 & 2 & 2012-08-07 & 0 & 2 & 1 & 36.0 \\
e23 & 2 & 2012-10-16 & -0.1 & 1 & 0.2 & 2.5 \\
e23 & 2 & 2012-10-16 & 0.29 & 0.39 & 0.06 & 3.5 \\
e23 & 1 & 1992-10-25 & 1.5 & 1 & 0.2 & 4.9 \\
e23 & 2 & 2012-10-16 & 0 & 0.18 & 0.05 & 4.9 \\
e23 & 3 & 2014-04-19 & 0.09 & 0.25 & 0.08 & 4.9 \\
e23 & 2 & 2012-10-16 & 0.1 & 0.2 & 0.04 & 5.9 \\
e23 & 3 & 2014-04-19 & 0.04 & 0.13 & 0.05 & 5.9 \\
e23 & 1 & 1992-10-25 & 1 & 0.35 & 0.06 & 8.4 \\
e23 & 2 & 2013-03-02 & 0.5 & 0.4 & 0.1 & 12.6 \\
e23 & 2 & 2013-03-02 & 0.27 & 0.28 & 0.09 & 14.1 \\
e23 & 1 & 1993-03-16 & -0.5 & 0.9 & 0.5 & 22.5 \\
e23 & 2 & 2012-05-31 & 0.5 & 1.3 & 0.8 & 25.0 \\
e23 & 2 & 2012-06-21 & 1 & 3 & 1 & 27.0 \\
e23 & 2 & 2012-08-07 & 2 & 3 & 2 & 29.0 \\
e23 & 2 & 2012-06-21 & 0 & 1 & 1 & 33.0 \\
e23 & 2 & 2012-08-07 & 1 & 2 & 1 & 36.0 \\
e3 & 2 & 2012-10-16 & 4.8 & 5.1 & 0.2 & 2.5 \\
e3 & 2 & 2012-10-16 & 6 & 6.3 & 0.06 & 3.5 \\
e3 & 1 & 1992-10-25 & 8 & 7.9 & 0.2 & 4.9 \\
e3 & 2 & 2012-10-16 & 7.3 & 7.6 & 0.05 & 4.9 \\
e3 & 3 & 2014-04-19 & 6.6 & 6.8 & 0.08 & 4.9 \\
e3 & 2 & 2012-10-16 & 7.1 & 7.3 & 0.04 & 5.9 \\
e3 & 3 & 2014-04-19 & 6.6 & 6.9 & 0.05 & 5.9 \\
e3 & 1 & 1992-10-25 & 6.9 & 6.3 & 0.06 & 8.4 \\
e3 & 2 & 2013-03-02 & 15 & 16 & 0.1 & 12.6 \\
e3 & 2 & 2013-03-02 & 14 & 14 & 0.09 & 14.1 \\
e3 & 1 & 1993-03-16 & 9.7 & 12 & 0.5 & 22.5 \\
e3 & 2 & 2012-05-31 & 10 & 12 & 0.8 & 25.0 \\
e3 & 2 & 2012-06-21 & 8 & 11 & 1 & 27.0 \\
e3 & 2 & 2012-08-07 & 5 & 9 & 2 & 29.0 \\
e3 & 2 & 2012-06-21 & 5 & 8 & 1 & 33.0 \\
e3 & 2 & 2012-08-07 & 3 & 7 & 1 & 36.0 \\
e4 & 2 & 2012-10-16 & 1.6 & 1.8 & 0.2 & 2.5 \\
e4 & 2 & 2012-10-16 & 2.3 & 2.6 & 0.06 & 3.5 \\
e4 & 2 & 2012-10-16 & 3.7 & 4 & 0.05 & 4.9 \\
e4 & 1 & 1992-10-25 & 3.6 & 3.6 & 0.2 & 4.9 \\
e4 & 3 & 2014-04-19 & 3.3 & 3.6 & 0.08 & 4.9 \\
e4 & 2 & 2012-10-16 & 4.2 & 4.4 & 0.04 & 5.9 \\
e4 & 3 & 2014-04-19 & 4.3 & 4.6 & 0.05 & 5.9 \\
e4 & 1 & 1992-10-25 & 2.8 & 2.5 & 0.06 & 8.4 \\
e4 & 2 & 2013-03-02 & 9.1 & 9.4 & 0.1 & 12.6 \\
e4 & 2 & 2013-03-02 & 8.8 & 9.2 & 0.09 & 14.1 \\
e4 & 1 & 1993-03-16 & 6.9 & 8.1 & 0.5 & 22.5 \\
e4 & 2 & 2012-05-31 & 8.5 & 9.5 & 0.8 & 25.0 \\
e4 & 2 & 2012-06-21 & 8 & 10 & 1 & 27.0 \\
e4 & 2 & 2012-08-07 & 4 & 6 & 2 & 29.0 \\
e4 & 2 & 2012-06-21 & 9 & 12 & 1 & 33.0 \\
e4 & 2 & 2012-08-07 & 3 & 6 & 1 & 36.0 \\
e5 & 2 & 2012-10-16 & 0.1 & 4.1 & 0.2 & 2.5 \\
e5 & 2 & 2012-10-16 & 4.9 & 6.5 & 0.06 & 3.5 \\
e5 & 3 & 2014-04-19 & 6.5 & 6.8 & 0.08 & 4.9 \\
e5 & 1 & 1992-10-25 & 4.8 & 5.7 & 0.2 & 4.9 \\
e5 & 2 & 2012-10-16 & 7.8 & 8.1 & 0.05 & 4.9 \\
e5 & 2 & 2012-10-16 & 8 & 8.5 & 0.04 & 5.9 \\
e5 & 3 & 2014-04-19 & 8.1 & 9 & 0.05 & 5.9 \\
e5 & 1 & 1992-10-25 & 6.9 & 6.8 & 0.06 & 8.4 \\
e5 & 2 & 2013-03-02 & 24 & 25 & 0.1 & 12.6 \\
e5 & 2 & 2013-03-02 & 24 & 25 & 0.09 & 14.1 \\
e5 & 1 & 1993-03-16 & 15 & 17 & 0.5 & 22.5 \\
e5 & 2 & 2012-05-31 & 30 & 30 & 0.8 & 25.0 \\
e5 & 2 & 2012-06-21 & 25 & 26 & 1 & 27.0 \\
e5 & 2 & 2012-08-07 & 18 & 20 & 2 & 29.0 \\
e5 & 2 & 2012-06-21 & 20 & 22 & 1 & 33.0 \\
e5 & 2 & 2012-08-07 & 15 & 16 & 1 & 36.0 \\
e6 & 2 & 2012-10-16 & 3.6 & 5.5 & 0.2 & 2.5 \\
e6 & 2 & 2012-10-16 & 4.1 & 4.2 & 0.06 & 3.5 \\
e6 & 3 & 2014-04-19 & 2.5 & 2.2 & 0.08 & 4.9 \\
e6 & 1 & 1992-10-25 & 4.1 & 4.8 & 0.2 & 4.9 \\
e6 & 2 & 2012-10-16 & 2.5 & 2.6 & 0.05 & 4.9 \\
e6 & 3 & 2014-04-19 & 0.64 & 1.7 & 0.05 & 5.9 \\
e6 & 2 & 2012-10-16 & 1.5 & 1.6 & 0.04 & 5.9 \\
e6 & 1 & 1992-10-25 & 4.2 & 4.1 & 0.06 & 8.4 \\
e6 & 2 & 2013-03-02 & 3.4 & 3.7 & 0.1 & 12.6 \\
e6 & 2 & 2013-03-02 & 2.7 & 2.9 & 0.09 & 14.1 \\
e6 & 1 & 1993-03-16 & 0.9 & 2.1 & 0.5 & 22.5 \\
e6 & 2 & 2012-05-31 & 1.3 & 2.3 & 0.8 & 25.0 \\
e6 & 2 & 2012-06-21 & 1 & 3 & 1 & 27.0 \\
e6 & 2 & 2012-08-07 & 1 & 2 & 2 & 29.0 \\
e6 & 2 & 2012-06-21 & 1 & 2 & 1 & 33.0 \\
e6 & 2 & 2012-08-07 & 1 & 1 & 1 & 36.0 \\
e7 & 2 & 2012-10-16 & 0.5 & 0.7 & 0.2 & 2.5 \\
e7 & 2 & 2012-10-16 & 0.03 & 0.18 & 0.06 & 3.5 \\
e7 & 2 & 2012-10-16 & 0.09 & 0.11 & 0.05 & 4.9 \\
e7 & 3 & 2014-04-19 & 0.29 & 0.18 & 0.08 & 4.9 \\
e7 & 1 & 1992-10-25 & 0.5 & 0.7 & 0.2 & 4.9 \\
e7 & 2 & 2012-10-16 & 0.06 & 0.09 & 0.04 & 5.9 \\
e7 & 3 & 2014-04-19 & 0.14 & 0.11 & 0.05 & 5.9 \\
e7 & 1 & 1992-10-25 & 0.37 & 0.29 & 0.06 & 8.4 \\
e7 & 2 & 2013-03-02 & 0 & 0.2 & 0.1 & 12.6 \\
e7 & 2 & 2013-03-02 & 0.02 & 0.17 & 0.09 & 14.1 \\
e7 & 1 & 1993-03-16 & -0.1 & 1.7 & 0.5 & 22.5 \\
e7 & 2 & 2012-05-31 & 0.8 & 1.2 & 0.8 & 25.0 \\
e7 & 2 & 2012-06-21 & 1 & 2 & 1 & 27.0 \\
e7 & 2 & 2012-08-07 & 2 & 6 & 2 & 29.0 \\
e7 & 2 & 2012-06-21 & - & - & 1 & 33.0 \\
e7 & 2 & 2012-08-07 & - & - & 1 & 36.0 \\
e8n & 2 & 2012-10-16 & 1.3 & 1.4 & 0.2 & 2.5 \\
e8n & 2 & 2012-10-16 & 0.86 & 1.1 & 0.06 & 3.5 \\
e8n & 1 & 1992-10-25 & 1.5 & 1.5 & 0.2 & 4.9 \\
e8n & 2 & 2012-10-16 & 0.81 & 1.1 & 0.05 & 4.9 \\
e8n & 3 & 2014-04-19 & 0.94 & 1.2 & 0.08 & 4.9 \\
e8n & 2 & 2012-10-16 & 0.92 & 1.1 & 0.04 & 5.9 \\
e8n & 3 & 2014-04-19 & 0.77 & 1 & 0.05 & 5.9 \\
e8n & 1 & 1992-10-25 & 1.6 & 1.1 & 0.06 & 8.4 \\
e8n & 2 & 2013-03-02 & 2.3 & 2.6 & 0.1 & 12.6 \\
e8n & 2 & 2013-03-02 & 2.4 & 2.7 & 0.09 & 14.1 \\
e8n & 1 & 1993-03-16 & 1.2 & 2.4 & 0.5 & 22.5 \\
e8n & 2 & 2012-05-31 & 4.7 & 5.7 & 0.8 & 25.0 \\
e8n & 2 & 2012-06-21 & 4 & 6 & 1 & 27.0 \\
e8n & 2 & 2012-08-07 & 3 & 5 & 2 & 29.0 \\
e8n & 2 & 2012-06-21 & 4 & 6 & 1 & 33.0 \\
e8n & 2 & 2012-08-07 & 4 & 7 & 1 & 36.0 \\
e8s & 2 & 2012-10-16 & 1.3 & 1.3 & 0.2 & 2.5 \\
e8s & 2 & 2012-10-16 & 0.86 & 1.1 & 0.06 & 3.5 \\
e8s & 2 & 2012-10-16 & 0.89 & 1.2 & 0.05 & 4.9 \\
e8s & 3 & 2014-04-19 & 0.94 & 1.2 & 0.08 & 4.9 \\
e8s & 1 & 1992-10-25 & 1.5 & 1.5 & 0.2 & 4.9 \\
e8s & 2 & 2012-10-16 & 1.2 & 1.4 & 0.04 & 5.9 \\
e8s & 3 & 2014-04-19 & 1.1 & 1.3 & 0.05 & 5.9 \\
e8s & 1 & 1992-10-25 & 1.6 & 1.1 & 0.06 & 8.4 \\
e8s & 2 & 2013-03-02 & 2.3 & 2.6 & 0.1 & 12.6 \\
e8s & 2 & 2013-03-02 & 2.1 & 2.4 & 0.09 & 14.1 \\
e8s & 1 & 1993-03-16 & 0.8 & 2.2 & 0.5 & 22.5 \\
e8s & 2 & 2012-05-31 & 3.7 & 4.8 & 0.8 & 25.0 \\
e8s & 2 & 2012-06-21 & 3 & 4 & 1 & 27.0 \\
e8s & 2 & 2012-08-07 & 1 & 3 & 2 & 29.0 \\
e8s & 2 & 2012-06-21 & 1 & 2 & 1 & 33.0 \\
e8s & 2 & 2012-08-07 & 2 & 5 & 1 & 36.0 \\
e9 & 2 & 2012-10-16 & 2.6 & 3 & 0.2 & 2.5 \\
e9 & 2 & 2012-10-16 & 2.2 & 2.5 & 0.06 & 3.5 \\
e9 & 3 & 2014-04-19 & 1.9 & 1.9 & 0.08 & 4.9 \\
e9 & 2 & 2012-10-16 & 1.9 & 2.1 & 0.05 & 4.9 \\
e9 & 1 & 1992-10-25 & 2.4 & 2.1 & 0.2 & 4.9 \\
e9 & 2 & 2012-10-16 & 1.5 & 1.6 & 0.04 & 5.9 \\
e9 & 3 & 2014-04-19 & 1.3 & 1.5 & 0.05 & 5.9 \\
e9 & 1 & 1992-10-25 & 1.8 & 1.2 & 0.06 & 8.4 \\
e9 & 2 & 2013-03-02 & 2.8 & 2.8 & 0.1 & 12.6 \\
e9 & 2 & 2013-03-02 & 2.2 & 2.5 & 0.09 & 14.1 \\
e9 & 1 & 1993-03-16 & 0.3 & 2.4 & 0.5 & 22.5 \\
e9 & 2 & 2012-05-31 & 1.9 & 2.6 & 0.8 & 25.0 \\
e9 & 2 & 2012-06-21 & 3 & 5 & 1 & 27.0 \\
e9 & 2 & 2012-08-07 & 2 & 4 & 2 & 29.0 \\
e9 & 2 & 2012-06-21 & 2 & 3 & 1 & 33.0 \\
e9 & 2 & 2012-08-07 & 3 & 6 & 1 & 36.0 \\
\hline

\end{longtable}
\twocolumn

\clearpage
\section{Point Source Photometry Catalog}
\label{sec:SEDs}
The full version of the point source photometry catalog is available in digital
form from
\url{https://github.com/keflavich/paper_w51_evla/blob/master/tables/EVLA_VLA_PointSourcePhotometry.ipac}.
A sample is shown in Table \ref{tab:contsrcs}.
\todo{To the journal: we would like this to be hosted by CDS.}

Figure \ref{fig:d4sed} shows cutout images with apertures for the source d4e.
Additional figures for all candidates are included in a supplemental document.

\Figure{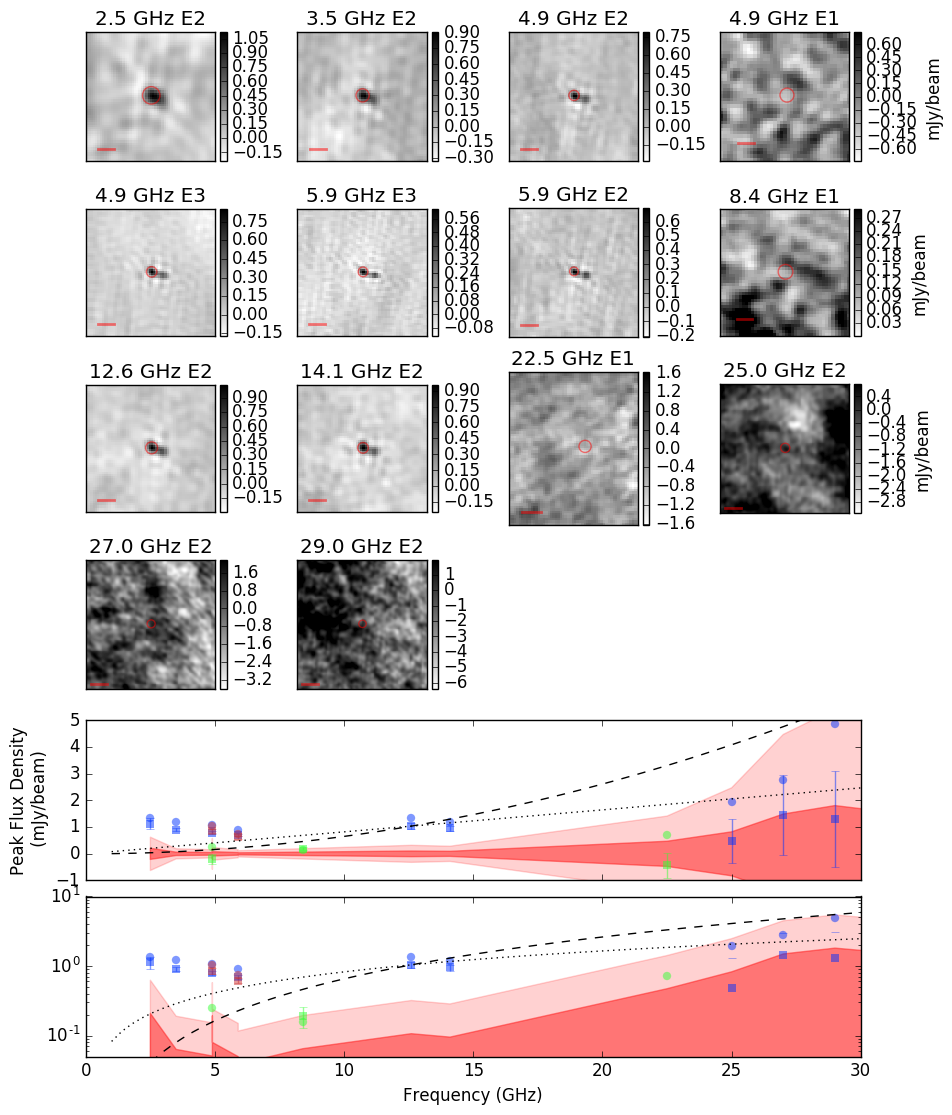}
{Cutouts of the source W51 d4e and its SED.
The red circle shows the source aperture used to measure the peak and total
flux density.  The red line shows a 1\arcsec scalebar.  The Epochs are indicated as E1,
E2, and E3.
In the SED plot, the dark and light red show the 1-$\sigma$ and 3-$\sigma$
noise levels.  The SED plot is shown twice, first linear-linear, and second
linear-log.
Green points are from Epoch 1, blue points are Epoch 2, and red points are
Epoch 3.  Squares show the peak flux density, circles show the
`background-subtracted' flux density - i.e., the peak minus the minimum value
in the map.
The dotted curve shows a linear SED ($S_\nu \propto \nu$) and the dashed shows
a power-law curve with $\alpha=2$, i.e. $S_\nu \propto \nu^2$.  Both curves are
normalized to go through the 12.1 GHz Epoch 2 data point.  In this case, all of
the Epoch 1 (green) points are within the noise, while all of the Epoch 2 and 3
(blue and red) points are detected at $>5$-$\sigma$ below 15 GHz.
}
{fig:d4sed}{0.5}{6.5in}

\section{Observation metadata}
\label{sec:obsmeta}
The observation metadata is summarized in Table \ref{tab:obs_meta}.

\begin{table*}[htp]
\caption{Observation Metadata}
\begin{tabular}{llllllll}
\label{tab:obs_meta}
Band & Program & TOS & Config & Reference & BW & Date & Notes \\
$\mathrm{}$ & $\mathrm{}$ & $\mathrm{s}$ & $\mathrm{}$ & $\mathrm{}$ & $\mathrm{MHz}$ & $\mathrm{}$ & $\mathrm{}$ \\
\hline
29.0 GHz Epoch 2 & 12A-274 & 5481 & B & Goddi2015a,Goddi2016a & 64 & 12-Aug-07 & - \\
33.0 GHz Epoch 2 & 12A-274 & 5481 & B & Goddi2015a,Goddi2016a & 64 & 12-Aug-07 & - \\
36.0 GHz Epoch 2 & 12A-274 & 5481 & B & Goddi2015a,Goddi2016a & 64 & 12-Aug-07 & - \\
2.5 GHz Epoch 2 & 12B-365 & 778 & A & - & 1024 & 12-Dec-09 & - \\
3.5 GHz Epoch 2 & 12B-365 & 778 & A & - & 1024 & 12-Dec-09 & - \\
2.5 GHz Epoch 2 & 12B-365 & 658 & A & - & 1024 & 12-Dec-24 & - \\
3.5 GHz Epoch 2 & 12B-365 & 658 & A & - & 1024 & 12-Dec-24 & - \\
29.0 GHz Epoch 2 & 12A-274 & 5585 & B & Goddi2015a,Goddi2016a & 64 & 12-Jun-21 & - \\
33.0 GHz Epoch 2 & 12A-274 & 5585 & B & Goddi2015a,Goddi2016a & 64 & 12-Jun-21 & - \\
36.0 GHz Epoch 2 & 12A-274 & 5585 & B & Goddi2015a,Goddi2016a & 64 & 12-Jun-21 & - \\
25.0 GHz Epoch 2 & 12A-274 & 3829 & B & Goddi2015a,Goddi2016a & 64 & 12-May-31 & - \\
27.0 GHz Epoch 2 & 12A-274 & 3829 & B & Goddi2015a,Goddi2016a & 64 & 12-May-31 & - \\
5.9 GHz Epoch 2 & 12B-365 & 777 & A & - & 1024 & 12-Nov-17 & - \\
4.9 GHz Epoch 2 & 12B-365 & 777 & A & - & 1024 & 12-Nov-17 & - \\
5.9 GHz Epoch 2 & 12B-365 & 658 & A & - & 1024 & 12-Nov-24 & - \\
4.9 GHz Epoch 2 & 12B-365 & 658 & A & - & 1024 & 12-Nov-24 & - \\
5.9 GHz Epoch 2 & 12B-365 & 2154 & A & - & 1024 & 12-Oct-16 & - \\
4.9 GHz Epoch 2 & 12B-365 & 2154 & A & - & 1024 & 12-Oct-16 & - \\
5.9 GHz Epoch 2 & 12B-365 & 777 & A & - & 1024 & 12-Oct-29 & - \\
4.9 GHz Epoch 2 & 12B-365 & 777 & A & - & 1024 & 12-Oct-29 & - \\
4.9 GHz Epoch 3 & 13A-064 & 4069 & C & - & 1024 & 13-Jun-04 & - \\
5.9 GHz Epoch 3 & 13A-064 & 4069 & C & - & 1024 & 13-Jun-04 & - \\
14.1 GHz Epoch 2 & 13A-064 & 4149 & D & - & 1024 & 13-Mar-02 & - \\
12.6 GHz Epoch 2 & 13A-064 & 4149 & D & - & 1024 & 13-Mar-02 & - \\
14.1 GHz Epoch 2 & 13A-064 & 14704 & B & - & 1024 & 13-Oct-01 & - \\
12.6 GHz Epoch 2 & 13A-064 & 14704 & B & - & 1024 & 13-Oct-01 & - \\
4.9 GHz Epoch 3 & 13A-064 & 15952 & A & - & 1024 & 14-Apr-19 & - \\
5.9 GHz Epoch 3 & 13A-064 & 15952 & A & - & 1024 & 14-Apr-19 & - \\
4.9 GHz Epoch 1 & AD0128 & 6830 & B & Mehringer1994a & 25 & 84-Feb-06 & - \\
4.9 GHz Epoch 1 & AM0367 & 10410 & C & Mehringer1994a & 100 & 92-Apr-23 & - \\
4.9 GHz Epoch 1 & AM0367 & 6240 & D & Mehringer1994a & 100 & 92-Aug-04 & - \\
4.9 GHz Epoch 1 & AM0367 & 4800 & D & Mehringer1994a & 100 & 92-Aug-04 & - \\
8.4 GHz Epoch 1 & AP0242 & 2640 & B & Gaume1993a & 50 & 92-Oct-24 & Uncertain provenance \\
4.9 GHz Epoch 1 & AM0374 & 6860 & A & Mehringer1994a & 50 & 92-Oct-25 & - \\
4.9 GHz Epoch 1 & AM0374 & 3350 & A & Mehringer1994a & 50 & 92-Oct-26 & - \\
22.5 GHz Epoch 1 & AM0374 & 6360 & B & Gaume1993a & 100 & 93-Mar-16 & - \\
\hline
\end{tabular}

\end{table*}

\section{Velocity Field in W51e2}
\label{sec:vfield}
Many works have examined the velocity field of the gas in W51e2
\citep{Zhang1997a,Keto2008b,Shi2010b,Shi2010a,Goddi2016a}.  For comparison, we
show the velocity field of a $6\arcsec\times6\arcsec$ region centered on W51e2
in Figure
\ref{fig:w51e2velofield}.

\FigureTwo
{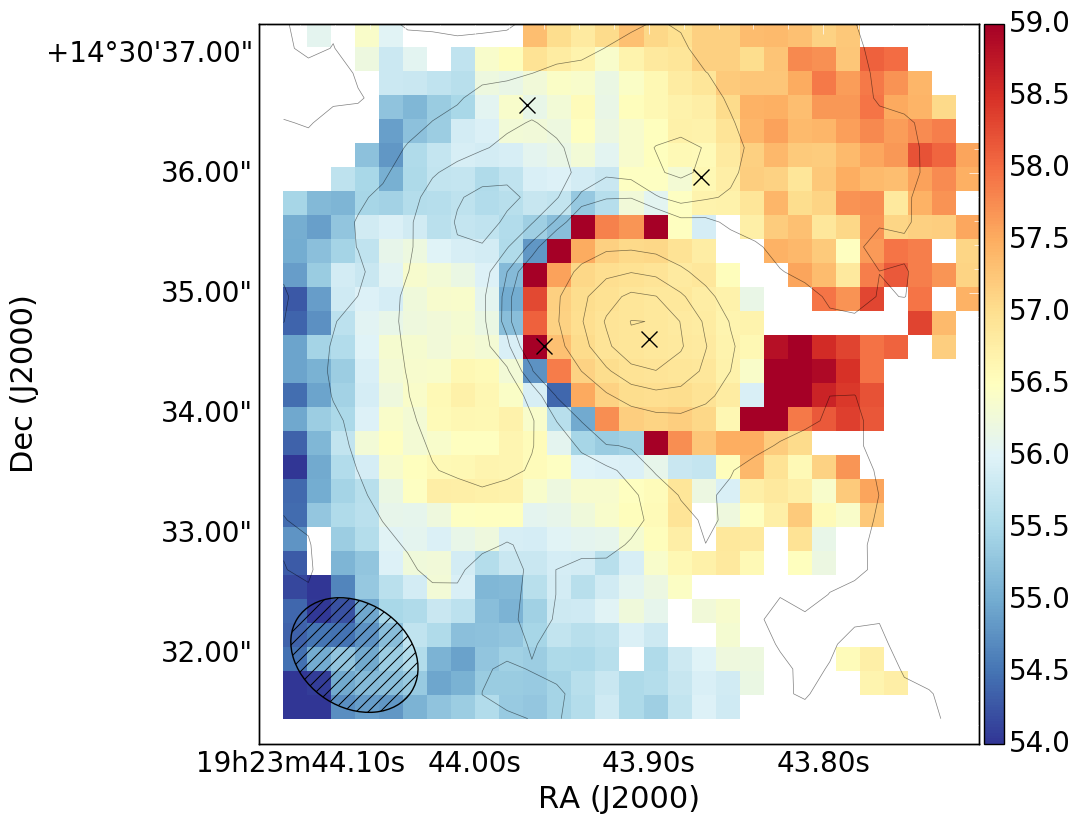}
{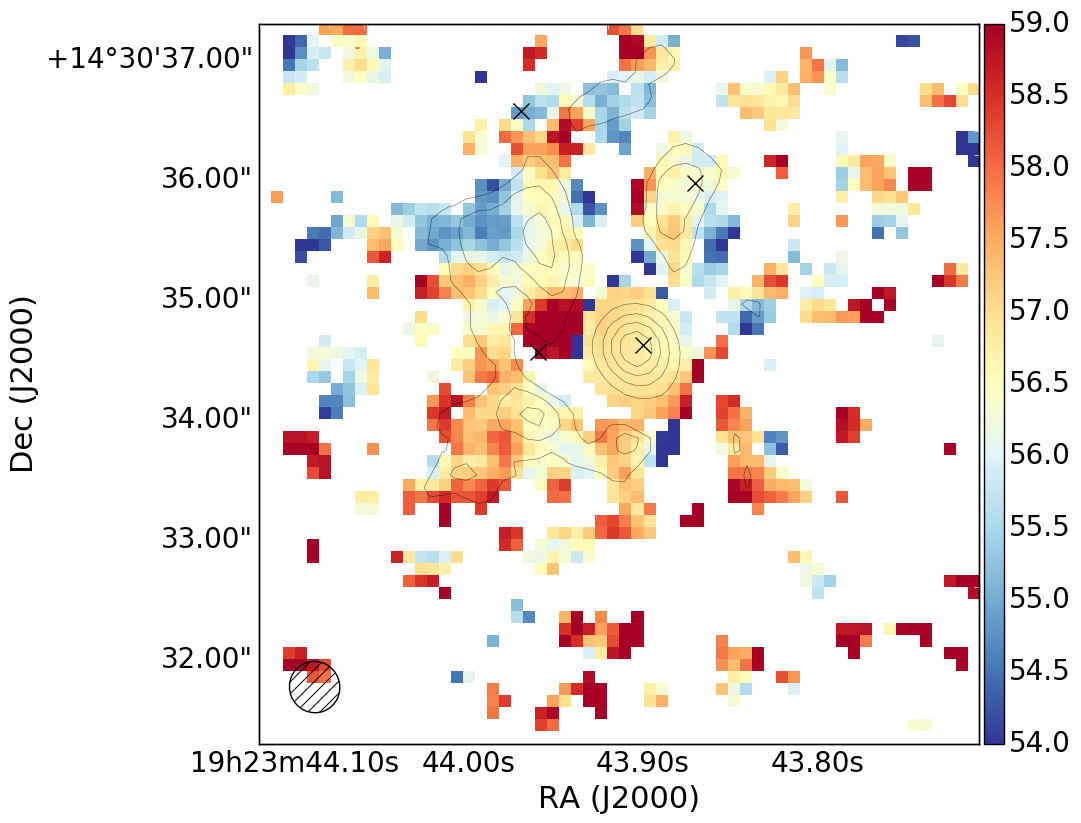}
{Velocity (moment 1) maps of \ortho \twotwo in the W51e2 region using (a) the
naturally weighted map and (b) the Briggs $w=0.5$ weighted map.  The X's mark
\citet{Shi2010a} cores.   
The contours show the moment 0 maps integrated from 50 to 63 \kms at
[-0.4,-0.3,-0.2,-0.1, 0.010, 0.020, 0.030, 0.040] Jy \kms (natural) and
[-0.28,-0.21,-0.14,-0.07, 0.007, 0.0105, 0.014] Jy \kms (Briggs).
}
{fig:w51e2velofield}{1}{3.5in}

\end{document}